\documentclass{ieeeaccess}
\usepackage{cite}
\usepackage{amsmath,amssymb,amsfonts}
\usepackage{algorithmic}
\usepackage{graphicx}
\usepackage{textcomp}

\usepackage{siunitx}
\usepackage{booktabs}
\usepackage{nicefrac}
\usepackage{multirow}
\usepackage{gensymb}
\usepackage{enumitem}
\usepackage{tablefootnote}
\usepackage{diagbox}
\usepackage{amssymb}
\usepackage{pifont}
\usepackage[dvipsnames]{xcolor}

\DeclareSIUnit \dBm {dBm}
\DeclareSIUnit \dB {dB}
\DeclareSIUnit \dBi {dBi}
\DeclareSIUnit \Kbps {Kbps}
\DeclareSIUnit \Mbps {Mbps}
\DeclareSIUnit \Gbps {Gbps}
\DeclareSIUnit \Tbps {Tbps}
\DeclareSIUnit \kBps {kBps}
\DeclareSIUnit \MBps {MBps}
\DeclareSIUnit \GBps {GBps}
\DeclareSIUnit \TBps {TBps}

\newcommand{\cmark}{\ding{51}}
\newcommand{\xmark}{\ding{55}}
\newcommand{\ocircle}{\ding{109}}%
\newcolumntype{N}{>{\centering\arraybackslash}m{3cm}}
\newcolumntype{K}{>{\centering\arraybackslash}m{0.8cm}}
\newcolumntype{P}[1]{>{\centering\arraybackslash}p{#1}}
\newcolumntype{L}[1]{>{\arraybackslash}p{#1}}
\newcolumntype{M}[1]{>{\centering\arraybackslash}m{#1}}

\def\BibTeX{{\rm B\kern-.05em{\sc i\kern-.025em b}\kern-.08em
    T\kern-.1667em\lower.7ex\hbox{E}\kern-.125emX}}
\begin{document}
\bstctlcite{IEEEexample:BSTcontrol}
\history{Date of publication xxxx 00, 0000, date of current version xxxx 00, 0000.}
\doi{10.1109/ACCESS.2023.0322000}

\title{AI-Native Multi-Access Future Networks - The REASON Architecture}
\author{\uppercase{Konstantinos Katsaros}\authorrefmark{1},~\IEEEmembership{Member,~IEEE},
\uppercase{Ioannis Mavromatis}\authorrefmark{1},
\uppercase{Kostantinos Antonakoglou}\authorrefmark{1},
\uppercase{Saptarshi Ghosh}\authorrefmark{1},
\uppercase{Dritan Kaleshi}\authorrefmark{1},
\uppercase{Toktam Mahmoodi}\authorrefmark{2},
\uppercase{Hamid Asgari}\authorrefmark{3},
\uppercase{Anastasios Karousos}\authorrefmark{4},
\uppercase{Iman Tavakkolnia}\authorrefmark{5},
\uppercase{Hossein Safi}\authorrefmark{5},
\uppercase{Harald Hass}\authorrefmark{5},
\uppercase{Constantinos Vrontos}\authorrefmark{6},
\uppercase{Amin Emami}\authorrefmark{6},
\uppercase{Juan Parra Ullauri}\authorrefmark{6},
\uppercase{Shadi Moazzeni}\authorrefmark{6},
and~\uppercase{Dimitra Simeonidou}\authorrefmark{6}}

\address[1]{Digital Catapult, NW1 2RA, London, UK}
\address[2]{Kings College London, Department of Engineering, Strand Building, Strand Campus, Strand, London, WC2R 2LS, UK}
\address[3]{Thales Research and Technology, 350 Longwater Avenue, Reading, Berkshire RG2 6GF, UK}
\address[4]{Real Wireless, Pulborough, West Sussex, RH20 4XB, United Kingdom}
\address[5]{University of Cambridge, Department of Engineering, Trumpington Street, Cambridge CB2 1PZ, UK}
\address[6]{University of Bristol, Faculty of Engineering, Woodland Road, Clifton, Bristol, BS8 1UB, UK}
\tfootnote{This work is a contribution by Project REASON, a UK Government funded project under the Future Open Networks Research Challenge (FONRC) sponsored by the Department of Science Innovation and Technology (DSIT).}

\markboth
{K. Katsaros \headeretal: Preparation of Papers for IEEE TRANSACTIONS and JOURNALS}
{K. Katsaros \headeretal: Preparation of Papers for IEEE TRANSACTIONS and JOURNALS}
\corresp{Corresponding authors: \{Kostas.Katsaros, Ioannis.Mavromatis\}@digicatapult.org.uk.}

\begin{abstract}
{
The development of the sixth generation of communication networks (6G) has been gaining momentum over the past years, with a target of being introduced by 2030. Several initiatives worldwide are developing innovative solutions and setting the direction for the key features of these networks. Some common emerging themes are the tight integration of AI, the convergence of multiple access technologies and sustainable operation, aiming to meet stringent performance and societal requirements.

To that end, we are introducing REASON - Realising Enabling Architectures and Solutions for Open Networks. The REASON project aims to address technical challenges in future network deployments, such as E2E service orchestration, sustainability, security and trust management, and policy management, utilising AI-native principles, considering multiple access technologies and cloud-native solutions.

This paper presents REASON's architecture and the identified requirements for future networks. The architecture is meticulously designed for modularity, interoperability, scalability, simplified troubleshooting, flexibility, and enhanced security, taking into consideration current and future standardisation efforts, and the ease of implementation and training. It is structured into four horizontal layers: Physical Infrastructure, Network Service, Knowledge, and End-User Application, complemented by two vertical layers: Management and Orchestration, and E2E Security. This layered approach ensures a robust, adaptable framework to support the diverse and evolving requirements of 6G networks, fostering innovation and facilitating seamless integration of advanced technologies.}
\end{abstract}


\titlepgskip=-21pt

\maketitle

\section{Introduction}
\label{sec:introduction}
\PARstart{T}{he} forthcoming generation of communication networks, colloquially referred to as Sixth Generation (6G), represents a transformative leap beyond the existing paradigms of mobile communications~\cite{6gAndBeyond}. This leap is essential to meet burgeoning demands for individual users and vertical industries, covering both indoor and outdoor environments and spanning multiple sectors and applications like automotive, manufacturing, public safety, eHealth, immersive media, and more~\cite{6gUseCases}.

These applications shift away from conventional Key Performance Indicators (KPIs), like faster data rates, reduced latency, or broader coverage, and move towards more complex classes of enablers for advanced services~\cite{shiftKPitoKVI}. For example, the Fifth Generation (5G) Infrastructure Association (5G IA)~\cite{5gIA} provides a comprehensive list discussing ``integrated sensing and communications'', ``cognition and connected intelligence'', ``trustworthy and sustainable infrastructures'', and more. Similarly, other initiatives, like Hexa-X~\cite{HexaX}, discuss enablers like ``network of networks'', ``global service coverage'', ``extreme experiences'', etc. A comprehensive comparison of the different visions can be found at~\cite{6gVisions}. Overall, all visions converge to a future network that architects a seamlessly interconnected world, where the integration of digital, physical, and human systems unfolds new dimensions of experience, efficiency, and societal transformation.

\begin{table}[!t]
\renewcommand{\arraystretch}{1}
\centering
    \caption{List of Acronyms.}
    \begin{tabular}{lp{0.68\columnwidth}}
        \toprule
        \textbf{Acronym} & \textbf{Description} \\
        \midrule
            5G & 5th Generation \\
            5GPPP & 5G Infrastructure Public Private Partnership \\
            6G & 6th Generation \\
            AI & Artificial Intelligence \\
            AIaaS & AI-as-a-Service \\
            AT &  Access Technology \\
            CIA & Confidentiality, Integrity, and Availability \\
            CI/CD & Continuous Integration / Continuous Delivery \\
            CNF & Container Network Function \\
            CPaaS & Communications-Platform-as-a-Service \\
            DAO & Distributed Autonomous Organisation \\
            DLT & Distributed Ledger Technology \\
            DT & Digital Twin \\
            E2E & End-to-End \\
            eMBB & enhanced Mobile Broadband \\
            FDRL & Federated Deep Reinforcement Learning \\
            FL & Federated Learning \\
            FSO & Free Space Optical \\
            GEO & Geostationary Orbit \\
            IaC & Infrastructure-as-Code \\
            IMT & International Mobile Telecommunications \\
            IoT & Internet of Things \\
            IT & Information Technology \\
            ITU-T & ITU Telecommunication \\
            KPI & Key Performance Indicators \\
            KVI & Key Value Indicators \\
            LEO & Low Earth Orbit \\
            LiFi & Light Fidelity \\
            MANO & Management and Orchestration \\
            \multirow{2}{*}{mATRIC} & Multi-access Technology \\
            & Real-Time Intelligent Controller \\
            MEO & Medium Earth Orbit \\
            mMTC & massive Machine Type Communication \\
            MLOps & Machine Learning Operations \\
            NaC & Network-as-code \\
            NF & Network Function \\
            NFV & Network Function Virtualisation \\
            NGMN & Next Generation Mobile Networks Alliance \\
            NIO & Network Intelligence Orchestration \\
            NPN & Non-Public Network \\
            NRE & Network Resource Elasticity \\
            NTN & Non-Terrestrial Network \\
            OPEX & Operational Cost \\
            O-RAN & Open Radio Access Network \\
            OT & Operational Technology \\
            QoS & Quality of Service \\
            QoE & Quality of Experience \\
            RAN & Radio Access Network \\
            RAT & Radio Access Technologies \\
            \multirow{2}{*}{REASON} & Realising Enabling Architectures and \\
            & Solutions for Open Networks \\
            RINA & Recursive InterNetwork Architecture \\
            RF & Radio Frequency \\
            SBA & Service-based Architecture \\
            SDG & Sustainable Development Goals \\
            SLA & Service-Level Agreement \\
            SFC & Service Function Chain \\
            UPF & User Plane Function \\
            URLLC & Ultra-Reliable and Low-Latency Communication \\
            VIM & Virtual Infrastructure Manager \\
            VM & Virtual Machine \\
            VNF & Virtual Network Function \\
            XAI & Explainable AI \\
        \bottomrule
    \end{tabular}\label{tab:acronyms}
\end{table}

However, shifting from traditional KPIs to a more complex set of enablers is only part of the wider paradigm shift shown in 6G ecosystems. We witness a transition from ``govern by performance'' to ``govern by value''~\cite{shiftKPitoKVI}. In other words, what is needed is not an indication of performance but an indication of value for the developed new ecosystems. This is how the term Key Value Indicators (KVIs) was born~\cite{kviReport}. 6G is expected to not only bring innovative solutions for the above complex problems but also address other societal challenges, like environmental sustainability, ethical implications of technological advancements, and more~\cite{kvi6G}. A comprehensive, collaborative, and multifaceted approach is imperative to ensure progress that benefits all sectors of society and positively contributes to the global community.

The architecture design is one of the most important aspects of each new generation of a communications network system. It is considered its foundation and describes what and how services are provided and how they are integrated. Based on that, this paper presents Realising Enabling Architectures and Solutions for Open Networks (REASON) project\footnote{REASON Project: https://reason-open-networks.ac.uk/} and its envisaged architecture. REASON aims to address the above technical and non-technical challenges and deliver an End-to-End (E2E) reference Open Network blueprint, considering all segments and functions of a future network. REASON will pursue to influence the future technology roadmap, making openness, interoperability and Artificial Intelligence (AI)-native capabilities the default standards in network architectures and systems.

Reference architectures are usually driven by three factors~\cite{referenceArchitecture6G}, i.e., the ``new scenarios and requirements to be delivered'', the ``new technologies that need to be integrated'', and the ``inspiration from existing systems''. This is the scope of this paper. Analysing the challenges identified and several use cases defined, we will provide a comprehensive set of requirements to drive future 6G networks. These requirements will later be used to define a novel reference architecture that enables principles like \textit{interoperability}, \textit{agility}, \textit{sustainability}, \textit{resilience}, and \textit{security}, all key to future Industrial, Entertainment and Smart Cities applications~\cite{keyPrinciples}. Our use case proposition is also expected to demonstrate not only the capabilities of the REASON architecture but also to motivate future scenarios with strong market demand and commercial opportunities.

As an expected outcome, the REASON architecture aims to enable \textit{multi-technology access network} integration to meet the emerging 6G KPIs. Moreover, it will promote \textit{network densification} to ensure support of the described 6G use cases. The architecture will be \textit{AI-native}, enabling service and network optimisations in an E2E fashion. \textit{Cognitive tools} and \textit{task-based networking} are considered integral parts of the architecture, and we will discuss how they are incorporated into our system. Finally, the aspects of \textit{energy-awareness}, \textit{E2E monitoring}, seamless \textit{edge-cloud computing continuum}, \textit{openness}, \textit{interoperability} and \textit{security} will drive many of our architectural decisions, and we explain how the proposed design addresses them.

This paper is structured as follows. Sec.~\ref{sec:background} compares 5G and expected 6G capabilities and comprehensively compares REASON with other future networking projects. The technical and societal challenges identified in future 6G networks and some expected usage scenarios are presented in Sec.~\ref{sec:challenges}. Sec.~\ref{sec:reason_usecases} provides insight into the use cases identified within REASON. Building upon the challenges and use cases, Sec.~\ref{sec:requirements} presents all the requirements that should be addressed from future network architecture and implementation. Following the requirements, the envisaged REASON architecture is detailed in Sec.~\ref{sec:reference_architecture}. Finally, Sec.~\ref{sec:conclusion} provides our final remarks and some future directions within the REASON project. A table summarising all the acronyms can be found in Table~\ref{tab:acronyms}.

\section{Background}\label{sec:background}

As discussed, both KPIs and KVIs will guide the design of 6G systems. Some tentative recommendations were initially presented in this white paper~\cite{rajatheva2020white}, published in 2020. ITU Telecommunication (ITU-T) Standardisation Sector has recently published the official recommendations for 6G in the document named ``International Mobile Telecommunications (IMT) towards 2030 and beyond (IMT-2030)''~\cite{ITU2030}. This report summarises the recommendations for KPIs and KVIs in contrast with 5G and compares them to the capabilities introduced in the IMT-2020~\cite{imt2020} report last updated in January 2021. A more recent white paper from 5GPPP (published in December 2022)~\cite{5gppp2022architecture} extends the above 6G landscape, providing more information on the use case families and the technological enablers for 6G. As the 6G landscape is not just about advancing technology but also focuses on societal transformations, this section touches upon KPIs and KVIs that drive various decisions to be considered for all future networking systems.

\subsection{KPIs in Future Networks}

As discussed in the above-mentioned reports~\cite{ITU2030,5gppp2022architecture}, 6G systems are expected to provide enhanced capabilities necessary to support the new scenarios and use cases introduced. Starting with the quantifiable KPIs, a summary is provided in Table~\ref{tab:kpis}. We identify three key areas of KPIs, these being:
\begin{itemize}
    \item \textbf{Capacity/Throughput}: This category focuses on KPIs related to the system throughput, like the theoretical and perceived data rate, spectral efficiency, etc. The user-perceived values refer to the guaranteed (ubiquitous) service achieved across the considered target coverage area of a mobile device~\cite {rajatheva2020white}.
    \item \textbf{Network Reliability}: This category captures traditional network metrics such as jitter, latency and reliability, critical for any service operation. It is worth mentioning that the jitter is a new KPI introduced in 6G, which was absent in previous releases.
    \item \textbf{System KPIs}: Lastly, this category captures more systemic KPIs, such as energy efficiency, device mobility, positioning, etc. Mobility refers to the seamless transfer between radio nodes belonging to different layers and/or Radio Access Technologies (multi-layer/multi-RAT). 6G introduces the energy efficiency KPI, which was not considered before, to strengthen the long-term sustainability of future networks. Moreover, the positioning, even though considered in 5G, is currently being enhanced and quantified leveraging the reduced latency and jitter and the intelligence introduced in the network~\cite{positioning}.
\end{itemize}

\begin{table}[t]
    \centering
    \caption{The evolution from 5G to 6G systems~\cite{ITU2030}.}
    \begin{tabular}{lll}
    \toprule
    \textbf{Key Performance Indicator} & \textbf{5G} & \textbf{6G} \\
    \midrule
    \multicolumn{3}{l}{\textbf{Capacity/Throughput}} \\
    Peak Data Rate (\SI{}{\Gbps}) & $20$ & $1000$  \\
    User Perceived Data Rate (\SI{}{\Gbps}) & $0.1$ & $1$  \\
    Peak Spectral Efficiency (\SI{}{\bit/\second/\hertz}) & $30$ & $60$  \\
    User Perceived Spectral Efficiency (\SI{}{\bit/\second/\hertz}) & $0.3$ & $3$  \\
    Maximum Channel Bandwidth (\SI{}{\giga\hertz}) & $1$ & $100$  \\
    Area Traffic Capacity (\SI{}{\Mbps/\meter}$^2$) & $10$ & $1000$  \\
    Connection Density (\SI{}{devices/\kilo\meter}$^2$) & $10^6$ & $10^8$  \\
    \midrule
    \multicolumn{3}{l}{\textbf{Network Reliability}} \\
    End-to-end Latency (\SI{}{\milli\second}) & $1$ & $0.1$  \\
    Jitter (\SI{}{\milli\second}) & N/A & $10^{-3}$  \\
    Reliability (Packet Error Rate) & $10^{-5}$ & $10^{-7}$  \\
    \midrule
    \multicolumn{3}{l}{\textbf{System KPIs}} \\
    Energy Efficiency (\SI{}{\tera\bit/\joule}) & N/A & $1$  \\
    Mobility (\SI{}{\kilo\meter/\hour}) & $500$ & $1000$  \\
    Positioning (\SI{}{\cm}) & N/A & $1-10$  \\
    \bottomrule
    \end{tabular}
    \label{tab:kpis}
\end{table}

 \begin{figure}[t]
    \centering
    \includegraphics[width=0.7\columnwidth]{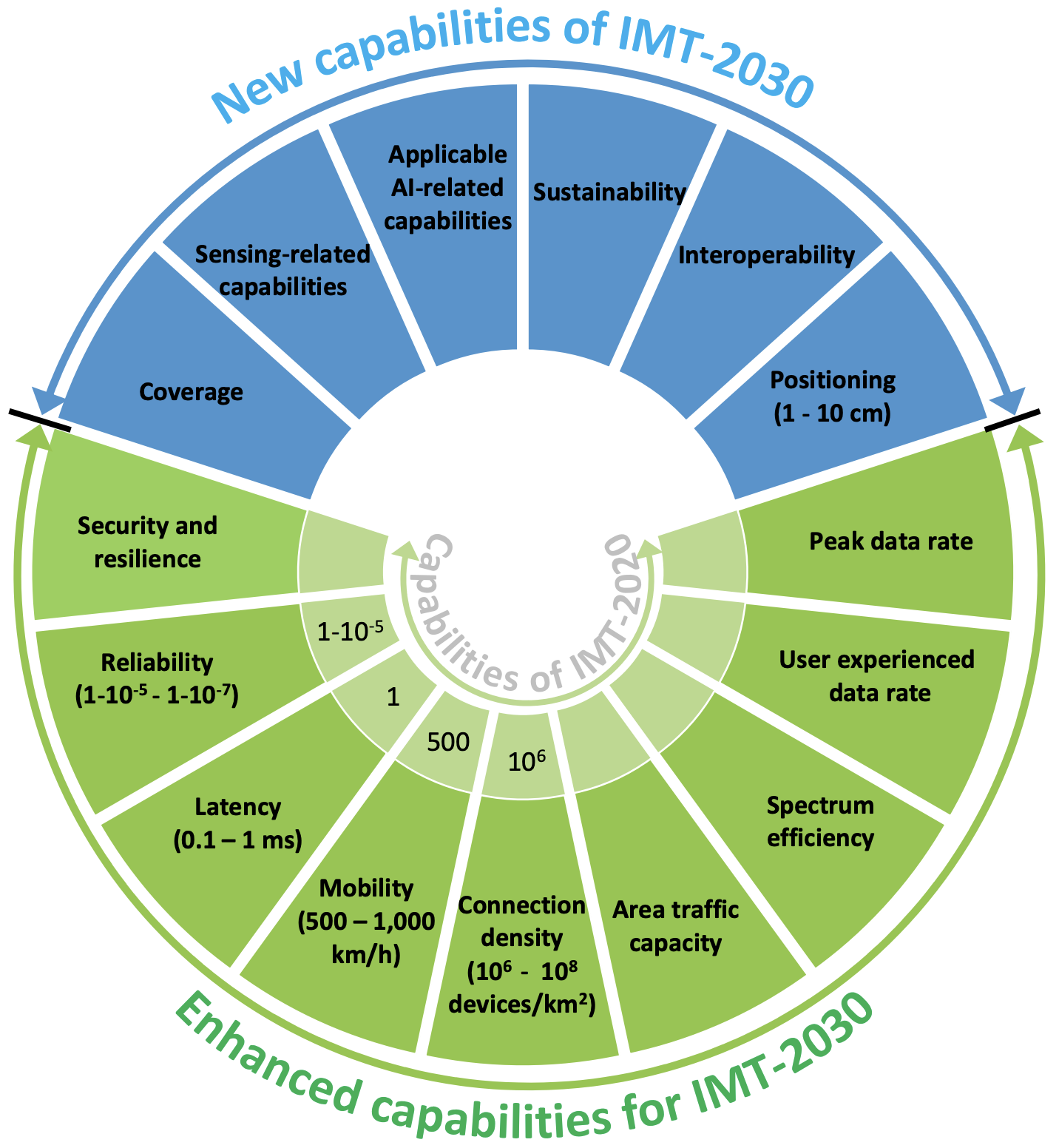}
    \caption{IMT-2030 Capabilities beyond 5G~\cite{ITU2030}.}
    \label{fig:IMT2030Cap}
\end{figure}

Non-quantifiable capabilities are also offered in the proposed future 6G networks, such as \textit{sensing} (range/velocity/angle estimations, object detection, localisation, and more), which is related to the positioning KPI mentioned earlier; \textit{security}, i.e., confidentiality, integrity, and availability of data and information and hardening of networks; \textit{resiliency}, i.e., targeting continuous operation even during disturbances; \textit{sustainability}, focusing on the minimisation of greenhouse gas emissions through the entire lifecycle of a 6G network (the energy efficiency KPI mentioned earlier is a quantifiable metric of sustainability); \textit{interoperability} across radio interfaces and other fixed and satellite networks; and finally AI-native capabilities, across an E2E system. Overall, IMT-2030 defines 15 capabilities required by 6G systems. These capabilities are summarised in Fig.~\ref{fig:IMT2030Cap}. Following the above, the next section lists various important KVIs identified for 6G and future networks and groups them into six thematic areas.

\subsection{KVIs in Future Networks}
The evolution of 6G networks will significantly contribute to societal, economic, and ethical advancements by addressing critical challenges faced in modern life. KVIs go beyond performance metrics, aiming to assess the broader implications of network technologies. In this section, we explore some of the most impactful KVIs, provide real-world examples that illustrate their potential benefits and capture some ethical dimensions/considerations that should be kept in mind while designing a future network.

\paragraph{Digital Inclusion and Equity}
A key KVI for 6G is the ability to promote \textbf{universal access} to high-quality, affordable connectivity. Bridging the digital divide is essential for equal opportunities in education, employment, and healthcare. This concept of digital inclusion not only addresses underserved populations in rural areas but also extends to marginalised urban communities. For example, a deployment of 6G infrastructure in remote regions (e.g., rural Africa, Himalayan areas, etc.) could enable real-time access to telemedicine, benefiting from AI-powered diagnostics and consultations with specialists located in distant urban centres, reducing the disparity in healthcare services between urban and rural populations. Considering the ethical dimensions, 6G connectivity accessible to all raises ethical considerations around the fair distribution of resources, with governments and private sectors needing to collaborate on policies that prevent exclusion based on socio-economic status. Within the REASON project, we investigate ways to support such causes, using multiple Access Technologies (AT) that best fit the needs and constraints (economical, environmental, etc.) and looking to enable up-skilling, thus reducing the digital divide.

\paragraph{Environmental Sustainability}
Sustainability is one of the most pressing global challenges, and 6G can play a pivotal role in reducing the environmental footprint of communications technologies. The integration of \textbf{energy-efficient designs} and \textbf{intelligent resource management} systems will drive the reduction of energy consumption and greenhouse gas emissions across networks. For example, through network energy efficiency improvements, 6G could enable the deployment of green smart cities where IoT devices optimise electricity usage, waste management, and transportation systems. Smart grid deployments, tightly integrated with 6G, could drastically reduce energy waste, particularly in urban centres, with the aim of carbon neutrality. Ethical considerations of such deployments are that such systems, while designed to prioritise long-term environmental benefits, should not cause undue harm to other sectors (e.g., increased e-waste from obsolete technologies). REASON is investigating several methods to reduce energy consumption. More details in Sec.~\ref{sec:energy_sustainability}.

\paragraph{Economic Productivity and Innovation}
6G networks will drive \textbf{economic growth} by supporting new industries, services, and business models. The ability of 6G to enable seamless communication and data exchange at unprecedented scales will lead to transformative changes in sectors such as manufacturing, retail, and logistics, enhancing productivity and innovation. For example, in the context of Industry 4.0, a future factory powered by 6G could employ AI-driven robots that communicate in real-time to optimise production processes, reducing downtime and increasing output. Real-time data analysis facilitated by ultra-low-latency networks could allow manufacturers to fine-tune their production lines based on market demand, reducing waste and operational costs. While these advancements will create high-skilled jobs and foster innovation, ethical considerations highlight the need for workforce retraining and adaptation in economies transitioning to automation-driven industries. REASON focuses on UK-born technologies, providing technological capability, sovereignty, and economic growth.

\paragraph{Ethical AI and Trustworthy Networks}
The integration of AI and ML in 6G networks will require robust ethical frameworks to govern their use. KVIs, in this area, focus on the \textbf{trustworthiness} of AI, ensuring that these systems are \textbf{transparent}, \textbf{fair}, and \textbf{accountable}. Consider, for example, a 6G-enabled AI-driven autonomous vehicles fleet. The capability to make real-time decisions that ensure passenger safety and ethical traffic management will be critical for such a system. In large cities such as New York or London, autonomous fleets powered by 6G networks could reduce traffic congestion and accidents, but the ethical programming of AI systems to make split-second decisions in potentially life-threatening situations needs careful regulation. Multi-modality and foundational models could tackle this problem~\cite{cheeky_citation_2}, ensuring that decisions taken can leverage multiple sensors and data streams and enhancing the trustworthiness of an ML pipeline. Designing AI systems that avoid biases, protect user privacy, and ensure fair treatment of all societal groups will be a priority, particularly in sectors like law enforcement or healthcare, where AI’s decisions could disproportionately affect vulnerable populations. Some of these areas are being considered in REASON, with details being provided in Secs.\ref{subsec:ai_ml} and~\ref{subsec:security}.

\paragraph{Resilience and Security}
The \textbf{resilience} of networks to cyber-attacks and natural disasters is another critical KVI. Future 6G systems must be designed to withstand disruptions while maintaining service continuity. For example, in countries prone to earthquakes (e.g., Japan), a resilient 6G network could ensure that emergency services remain operational during and after natural disasters. Autonomous drones and real-time video surveillance facilitated by 6G could support disaster response efforts, delivering medical supplies and performing search-and-rescue operations, even when traditional infrastructure is compromised. Considering the societal impact, resilient communication networks will enhance public safety and improve disaster recovery. However, they must also prioritise ethical considerations such as data security and privacy in the face of increasingly sophisticated cyber threats.

\paragraph{Human-Centric Services and Well-being}
Finally, 6G should facilitate more personalised, \textbf{human-centric} services, fostering well-being by improving how individuals interact with digital systems in everyday life. This KVI focuses on enhancing \textbf{user experience} in a manner that aligns with personal, social, and ethical values. For example, in education, immersive technologies powered by 6G could offer students virtual, interactive learning environments and, most importantly, access to world-renowned experts and peers, even for students in underdeveloped countries. As educational access becomes more democratised, ethical concerns will arise about ensuring fair access to these tools across diverse populations and addressing the mental health impacts of prolonged exposure to virtual environments.

\bigbreak

The above KVIs and KPIs should be considered by any future network architecture. As 6G is not just about advancing technology but also about transforming society, economies, and ethical frameworks, it has the potential to fundamentally reshape the world in ways that provide tangible value to everyday lives. In the following sections, we will discuss in more detail how the REASON architecture aims to embed these principles in its design and provides the above capabilities in an E2E fashion.

\subsection{REASON and other State-of-the-Art Projects}\label{subsec:sota}
REASON intends to provide a novel architecture that aligns with the 6G landscape introduced in the previous section and moves the research community a step closer to achieving such an ambitious goal. Before we detail how REASON intends to achieve that, we compare REASON with other indicative Beyond 5G and 6G research initiatives and their key features. We target three prominent areas for our investigation, i.e., ``Native-AI'', ``Multi-Access Technology'' connectivity and ``Network Automation / Orchestration'', which are the key contributions from REASON. Table~\ref{tab:research_initiatives} showcases the projects analysed and their delivery timeframe. Also, a broader overview and comparison of the listed projects are presented in Sec.~\ref{subsec:init_comparison}, paving the way for the more in-depth analysis of the two above-mentioned areas that follow in Secs.~\ref{subsec:nativeAI} and~\ref{subsec:automation_orchestration}.

Many initiatives have presented innovations aligning with the 6G landscape. Our investigation will not be an extensive list, but we will rather focus on projects that:
\begin{itemize}
    \item Will be delivered within a similar timeline (roughly) as REASON.
    \item Are active for at least one year, so they have already produced some results.
    \item Have introduced advancements in the above-mentioned areas.
\end{itemize}
Moreover, while the tables presented provide a comprehensive comparison against all the initiatives and REASON, for each feature chosen, we describe the projects with the most prominent/interesting approaches in the text. We do so so it is easier for the reader to identify the novel features introduced in other projects and how they compare with REASON. These features identified pave the way for our architecture design presented in Sec.~\ref{sec:reference_architecture}.

\begin{table}[t]
    \renewcommand{\arraystretch}{1.15}
    \centering
    \caption{5G and Beyond Research Initiatives}
    \label{tab:research_initiatives}
    \begin{tabular}{r||P{0.15\columnwidth}P{0.15\columnwidth}P{0.15\columnwidth}P{0.11\columnwidth}}
\multirow{2}{*}{\textbf{Project Name}} & \textbf{Start (dd/mm/yy)} & \textbf{End (dd/mm/yy)} & \multirow{2}{*}{\textbf{Funding}} \\ \hline
        \textbf{HEXA-X: I \& II}\tablefootnote{HEXA-X: I \& II: https://hexa-x-ii.eu/} &  01/01/21 & 30/06/25 & EU \\
        \textbf{HORSE}\tablefootnote{HORSE-6G: https://www.horse-6g.eu/} & 01/01/23 & 31/12/25 & EU \\
        \textbf{ADROIT6G}\tablefootnote{ADROIT6G: https://adroit6g.eu/} & 01/01/21 & 31/12/25 & EU \\
        \textbf{DAEMON}\tablefootnote{DAEMON: https://h2020daemon.eu/}  & 01/01/21 & 31/03/24 & EU \\
        \textbf{DESIRE6G}\tablefootnote{DESIRE6G: https://desire6g.eu/} & 01/01/23 & 31/12/25 & EU \\
        \textbf{RIGOROUS}\tablefootnote{RIGOROUS: https://rigourous.eu/} & 01/01/21 & 31/12/25 & EU \\
        \textbf{ACROSS}\tablefootnote{ACROSS: https://across-he.eu/} &  01/01/21 & 31/12/25 & EU \\
        \textbf{ETHER}\tablefootnote{ETHER: https://www.ether-project.eu/} &  01/01/21 & 31/12/25 & EU \\
        \textbf{NANCY}\tablefootnote{NANCY: https://nancy-project.eu/} & 01/01/21 & 31/12/25 & EU \\
        \textbf{PREDICT6G}\tablefootnote{PREDICT-6G: https://predict-6g.eu/} & 01/01/23 & 30/06/25 & EU \\

        \textbf{REASON} & 01/02/23 & 28/02/25 & UK
    \end{tabular}
\end{table}

\subsubsection{Initial Comparison}\label{subsec:init_comparison}
HORSE~\cite{horse_MLOPS} and NANCY~\cite{nancy2023architecture} address challenges towards 6G infrastructure operation and service management. Similarly, ANDROIT-6G proposes an AI-powered, fully cloud-native network~\cite{androit6GExplainAI}, focusing primarily on distributed deployments and optimisations. REASON inherits the good cloud-native practices presented in ANDROIT-6G. Moreover, while using AI/ML is integral for the above projects, they treat AI/ML as an ``add-on'' for optimisations. REASON's approach for an AI-Native implementation pushes the boundaries of seamless AI/ML integration by addressing ethical concerns while ensuring trustworthiness, resilience, and verifiability across the entire network.

On the other hand, DAEMON~\cite{daemon6g}, ACROSS~\cite{across}, and PREDICT6G~\cite{predictAIPlane} describe an AI-Native architecture, developing various solutions around scheduling, energy-aware orchestration and more. DAEMON and PREDICT6G focus on single-domain deployments, while REASON targets cross-domain functionality and enhanced E2E monitoring capabilities. These features are already discussed within ACROSS, but REASON stands out compared to that by intelligently addressing the prediction of resource requirements while also introducing more mechanisms for resource allocation.

DESIRE-6G proposes a novel unified programmable data plane and fine-grained monitoring capabilities~\cite{desire6gUnifiedDataPlane}. ETHER develops solutions across a unified Radio Access Network (RAN). REASON builds upon both ideas and extends the programmable data plane with a tight integration with a multi-access RAN Intelligent Controller (RIC). ETHER, compared to REASON, introduces a 3D multi-layer network architecture with satellite communications playing a pivotal role in their solutions. REASON, while acknowledging the necessity for satellite communications in future networks, at its current iteration, prioritises a more flat architecture utilising technologies such as WiFi or Light Fidelity (LiFi).

RIGOROUS focuses a lot on the security of the network, encryption practices, privacy-based deployments, and detection of anomalies~\cite{rigorous2023design}. REASON takes a different approach to security. We focus primarily on providing trustworthy ML pipelines, tightly integrating ML verifiability tools and explainability in our ML operations. Of course, we endeavour to learn from RIGOROUS security considerations and integrate them into our solutions.

Finally, HEXA-X is the more mature project showing solutions and advancements in most of the above-mentioned areas. As HEXA-X is currently at its second iteration, REASON has studied in detail the findings of HEXA-X I and closely monitors the advancements from HEXA-X II, getting invaluable information and inspiration from both projects. A key differentiating factor between REASON and HEXA-X is the AI-enabled multi-access RIC functionalities developed within REASON. We believe such functionality is crucial for supporting the heterogeneous network landscape anticipated in future 6G deployments.

\subsubsection{Native AI}\label{subsec:nativeAI}
AI-Native refers to systems, products or services where AI is crucial for their operation. This differs from AI-based solutions, where AI is usually offered as an ``add-on''. A great example of that is YouTube, where AI is used for video recommendations or content tagging but not for enhancing the user experience~\footnote{YouTube – A Super Platform Driven by AI: https://tinyurl.com/youtube-ai-platform}. Based on that, we analysed various aspects of intelligence and cognitive features and how such features are orchestrated/managed within a future network. Briefly, the domains analysed include a variety of enablers and features, such as:
\begin{itemize}
    \item AI Orchestration (includes AI Lifecycle Management)
    \item Machine Learning Operations (MLOps)
    \item Cognitive Capabilities
    \item Digital Twins for Network Intelligence
    \item Trustworthy AI
    \item Monitoring of AI services
    \item Intelligent Resource Prediction
\end{itemize}

\begin{table*}[t]
    \renewcommand{\arraystretch}{1.15}
    \centering
    \caption{Key features around the Native AI and comparing the State-of-the-art projects with REASON.}
    \label{tab:ai_plane_sota}
    \begin{tabular}{|M{0.09\textwidth}||M{0.1\textwidth}|M{0.1\textwidth}|M{0.1\textwidth}|M{0.1\textwidth}|M{0.1\textwidth}|M{0.1\textwidth}|M{0.1\textwidth}|}
        \hline
        \diagbox[innerwidth=0.09\textwidth,font=\footnotesize]{Enablers}{Features} & \textbf{AI Orchestration} & \textbf{MLOps} & \textbf{Cognitive Funcionality} & \textbf{DTs for  Net. Intelligence} & \textbf{Trustworthy} & \textbf{Monitoring}  & \textbf{ML Resource Prediction} \\ \hline\hline
        HEXA-X   & \cmark & \cmark & \cmark & \cmark & \cmark & \cmark & \ocircle\\
        HORSE   & \cmark & \cmark & \cmark & \cmark & \cmark & \cmark & \xmark \\
        ANDROIT6G  & \cmark & \xmark & \cmark & \xmark & \xmark & \cmark & \xmark \\
        DAEMON   & \cmark & \cmark & \cmark & \xmark & \xmark & \cmark & \xmark \\
        DESIRE6G & \cmark & \cmark & \cmark & \xmark & \xmark & \cmark & \xmark \\
        RIGOROUS   & \cmark & \xmark & \xmark & \cmark & \xmark & \cmark & \xmark \\
        ACROSS  & \cmark & \xmark & \cmark & \cmark & \cmark & \cmark & \xmark \\
        ETHER  & \xmark & \xmark & \xmark & \xmark & \xmark & \cmark  & \xmark \\
        NANCY  & \xmark & \xmark & \cmark & \xmark & \cmark & \cmark & \xmark \\
        PREDICT6G & \cmark & \cmark & \xmark & \cmark & \xmark & \cmark & \ocircle  \\
        \hline \hline
        \textbf{REASON}   & \cmark & \cmark & \cmark & \cmark & \cmark & \cmark & \cmark  \\
        \hline
    \end{tabular}
    \begin{tabular}{M{0.9\columnwidth}}
        \centering \cmark\ Feature Present \quad \xmark\ Feature Absent \quad \ocircle \, Partially Addressed \\
        \end{tabular}
\end{table*}

More specifically, AI orchestration refers to the automated management, coordination, and optimisation of AI model lifecycles, including versioning, training, deployment, monitoring, and retirement~\cite{aiorchestration}. MLOps constitutes a collection (structured framework) of practices and principles designed to streamline and automate the entire ML lifecycle~\cite{mlops}. Cognitive capabilities refer to the processes and activities of acquiring, processing and using information that can enhance the verification, explainability, reasoning, etc~\cite{androit6GExplainAI}. Regarding the DTs, we refer to isolated environments that support by-directional interaction with the physical system and can be used to validate network intelligence scenarios~\cite{dtNetworkIntelligence}. Trustworthy AI looks into ethical alignment and regulation compliance for fairness, privacy, sustainability, and safety~\cite{responsibleai}. The monitoring of AI builds a framework for collecting data across all ML pipelines and feeds actionable intelligence into AI orchestration to inform dynamic adjustments~\cite{mlops}. Finally, intelligent resource prediction refers to strategies for efficiently and effectively predicting and allocating resources for running AI models within a network~\cite{mei2023gpu}. A comparison based on the above can be found in Table~\ref{tab:ai_plane_sota}.

Starting with \textbf{AI Orchestration}, all projects but ETHER and NANCY propose some form of orchestration framework for AL/ML models. For example, HEXA-X I\&II integrate an AI analytics framework, with AI-as-a-Service (AIaaS) and cross-domain AI request handling via an open network interface~\cite{hexaXVision}, DAEMON designs a Network Intelligence Orchestration (NIO) that handles the most important Network Intelligence (NI)-related mechanisms such as lifecycle management, coordination, and data management~\cite{daemon6g}, and DESIRE6G develops an ML-function orchestrator for secure ML pipelines based on DLT~\cite{gonzalez2024deployment}. Overall, we see a spectrum of orchestration strategies, from open AI services to specialised task-oriented orchestration. Similarly, REASON will work on an AI orchestration plane considering a model's entire lifecycle (from inception to deprecation and retirement).

Regarding \textbf{MLOps}, only a subset of the projects incorporate MLOps practices into their ML pipelines. An example is HEXA-X I\&II, which incorporate their proposed AIaaS solution within a cloud-native-enabled CI/CD environment~\cite{labrador2022hexax}. PREDICT6G~\cite{giardina2023implementation} and HORSE~\cite{horse_MLOPS} propose a complete solution, developing a framework that manages the complete lifecycle of  AI/ML models, from design and training to deployment in production environments at different domains,  allowing their integration into the proposed Management \& Orchestration domain control loops. REASON will take that one step further, proposing novel ways of managing multiple AI workflows intelligently and exchanging them/chaining them to meet the QoS requirements of each use case. Moreover, a management and orchestration framework will be designed to accompany them.

\textbf{Cognitive Capabilities} are shown in various projects, focusing on Explainable AI (XAI) and continuous model validation. HEXA-X I\&II emphasises the use of an Explainable AI framework for predictive tasks and AI self-verification, with a focus on transparency and reliability in AI decisions~\cite{hexaXVision}, ANDROIT6G provides a framework that balances between performance and explainability~\cite{androit6GExplainAI} and DAEMON6G investigates XAI tools that couple the model operation with techniques that mine input data~\cite{FIANDRINO202247}. Some projects, like HORSE and RIGOROUS, omitted this feature entirely, prioritising other aspects of AI orchestration over cognitive insights. REASON takes the approach that AI should become explainable, robust, and verifiable, and it will build on ensuring correctness, reliability, and adherence to certain standards for all its AI/ML operations.

We see a diverse landscape in the \textbf{Digital Twin for Network Intelligence} category. HORSE focuses on DTs for detection and prediction and impact analysis~\cite{HORSE_architecture}, RIGOROUS takes a security approach and uses DTs for network security testing~\cite{rigorous2023design} while PREDICT6G focuses on predicting network analytics with DTs~\cite{giardina2023innovation}. Some projects, e.g., DESIRE6G and ANDROIT6G, do not consider DTs or just briefly touch upon them in their architecture without going into specific details. REASON will develop a state-of-the-art digital twin of a 6G network, combining various Access Technologies (ATs) and enabling offline AI model training while also including crucial modules for lifecycle management, explainability, and verifiability.

\textbf{Trustworthy AI} is a feature prominently highlighted in HEXA-X I\&II~\cite{hexaXVision}, included under the umbrella of AI/ML security and privacy and suggesting a holistic approach to AI trust. The rest of the projects do not tackle challenges around AI trustworthiness. Some projects like DESIRE6G consider technologies like Federated Learning (FL), so the trustworthiness of AI is in some sense touched upon. REASON will incorporate AI monitoring capabilities within its architecture and take trustworthiness and AI ethics into consideration for all the solutions provided.

In terms of \textbf{Monitoring}, all projects provide monitoring capabilities with regard to the AI/ML models. HEXA-X I\&II stands out with the most developed solution providing a comprehensive monitoring framework for software integration and the data exchanged~\cite{hexaXVision}. In contrast, RIGOROUS builds a monitoring framework capable of data processing on the fly (focusing on privacy-preserving)~\cite{rigorous2023design}. REASON extending on the above will integrate profiling capabilities within its MLOps pipeline while also considering energy consumption and cost of the different ML workflows.

Finally, \textbf{Intelligent Resource Prediction} is only captured by a handful of projects.  PREDICT6G~\cite{predictPrediction} relies on ML to forecast the resources required for a future service or deployment -- resources for ML are also considered -- but does not capture energy constraints or sustainability aspects related to ML. DAEMON6G, even though it touches upon capacity forecasting as part of their NIO~\cite{soto2024designing,daemon6g}, never describes how it will be delivered, so this may be reflected in future activities. HEXA-X focuses primarily on forecasting network resources~\cite{hexaXForecasting} for short-, mid- and long-term periods but does not consider the energy requirements or the resources for making these intelligent decisions. REASON will devise a detailed profiling framework that will not only be able to intelligently predict network and computing resources but also consider energy constraints and the expected QoS of all services. Additionally, REASON will focus on predicting and managing resources for AI workloads to assess and evaluate the costs of running AI within the network, which has not been addressed in the literature.

These projects form clusters based on their approach and emphasis: HEXA-X I\&II, PREDICT6G, and DAEMON appear as the most comprehensive, integrating multiple features across the board. RIGOROUS maintains a focused approach to security and performance, while ADROIT6G and DESIRE6G take a service-oriented stance and RIGOROUS hones in on security through AI and data monitoring. REASON aims to provide a more holistic approach to tackling all of the above within the same architecture.

\subsubsection{Network Automation / Orchestration}\label{subsec:automation_orchestration}
Secondly, we compare all projects with respect to their ``Network Automation / Orchestration'' and network management capabilities. The domains analysed are as follows:
\begin{itemize}
    \item Service-Level Agreement (SLA) / Security SLA support
    \item Dynamic resource allocation
    \item Zero-touch Service Management \& Orchestration
    \item Network Management
    \item Monitoring
    \item Security and Trust Management
    \item Policy Management
    \item Multiple Administrative Domain Support
    \item Multiple AT support RAN Intelligent Controller (RIC)
\end{itemize}

\begin{table*}[t]
    \renewcommand{\arraystretch}{1.15}
    \centering
    \caption{Key features around Network Automation / Orchestration and provided network management capabilities.}
    \label{tab:orchestration_sota}
    \begin{tabular}{|M{0.09\textwidth}||M{0.05\textwidth}|M{0.1\textwidth}|M{0.11\textwidth}|M{0.05\textwidth}|M{0.08\textwidth}|M{0.1\textwidth}|M{0.05\textwidth}|M{0.06\textwidth}|M{0.05\textwidth}|}
        \hline
        \diagbox[innerwidth=0.09\textwidth,font=\footnotesize]{Enablers}{Features} & \textbf{(Secure) SLA} & \textbf{Dyn. Resource Allocation} & \textbf{Service Manag. \& Orch.} & \textbf{Network Manag.} & \textbf{Monitoring} & \textbf{Security \& Trust Manag.} & \textbf{Policy Manag.} & \textbf{Multiple Domains} & \textbf{mATs RIC} \\ \hline\hline
        HEXA-X   & \cmark & \cmark & \cmark & \cmark & \cmark & \cmark & \cmark & \cmark & \xmark \\
        HORSE   & \xmark & \xmark & \xmark & \cmark & \xmark & \cmark & \cmark & \xmark & \xmark \\
        ADROIT6G  & \cmark & \xmark & \cmark & \xmark & \xmark & \cmark & \xmark & \xmark & \xmark \\
        DAEMON   & \xmark & \cmark & \cmark & \xmark & \xmark & \cmark & \cmark & \xmark & \xmark \\
        DESIRE6G & \cmark & \cmark & \cmark & \xmark & \xmark & \cmark & \cmark & \xmark & \xmark \\
        RIGOROUS   & \cmark & \xmark & \xmark & \cmark & \xmark & \cmark & \cmark & \cmark & \xmark \\
        ACROSS  & \cmark & \xmark & \cmark & \cmark & \cmark & \cmark & \cmark & \cmark & \xmark \\
        ETHER  & \xmark & \xmark & \xmark & \xmark & \xmark & \cmark & \xmark & \cmark & \xmark \\
        NANCY  & \xmark & \xmark & \cmark & \xmark & \cmark & \cmark & \cmark & \xmark & \xmark \\
        PREDICT6G & \cmark & \cmark & \xmark & \cmark & \xmark & \cmark & \cmark & \xmark & \xmark \\
        \hline \hline
        \textbf{REASON}   & \cmark & \cmark & \cmark & \cmark & \cmark & \cmark & \cmark & \cmark & \cmark \\
        \hline
    \end{tabular}
\end{table*}

SLA/Security SLA can guarantee performance and security standards, critical for maintaining reliability and user trust in increasingly complex 6G environments~\cite{slaSecuresla}. Dynamic resource allocation allows for real-time adjustments in network resources to optimise performance and efficiency, especially during fluctuating demands~\cite{resourceAllocation}. Zero-touch Service Management \& Orchestration enables automated network operations, reducing human intervention and operational costs while increasing agility and responsiveness~\cite{zeroTouch}. Network Management encompasses the comprehensive administration of network infrastructure, essential for maintaining optimal performance and troubleshooting issues in 6G networks~\cite{networkManagement}. Monitoring involves continuous observation and analysis of network performance and security to identify and mitigate potential issues preemptively. Security and Trust Management ensures robust protection against cyber threats and fosters trust among users and devices, which is foundational for the integrity of future networks~\cite{security6G}. Policy Management allows for implementing and enforcing network rules and regulations, crucial for maintaining order and compliance in a highly dynamic network environment~\cite{policyManagement}. Multiple Administrative Domain Support facilitates seamless collaboration and interoperability between different network domains, essential for the integrated and expansive nature of 6G networks~\cite{ai-multirat}. Finally, in terms of multi-AT support, we compare the project on how they merge multiple ATs, focusing on whether they support the use of intelligent controllers. Our analysis is summarised in Table~\ref{tab:orchestration_sota}.

For the \textbf{SLAs}, HEXA-X I\&II~\cite{hexaXVision} considers SLA for slicing, considering various metrics, energy consumption, and a monitoring framework by 3rd party infrastructure nodes. DESIRE6G takes that a step further and considers E2E SLA decomposition to partial SLAs assigned to multiple domains~\cite{slaDecomposition}. Finally, RIGOROUS considers a translation of intents to SLAs, considering the security aspect as well~\cite{rigorousSLA}. REASON will combine the above features, translating intents to SLAs and later intelligently decomposing or chaining them while running them across multiple domains.

We see diverse approaches to optimising a network when considering \textbf{Dynamic Resource Allocation}. For example, DAEMON presents an energy-aware placement of NFV in heterogeneous infrastructures~\cite{resourceAllocationDAEMON}, PREDICT6G focuses on fairness and scalability for their network and computing resource scaling~\cite{fairAndScalable}, and ACROSS discusses Network Resource Elasticity (NRE) dynamically adjusting network resources and configuration to meet QoS and reduce Operational Costs (OPEX)~\cite{across}. REASON will focus on forecasting and provisioning services to address scalability while optimising resources across multiple domains and ATs.

All projects consider some form of \textbf{Service Management and Orchestration}; however, we see that only a subset of the initiatives tackle the problem with a zero-touch approach. HEXA-X I\&II introduces a four-layer closed-loop automation (one per layer identified in their architecture) that also handles policy conflicts~\cite{labrador2023hexax}. HORSE introduces intent-based control loops focusing on security while automatically evaluating their impact~\cite{HORSE_architecture}. REASON will provide both cross-domain and intra-domain orchestration and management functionality, tightly linked with an AI plane to enable a zero-touch deployment.

With regards to the \textbf{Network Management}, some initiatives like DAEMON couple the functionality from the service orchestration and network management (e.g., DAEMON supports federated multi-domain management and multiple virtualisation environments driven by their proposed Network Intelligent Orchestrator~\cite{daemon6g}) while others like PREDICT6G provide decouple these functions providing a framework complementary to the orchestrator (e.g., PREDICT6G has a separate function block for the path computation~\cite{giardina2023innovation}). REASON ties the network management within the two proposed orchestration frameworks (one at the domain level and one at the cross-domain level) while considering a multi-access network provided within each domain.

Following the AI monitoring capabilities described in Sec.~\ref{subsec:nativeAI}, many projects provide domain-wide or E2E-wide monitoring capabilities. For example, PREDICT6G collects measurement within a single domain, measuring various KPIs from the running services~\cite{giardina2023implementation}. NANCY takes that a step further and provides an E2E (within a single domain) monitoring of the service KPIs. It also monitors available computing resources (e.g., CPU utilisation) while integrating with a cloud-native monitoring framework. REASON will build upon such an approach and provide a cross-domain E2E monitoring framework, where resources, KPIs, and raw data are monitored both at the domain and cross-domain levels, are being profiled and accessed, and are later fed to a zero-touch orchestrator.

From the \textbf{Security and Trust Management} standpoint, it is evident that all projects take security very seriously and utilise a multitude of solutions to incorporate the three principles of information security (Confidentiality, Integrity Availability (CIA)) in their architectures. To name a few examples, ANDROIT6G takes a more communication network approach identifying known and unknown attacks between devices~\cite{androit6gSecurity}, HEXA-X I\&II~\cite{hexaXVision} focuses on trust level monitoring and management using Distributed Ledger Technology (DLT) and multiple projects employ Federated Learning (FL) to ensure the privacy of the data (NANCY~\cite{nancy2023architecture}, HORSE~\cite{HORSE_architecture}, etc.). Similarly, REASON will employ FL for privacy preservation of the data generated and run ML models locally at the domains/edges, exposing the model parameters to the cross-domain orchestrator. Moreover, we will incorporate a trust and security management framework that oversees various decisions within the system, provides secure SLA, considers security requirements in the user intents, and more.

Moving on to the \textbf{Policy Management}, there is again an abundance of diverse solutions proposed from the different projects. DAEMON proposes a policy interpreter~\cite{daemon6g}, DESIRE6G builds a policy framework module with an engine that enables network operators to define and enforce policies for network services~\cite{desire6g2023initial}, and RIGOROUS introduces a security and policy enforcement function in the proposed management domain~\cite{rigorous2023design}. REASON will again holistically tackle that by introducing an E2E policy management framework that covers intent-driven service management, network management, and trust and security management.

Lastly, from the above-mentioned projects, only a subset of them presents \textbf{Multi-domain Functionality} (i.e., HEXA-X I\&II, RIGOROUS, ACROSS, and ETHER), whereas the rest just focus on a single domain. REASON will work across multiple domains, providing a cross-domain orchestration functionality that is capable of managing and administering multiple domains (i.e., delivering services exploiting multiple domain and network capabilities).

Finally, in terms of multi-AT support, we compare the project on how they merge multiple ATs. HEXA-X proposes a multi-spectrum sharing mechanism, operating at the hardware level~\cite{hexaXSpectrum}. DAEMON, on the other hand, even though it pitches the use of a RIC, does not mention the use of different ATs~\cite{daemon6g}. Finally, ETHER merged terrestrial and non-terrestrial networks~\cite{ether} utilising medium-transparent mechanisms such as Fibre-Wireless~\cite{etherFiWi}. REASON stands out in that area, converging all ATs using an intelligent multi-AT real-time controller inspired by O-RAN~\cite{oRan}. This controller (described in more detail in Sec.~\ref{par:matric}) enables the utilisation of multiple technologies for a given service, seamlessly managing the resources across multiple ATs.

As in Sec.~\ref{subsec:nativeAI}, we see HEXA-X (due to its scale) addressing most of the above-mentioned areas of investigation, while other projects focus on specific areas. For example, PREDICT6G, DESIRE6G, and ANDROIT6G focus on resource allocation and handling this with an SLA approach, whereas NANCY, HORSE, or DAEMON take a more policy-driven approach. It is evident that all projects consider security and privacy paramount, and service orchestration and network management as two of the most prominent areas for 6G and future networks. Again, as in Sec.~\ref{subsec:nativeAI}, REASON aims to incorporate all the above features within its proposed architecture. In the following sections, we will focus on the technical and non-technical challenges identified from our state-of-the-art analysis and discuss how REASON aims to address them.

\section{Challenges and Usage Scenarios}\label{sec:challenges}

The above-presented capabilities pave the way for developing novel applications that can enhance the quality of life of many people and bring economic benefits to their key stakeholders while maintaining the necessary societal and ethical considerations required for future network deployments. In this section, we capture several challenges identified and six usage scenarios (as defined by ITU-R) that lay the foundation for the use cases we present in the following sections and the technical enablers proposed in our proposed architecture.

\subsection{Technical Challenges}
\label{technical_challenges}
Starting with the technical challenges, as discussed, different classes of enablers are expected, extending on the traditional network KPIs and moving towards advanced services, such as sensing, energy efficiency, security/trustworthiness, etc.

Even though individual solutions currently exist, there is a gap in providing such services in an E2E fashion. As identified in Sec.~\ref{subsec:automation_orchestration}, even though orchestration capabilities are currently provided by other projects and initiatives, there is a gap in how an E2E, cross-domain, zero-touch system can be achieved. Moreover, this, combined with the ``intelligence'' required for the automation of such a system, makes current solutions inadequate. For example, current Management and Orchestration (MANO) frameworks do not extend to the devices at the far end of the network. Solutions such as ETSI's Network Function Virtualisation (NFV) MANO framework~\cite{ETSI_NFV}, even though they consider edge devices through specialised Virtual Infrastructure Managers (VIMs), the solutions are still controlled via a centralised orchestrator and are not suitable for the extreme edge devices~\cite{e2eOrchestration_no1}. Similarly, many works only focus on single-domain orchestration (or even specific sub-domains within that), focusing on the radio, transport or core levels, resulting in sub-optimal allocation and utilisation of resources~\cite{orchestration_singlelayer}.

Moreover, service continuity, meeting stringent requirements for Ultra-Reliable Low Latency Communications (URLLC), and enhanced mobility are more technical challenges that are not currently met by existing solutions. Processing at the edge and multi-connectivity scenarios (where the same application could communicate simultaneously through different ATs managed by different operators) have been discussed in the past~\cite{multi-RAT}, but without meeting the requirements of 6G usage scenarios~\cite{Sylla2022-kr}, (as presented in Sec.~\ref{sec:background} and Table~\ref{tab:kpis}). It is necessary to develop solutions enabling the global management of edge devices at different levels and the access of different ATs across different operators. There are currently approaches towards this direction, either based on Software-Defined Networking (SDN)~\cite{sdn-multirat} or leveraging AI/ML~\cite{ai-multirat}; however, there is still quite a lot of room for improvement. Of course, following the above, questions arise around the interoperability, security, and sustainability of such solutions and are currently being investigated by the research community~\cite{interoperability6G,security6G,sustainability6G}.

Building upon the necessity for a unified E2E solution and the KVIs and KPIs introduced in Sec~\ref{sec:background}, we devise the REASON architecture presented in this paper. To start, we propose the concepts and features we expect to be part of all future network deployments. More specifically, we envision:

\begin{itemize}
    \item \textbf{Multi-access technologies}: any potential wireless and wired, radio and optical, licensed and licence-exempt, terrestrial and non-terrestrial (classical and quantum) technology could be used as an AT. The architecture should provide ways to seamlessly handover between technologies or operate them in parallel~\cite{cheeky_citation_1}. This is important to enable mobility, accommodate for lack of resources, etc. Moreover, sensing capabilities must be available from multiple technologies, providing valuable insights into where objects are and how things are moving to optimise the network accordingly~\cite{integratedSensing} or enabling multi-modal learning in a distributed fashion~\cite{cheeky_citation_2}.
    \item \textbf{Open architecture}: Native open network principles, e.g., as described in O-RAN~\cite{oRan} must be established (more details found in Sec.~\ref{subsec:open_network}). All the required interfaces and protocols should be defined and developed to ensure the interoperability and openness of the solutions. Moreover, data consumption and the free flow of data across disparate systems and platforms should be provided by the architecture.
    \item \textbf{Native AI}: The network should be AI-driven, providing both the infrastructure and capabilities for running AI functions and optimising in an AI-native way (as per the definition in Sec.~\ref{subsec:nativeAI}). New cognitive and AI planes should be defined within the overall architecture, which uses network monitoring data and distributed analytics across the entire network stack to enhance performance, advance E2E service optimisation, facilitate agility and ensure full E2E trustworthiness of all network and data delivery functions. In this context, AI will also build resilience in future networks. Ethical AI deployment is paramount for such deployments. Verifiable AI technologies and practices are required to ensure the correctness, reliability and adherence to certain ethical standards and regulations. Finally, traditional ML approaches such as data drift detection~\cite{le3dDataDrift} or Concept Drift~\cite{FLAME} can enhance the accuracy and effectiveness of an ML model.

    Add-on AI, used as a supplementary feature rather than a core element, often limits its potential to enhance the user experience and tends to focus on local rather than global optimisation. In contrast, AI-native systems face challenges in deeply integrating AI-driven functions within the architecture, achieving seamless optimisation, and addressing ethical concerns while ensuring trustworthiness, resilience, and verifiability across the entire network. It is essential to analyse where AI is needed, its effects on the network, and its overall impact on the end-user experience.
    \item \textbf{Cloud to edge Computing}: A fully flexible approach is required, demanding allocating workloads to wherever computing and storage are located. Data and processes must flow seamlessly between edges, clouds, data centres, and users, which can be in various work locations and environments.
    \item \textbf{Densification}: New concepts should be proposed to support unprecedented network densification. Different deployment strategies must be incorporated into the network design, allowing for optimised coverage and capacity.
    \item \textbf{Sustainability and security}: Sustainability and security are fundamental themes that run through all architectural blocks. Energy consumption optimisation should be considered at the system level, across hardware and software layers, from edge-to-edge and domain-to-domain, delivering realistic co-optimisation insights. Security will address challenges arising from such a system-level approach and leverage the physical layer and cyber capabilities to harden the security of such heterogeneous, distributed infrastructures.
\end{itemize}

The REASON project aims to address and develop novel technologies and solutions across all of the technical challenges presented above. The aim is for future open communication networks to alleviate current bottlenecks in delivering optimised E2E multi-technology, multi-vendor networks. The REASON architecture can act as a blueprint for future E2E network architectures, which will set the framework for innovation and new developments across the entire technology stack driven by common architectural principles.

\subsection{Societal Challenges}\label{sub:societal}
6G and future networks will address key societal and ethical challenges not considered in previous generations. These societal challenges are integral to the use cases and should be taken into consideration for the architecture design process. Motivated by the United Nations (UN), we consider the UN Sustainable Development Goals (SDGs)\footnote{UN Sustainable Development Goals: https://sdgs.un.org/goals} and aim to align REASON with them. All 17 SDGs are relevant to REASON, but the following are especially linked with the REASON use cases and its proposed solutions. These SDGs are:
\begin{itemize}
    \item Goal 9: \textbf{Build resilient infrastructure, promote inclusive and sustainable industrialisation and foster innovation} - REASON is looking to propose a framework for future open network architectures that can drive innovation and new developments across the entire technology stack.
    \item Goal 10: \textbf{Reduce inequality within and among countries} - REASON is looking to develop innovative multi-access solutions that could be economical and sustainable for everyone, fulfilling the 4Cs (Coverage, Capacity, Cost, Consumption) requirements. Concepts like open disaggregation and implementation neutrality will be encouraged and are expected to lead to the wider adoption of open networking principles.
    \item Goal 11: \textbf{Make cities and human settlements inclusive, safe, resilient, and sustainable} - The REASON architecture should encompass Sustainable Cities and Communities, exploiting the multi-ATs and AI-powered monitoring and analytics to offer resiliency and backup systems to minimise disruptions during such events.
    \item Goal 13: \textbf{Take urgent action to combat climate change and its impacts} - One of the primary focus areas for REASON is the reduction of energy consumption from the E2E network (RAN, transport, edge, core).
\end{itemize}

Future networks should be human-centric, moving away from the static, unagile networks of the past. 6G should be technologically adaptable to address societal needs or ``Black Swan'' events effectively and efficiently. This concept is further explained in~\cite{6gVision}, discussing the societal and lifestyle changes that 6G and beyond communications can bring.  In REASON, we aim to tackle all the above challenges and provide solutions for sustainable network deployments that are energy efficient while providing enhanced coverage and throughput across a multitude of use cases and deployment scenarios.

\subsection{Usage Scenarios}\label{subsec:usage_scenarios}

Further to the capabilities and technological trends introduced earlier, IMT-2030~\cite{ITU2030} also identifies six usage scenarios for 6G.  Three of these are extensions of the 5G usage scenarios. Briefly, these are:

\begin{itemize}
    \item \textbf{Immersive Communication}: Extends the enhanced Mobile Broadband (eMBB) from IMT-2020, providing immersive experiences such as XR, telepresence, and holographic communications across various environments. It emphasises time-synchronised mixed traffic support and aims for enhanced spectrum efficiency, high reliability, and low latency for interactive engagements.
    \item \textbf{Hyper Reliable and Low-Latency Communication}: Builds on Ultra-Reliable and Low-Latency Communication (URLLC) from IMT-2020 to meet stringent reliability and latency requirements for critical operations, supporting use cases like industrial automation, emergency services, telemedicine, and power transmission monitoring, with needs for precise positioning and high connection density.
    \item \textbf{Massive Communication}: Extends massive Machine Type Communication (mMTC) from IMT-2020 for connecting massive numbers of devices or sensors, catering to Smart Cities, transportation, and various Internet-of-Things (IoT) applications. It focuses on high connection density, diverse data rates, low power consumption, and high security.
    \item \textbf{Ubiquitous Connectivity}: Aims to bridge the digital divide by enhancing connectivity, especially in rural, remote, and underserved areas. Typical use cases support IoT and mobile broadband communication to improve coverage and accessibility.
    \item \textbf{Artificial Intelligence and Communication}: Supports distributed computing and AI for automated driving, medical assistance, and other applications, requiring high traffic capacity, low latency, and integration of AI and compute functionalities for data processing and distributed AI across networks.
    \item \textbf{Integrated Sensing and Communication}: Leverages multi-dimensional sensing capabilities, offering spatial information for both connected and unconnected objects, supporting navigation, activity detection, and environmental monitoring. It requires precision in positioning and sensing capabilities for advanced applications.
\end{itemize}

These scenarios are similarly identified in the white paper published by Next Generation Mobile Networks Alliance (NGMN)~\cite{ngmn}. Both reports highlight \textit{sustainability}, \textit{security and resilience}, \textit{connecting the unconnected}, and \textit{ubiquitous intelligence} as the overarching aspects which drive the design principles behind each proposed architecture and are commonly applicable to all usage scenarios. NGMN further noted that use cases are ``provisional'' in the sense that a subset of the functionality could be served over existing advanced 5G networks. This strengthens the case for tackling the above challenges in a modular and flexible approach in response to the market and user demand. The above usage scenarios are governed not only by hard engineering KPIs but also by core values brought to society. Our proposed REASON use cases, briefly described in Sec.~\ref{sec:reason_usecases}, will consider these scenarios and capture the required functionality, constraints and actors to dictate various aspects of our proposed architecture later.

\section{REASON Use case Themes}\label{sec:reason_usecases}

Guided by the above-mentioned KPIs and KVIs (Secs.~\ref{sec:background} and~\ref{sec:challenges}) and the usage scenarios presented in Sec.~\ref{subsec:usage_scenarios}, we have selected five broad themes to analyse as part of REASON. Our list, even though not exhaustive, captures mature and non-mature use cases or driving innovations and differentiators. All proposed themes are intended to drive the technical requirements that should be captured for our envisioned architecture. These use cases were chosen as the most appropriate ones to demonstrate the proposed capabilities by REASON and should be accommodated by a common infrastructure and architecture while considering the challenges described in Sec.~\ref{sec:challenges}.

It is also worth stating that from our state-of-the-art analysis presented in Sec.~\ref{subsec:sota}, we observed that similar themes have been proposed by all the investigated initiatives. For example, HEXA-X I\&II use cases can be found in~\cite{wendt2023environmental}, ANDROIT6G's use cases in~\cite{jarvet2023adroit6g}, and DESIRE6G use cases in~\cite{lopezdasilva2023definition}. All projects propose use cases focusing on holographic and immersive experiences, digital twins, collaborative robotics and factory automation, and more. These use cases demonstrate the capabilities of each project. They also strengthen our case that our chosen use cases align with current approaches in 6G. Finally, while these use cases are relatively standard within the 6G ecosystem choosing, choosing a diverse range of use cases gives up the ability to demonstrate the REASON innovations intended to be delivered throughout the project across a spectrum of digital playgrounds while also driving our requirements analysis (Sec.~\ref{sec:requirements}).

\subsection{Metaverse at scale / Web 3.0}
The objective of this use case is to create an \textit{Internet Scale Metaverse}, offering immersive, distributed experiences at scale. The requirements of this use case include having high bandwidth (symmetric and over long distances), low latency, secure, and scalable digital infrastructure~\cite{metaverse}.

The metaverse is widely seen as the future of the Internet. It will transition users from interacting with websites and services through PCs and mobile devices (client/server) to a fully immersive, distributed, shared personal experience where multiple services are seamlessly integrated. Users can experience being in a different place from their actual location and experience augmented reality through various interactions (e.g., gestures,
touching objects, etc.).  These enhanced user experiences, through the seamless integration of services, can foster new forms of social and economic interactions for both professional and personal purposes.

Some of the key requirements to support the KVIs of this use case include:
\begin{itemize}
    \item \textbf{Digital Identity}: For users to own their identity and its associated values. If users are to move across multiple platforms and the metaverse, they will need a unique digital identity owned or controlled by them.
    \item \textbf{Tokenisation}: The use of decentralised ownership of assets and transaction of digital assets using Tokenisation will allow users to exchange value in the Metaverse.
    \item \textbf{Decentralisation}: To ensure security and privacy, the network must be decentralised and not controlled by any single entity.
    \item \textbf{Blockchain}: We envisage that a blockchain will be the foundation of support for the integrity and verifiability of transactions.
\end{itemize}

Moreover, some important KPIs include large bandwidth capacity (symmetric), which is necessary for high-quality video streaming, low and deterministic latency, high accuracy in positioning, and increased densification. Some preliminary studies and reviews suggest the values shown in Table~\ref{tab:metaverse}~\cite{metaverse1,metaverse2,metaverse3}.

\begin{table}[t]
\renewcommand{\arraystretch}{1.15}
    \centering
\caption{Metaverse Use Case KPIs}\label{tab:metaverse}
\begin{tabular}{r||l}

\textbf{KPI}                           & \textbf{Range}                                        \\ \hline \hline
Bounded Latency               & \SIrange{0.1}{20}{\milli\second} @$99\%$ end-to-end                      \\ \hline
Experienced User   Throughput & \SIrange{5}{100}{\Gbps}   symmetrical               \\ \hline
Network Capacity              & \SIrange{1}{100}{\nicefrac{\Tbps}{\kilo\meter\textsuperscript{2}}}                    \\ \hline
Connection/Device Density     & \SIrange{10}{100}{\nicefrac{}{\meter\textsuperscript{2}}}                               \\ \hline
Positioning                   & Sub-cm level accuracy,   i.e., \textless \SI{1}{\centi\meter} \\ \hline
Reliability                   & $\geq99.9999\%$ \\ \hline
\end{tabular}
\end{table}

\subsection{Distributed Digital Twin}
The objective of this use case is to digitise physical assets and create interconnected Digital Twins (DTs) for real-time monitoring, prediction, and optimisation across different industries. The challenges include integrating real-time AI/ML capabilities and operations, accurate positioning, sensing and actuating functionality for the network, and tight synchronisation. Moreover, some identified KPIs can be found in Table~\ref{tab:dt}~\cite{dt1,dt2}.

Digitalisation is a key element in creating 3D metaverse virtual spaces. It means creating a digital representation, or DT, of any of the industry’s physical assets~\cite{digitalTwin1,digitalTwin2}, for e.g.: hardware, software, materials, documentation, equipment, machinery, robot systems, computer systems etc.; in other words, any resources, processes, and/or industry systems. DT represent any of their physical assets, including Information Technology (IT) and Operational Technology (OT) infrastructures, as well as the communication network itself, used to interconnect all these digital and physical assets, allowing for real-time asset tracking and tracing.

A DT can collect, analyse, manipulate, simulate, emulate, and exchange data with the physical world, enabling improved monitoring, prediction, predictive maintenance, control, automation, optimisation, resilience, and security for the whole industrial process in the physical world. DTs allow industries to achieve improved operational efficiency, reduced costs, enhanced sustainability, and innovation in product and service offerings. Moreover, equipped with new XR devices, humans can interact with this cyber-physical world to monitor, visualise, collaborate, intervene, and interact by entering industrial metaverses. DTs can be the mediators that human users interact with to have their actions validated before they are implemented/deployed in the real world.

Some key requirements for this use case include:
\begin{itemize}
    \item \textbf{Interoperability}: Standards for seamless data exchange between different systems and digital twins.
    \item \textbf{AI/ML Integration}: Advanced analytics for predictive maintenance, operational optimisation, and decision-making support.
    \item \textbf{Network Slicing}: Customisable network capabilities to meet specific requirements of different digital twins.
\end{itemize}

\begin{table}[t]
\renewcommand{\arraystretch}{1.15}
    \centering
\caption{Distributed Digital Twins Use Case KPIs}\label{tab:dt}
\begin{tabular}{r||l}

\textbf{KPI}                           & \textbf{Range}                                        \\ \hline \hline
Bounded Latency               & \SIrange{0.1}{20}{\milli\second} @$99\%$ end-to-end                      \\ \hline
\multirow{2}{*}{Experienced User Throughput} & DL: \SIrange{0.1}{10}{\Gbps}  \\ \cline{2-2}  & UL: \SIrange{0.05}{5}{\Gbps} \\ \hline

Network Capacity              & \SIrange{10}{100}{\nicefrac{\Gbps}{\kilo\meter\textsuperscript{2}}}                    \\ \hline
Connection/Device Density     & \SIrange{1}{100}{\nicefrac{}{\meter\textsuperscript{2}}}                               \\ \hline
Positioning                   & cm level accuracy,   i.e., \textless \SI{10}{\centi\meter} \\ \hline
\end{tabular}
\end{table}

\subsection{Virtual Production}
This use case represents a scenario capable of supporting near-live content broadcast with minimal delay, two-way structured conversations for live news, multi-way natural conversations for live discussion or debate shows, and distributed remote live music performances~\cite{virtualProduction}. The main objective is to orchestrate local and remote low-latency synchronised media streams over a 6G network using an Edge Cloud infrastructure~\cite{streamEdgeCloud}. The jitter of the media streams should be considered, as well as the relative timing of each stream. This is critical for minimising buffering during frame synchronisation and editing in virtual production environments.

Virtual production leverages GPU and other computational resources in an Edge-Cloud continuum to mitigate timing variations, reducing the need for buffers and delays~\cite{virtualProductionGPU}. This can maintain a tight stream synchronisation and reduce editing and other production process delays. Some more challenges include minimal latency and jitter across the production chain, from video capture to post-production. Prioritisation among multiple concurrent streams to prevent congestion and ensure QoS is also a significant technical hurdle. The low latency virtual production use case requires KPIs for the network and Edge-Cloud, where the E2E Performance combines the network and compute performance. The E2E KPIs are determined by the level of interaction in a specific application, e.g. Near Live (content broadcast with a delay), Two-way structured conversation (Live News piece to camera), multi-way natural conversation (Live remote participants in a discussion or debate show), Distributed Remote Live Music (Live music lesson or orchestra). The latency requirements combine network/computer timing variation, buffering and frame alignment. The KPIs related to the network are summarised in Table~\ref{tab:virtual_network}~\cite{virtualProductionNetworks} and for the Edge/Cloud can be found in Table~\ref{tab:virtual_cloud}~\cite{streamEdgeCloud}.

Some key requirements for this use case include:
\begin{itemize}
    \item \textbf{Low Latency and Jitter Reduction:} Critical for maintaining tight stream synchronisation and reducing delays in production processes.
    \item \textbf{Stream Prioritisation:} Ensuring that streams are managed effectively, prioritising those requiring immediate processing to support real-time editing and production needs.
    \item \textbf{Edge Cloud Computing:} Utilising Edge Cloud resources for processing reduces the need for data to travel long distances, thereby minimising latency and jitter.
    \item \textbf{Network and Compute Performance:} A combination of network and computational capabilities that ensure E2E performance meets the stringent requirements of virtual production.
\end{itemize}

\begin{table}[t]
\renewcommand{\arraystretch}{1.15}
    \centering
\caption{Virtual Production Use Case KPIs - Network}\label{tab:virtual_network}
    \begin{tabular}{r||l}

        \textbf{KPI}                           & \textbf{Range}                                        \\ \hline \hline
        \multirow{4}{*}{Network Capacity} & HD Stream: \SIrange{5}{20}{\Mbps}  \\ \cline{2-2}
        & UHD Stream: \SIrange{20}{50}{\Mbps}  \\ \cline{2-2}
        & Audio Stream: \SIrange{0.048}{3}{\Mbps}  \\ \cline{2-2}
        & Ancillary Stream: \SI{64}{\Kbps}  \\ \cline{2-2} \hline
        E2E maximum latency & \SI{15}{\milli\second} / \SI{50}{\milli\second} / \SI{150}{\milli\second} / \SI{1700}{\milli\second}\tablefootnote{Latency and jitter depend on the type of service and the level of human interaction.}                    \\ \hline
        E2E maximum jitter    & \SI{1}{\milli\second} / \SI{25}{\milli\second} / \SI{50}{\milli\second} / N/A \\ \hline
        Synchronisation   & $<$\SI{1}{\micro\second} / $<$\SI{2}{\milli\second}\tablefootnote{Synchronisation of device reference clocks when there are multiple audio and video devices at the same location should be < 1 µs, achieved within 5 s of connection. For multi-channel audio and image synchronisation, time synchronisation between media streams should be < 10 µs~\cite{timeSync}. For a single audio/video device in 1 location, then synchronisation of < 2 ms is acceptable.}  \\ \hline
        Time alignment  &  $<$video frame

    \end{tabular}
\end{table}

\begin{table}[t]
\renewcommand{\arraystretch}{1.15}
    \centering
\caption{Virtual Production Use Case KPIs - Edge/Cloud}\label{tab:virtual_cloud}
\begin{tabular}{r||l}

\textbf{KPI}            & \textbf{Range}               \\ \hline \hline
Network Capacity        & \SI{10}{\Gbps}               \\ \hline
E2E maximum latency     & $<$\SI{50}{\milli\second}    \\ \hline
E2E maximum jitter      & $<$\SI{10}{\milli\second}    \\ \hline
Synchronisation   & $<$\SI{1}{\micro\second} / $<$\SI{2}{\milli\second}      \\ \hline
Stream switching time   & $<$2 frame periods, but frame aligned      \\ \hline
\end{tabular}
\end{table}

\subsection{Autonomous self-configuring factory}
The objective of this use case is to enable factories to increasingly rely on autonomous systems, including robots that collaborate with humans or other robots towards common goals~\cite{autonomousfactory}. The use of digital twinning, such as in urban micro-factories, aims to meet demand sustainably through hyper-local production. This necessitates high-bandwidth and ultra-low latency networks, which are achievable through next-generation private wireless deployments~\cite{digitalTwinning}.

Autonomous systems must navigate dense, dynamic environments, requiring precise localisation, high network reliability, and ultra-low latency. The synchronisation between digital and physical environments is pivotal, serving as the main network enabler. A summary of the KPIs is given in Table~\ref{tab:factory}.

Ensuring robust and secure communications in highly dynamic and potentially unpredictable factory environments is increasingly challenging. Various technologies, such as robotics, digital twins, and IoT devices, must seamlessly integrate to enable real-time, efficient operations~\cite{autonomousfactory}.

Some key requirements for this use case include:
\begin{itemize}
\item \textbf{High Bandwidth and Ultra-Low Latency:} Essential for supporting the real-time data needs of autonomous robots and digital twins, enabling them to make immediate decisions based on current factory conditions.
\item \textbf{Precise Localisation:} Critical for robot navigation and coordination within the factory environment, ensuring operation safety and efficiency.
\item \textbf{Network Reliability:} A highly reliable network is necessary to maintain continuous operations, even during hardware failures or environmental changes.
\item \textbf{Digital Twinning:} The ability to create and maintain a digital twin of the factory floor that is fully synchronised with its physical counterpart, enabling predictive maintenance and scenario planning.
\end{itemize}

\begin{table}[t]
\renewcommand{\arraystretch}{1.15}
    \centering
\caption{Autonomous Self-Configuring Factory Use Case KPIs}\label{tab:factory}
\begin{tabular}{|r|l|l|}
\hline
\textbf{KPI} & \textbf{Digital twin} & \textbf{Autonomous robotics} \\ \hline
Positioning & \SI{}{\centi\meter} accuracy & Sub-\SI{}{\centi\meter} accuracy \\ \hline
Orientation accuracy & degree (\degree) & \textless degree (\degree) \\ \hline
Data rate (minimum) & \SI{1}{\Gbps} & \textless \SI{1}{\Mbps} \\ \hline
Latency (maximum) & \textgreater \SI{20}{\milli\second} & \textless \SI{20}{\milli\second} \\ \hline
\end{tabular}
\end{table}

\subsection{Sustainability / Service Differentiation}
Sustainability is not so much a use case but more of a driver for 6G networks. Linked to UN SDG goal 13 described in Sec.~\ref{sub:societal}, which is about combating climate change and its impacts, telecoms infrastructure and, subsequently, the REASON platform must consider its impact in two complementary dimensions:

\begin{itemize}
    \item Controlling its impact on energy consumption and GreenHouse Gas (GHG) emissions consistent with trajectories agreed by the Intergovernmental Panel on Climate Change (IPCC) to stay below 1.5°C mean global temperature rise. Recommendation ITU-T L.1470 (ITU-T,2020)~\cite{ITU1470} takes a life cycle perspective and provides quantified trajectories of greenhouse gas emissions for the global ICT sector, including the mobile networks sub-sector. ITU-T L.1470 sets a normative upper limit for the ICT sector that supports the 1.5°C objective. Recommendation ITU-T L.1471 (ITU-T,2021)~\cite{ITU1471} provides specific net zero guidance for the ICT sector, embedding the trajectories of L.1470 as part of the net zero ambition. Moreover, general guiding definitions, principles, and recommendations for how organisations can achieve net zero greenhouse gas emissions are provided within ISO IWA 42:2022, which was launched at COP27. The development of performance requirements and evaluation criteria for energy consumption may refer to the following existing standards: (1) ETSI ES 203 228 V1.4.1 (2022-04) - ``Environmental Engineering (EE); Assessment of mobile network energy efficiency''~\cite{ETSI_environment}; and (2) ITU L.1331 (2022-01)- ``Assessment of mobile network energy efficiency''~\cite{ITU1331}.
    \item The impact on other sectors, both direct and secondary. ITU-T L.1480 sets out how to assess how the use of ICT affects other sectors (ITU-T,2022)\cite{ITU1480}.
\end{itemize}

One solution is to reduce energy consumption, from which wireless infrastructure providers would additionally benefit by reducing the cost of energy for powering their networks. However, given that deploying every new technology costs a lot of capital, there should be a clear benefit for the infrastructure providers in terms of service differentiation in doing so.

To provide service differentiation, differing approaches can be applied, (1) ranging from massive over-provisioning of resources to allow all potential differentiation of any given service to be achievable without special treatment (an ``extreme best efforts'' approach), to (2) a highly complex approach that tightly manages the desired KPI and Intent delivery through slicing and orchestration of suitable resources to allow all services to co-exist with minimal resource utilisation.

When the filter of Sustainability is applied to these approaches, (2) is where REASON will contribute. The allocation of resources to individual service delivery methods can be optimised through an E2E understanding of the overall network delivery requirements and the actionability of such knowledge through full network orchestration and life cycle management.

\section{Requirements}\label{sec:requirements}
REASON architecture must fulfil the requirements of the use cases described above, the management functions required by multiple ATs, edge computing and orchestration, and meet societal objectives. Starting with the requirements that will guide REASON's architecture design, the above-mentioned use cases have been analysed, and various insights have been extracted. While each use case is unique, analysing them along common dimensions identifies generalised patterns and fundamental platform requirements.

\subsection{Framework for Analysis}

To do so, we focus on the following:
\begin{itemize}
    \item \textbf{Functional Requirements}: We first analyse the given use-cases for the required capabilities, services, features, etc., needed to fulfil them. This shapes component capabilities and interface specifications.
    \item \textbf{Quality Attributes}: We then capture requirements related to the expected QoS, availability, reliability, scalability, survivability, maintainability, etc. For each use case, we capture the key KPIs envisioned to be achieved and provide quantitative values where possible.
    \item \textbf{Deployment Model}: Evaluate the topology and network requirements (core, edge, access), connectivity options from access network availability, compute infrastructure, etc. This frames integration and interoperability needs.
    \item \textbf{Life Cycle Workflow}: Understand orchestration workflows linking actors, activities, functions and information flows. This reveals platform automation potential and needs.
    \item \textbf{Interoperability}: Determine touchpoints with external systems and APIs for horizontal platform integration and vertical domain integration.
    \item \textbf{Sustainability}: As part of our KVIs analysis, we assess relevant sustainability objectives around circularity, carbon footprint, energy efficiency, etc., that manifest in technical requirements around hardware, algorithms and data flows.
    \item \textbf{Security \& Privacy}: Moreover, we identify assets, threats, vulnerabilities, etc., to define confidentiality, integrity and availability needs. Also, evaluate privacy and trust risks regarding personal or confidential data.
    \item \textbf{Technology Innovations}: Conducting a state-of-the-art analysis of the existing reference and platform architectures and all the related emerging innovations, e.g. in radio access, optical switching, AI, etc., we can guide our R\&D strategy, identify gaps in the literature, and evaluate all the identified requirements.
\end{itemize}

Two more aspects that can be considered are the ``ecosystem impact'', i.e., how REASON architecture could benefit key stakeholders to shape sustainability recommendations, future standards, the industry sector, etc. and the ``business value'', i.e., revenue opportunities and cost optimisation scenarios to build supporting business models. Due to the nature of the paper, these two aspects will not be considered for our requirements analysis and the architecture design.

\subsection{Requirements for Open Networking}\label{subsec:open_network}

Open networking is set to transcend traditional connectivity by integrating broader capabilities and functionalities, accommodating user and evolving application demands. Unlike today's networks, future systems like 6G will serve not just as communication hubs but as dynamic platforms that offer a flexible suite of services through an array of open APIs. These APIs will cater to various users and sectors, fostering innovation and improved network operations with potential applications in combined communication, sensing, and computing technologies. Business models around open networking will support diverse industry actors through customised network solutions that ensure performance and data confidentiality. Overall, the REASON architecture champions a paradigm of openness and interoperability, addressing emerging networking challenges and promoting dynamic reconfigurability across its structure. Briefly, the requirements are presented in the following Table~\ref{tab:open_networking}.

\begin{table}[t]
\renewcommand{\arraystretch}{1.1}
\centering
    \caption{Open Networking Requirements.}
    \begin{tabular}{lp{0.8\columnwidth}}
    \toprule
    \textbf{Req.} & \textbf{Requirement Description} \\ \midrule
    \textbf{ON1} & Should support and adopt Open RAN principles \\
    \textbf{ON2} & Must support interconnect of dynamic networks \\
    \textbf{ON3} & Must enable network and device mobility \\
    \textbf{ON4} & Should support Open interfaces and network function disaggregation \\
    \textbf{ON5} & Should define naming conventions above the IP layer \\
    \textbf{ON6} & Should address   routing challenges from network dis-integration \\
    \bottomrule
    \end{tabular}
    \label{tab:open_networking}
\end{table}

\begin{enumerate}[wide, labelwidth=!, labelindent=9pt, start=1,label={\bfseries ON\arabic*}]
    \item \textbf{- Open RAN principles:} The REASON architecture should encourage and enable the wider application of Open RAN principles: open disaggregation, standards-based compliance, demonstrated interoperability, and implementation neutrality. \label{req:on1}
    \item \textbf{- Interconnection of Networks:} REASON’s architecture must support the secure, open, technology-neutral interconnect of networks created and orchestrated to meet changing use case requirements for disaggregated network functions that can be dynamically reconfigured. Networks can be created on demand, virtualised, operated concurrently alongside each other, nested within each other, aggregated, concatenated, and partitioned. Typical examples of such technology are found in existing applications of Communications-Platform-as-a-Service (CPaaS), Infrastructure-as-Code (IaC), and Network-as-code (NaC).
    \item \textbf{- Network and Device Mobility:} The REASON architecture must provide the capability for device and network mobility seamlessly between and within ATs and between and within networks. It will require a well-defined identity of all assets and a coherent policy framework to support mobile access across all deployed instances.
    \item \textbf{- Open Interfaces and Network Function Dissagregation:}REASON architecture should adopt open and standardised interfaces exposed in a trusted manner to allow access across the infrastructure. These APIs can be consumed by applications disaggregated into micro-services executed in cloud-native environments (e.g., Kubernetes clusters), distributed across multiple edge locations interconnected between clouds in a virtual network namespace.
    \item \textbf{- Naming and Addressing:} The REASON architecture must make choices on naming conventions for functions above the IP transport protocol. The use of multi-ATs, some of which are not based on IP, necessitates a new inter-process communication and a single recurring set of protocols rather than specialised ones.
    \item \textbf{- Routing, network dis-integration and (network) mobility:} The REASON architecture must address the challenges that Web 3.0 and network dis-integration poses for routing in fixed networks~\cite{web3.0Routing}. Web 3.0 envisages the emergence of Distributed Autonomous Organisations (DAOs)~\cite{dao}, introducing concepts like ``semantic communications''~\cite{semantic}, ``organic networking''~\cite{organicNetworking}, etc.

\end{enumerate}

\subsection{Network Deployment and Scenarios Requirements}\label{subsec:network_deployment}
This section focuses on the requirements for diverse deployment scenarios and technological integrations. The architecture's adaptability across various environmental settings and its compatibility with an array of ATs are central to its design. These requirements ensure the network remains robust, versatile, and efficient in varying conditions and technological landscapes. Table~\ref{tab:network_deployment} briefly summarises these requirements.

\begin{table}[t]
\renewcommand{\arraystretch}{1.1}
\centering
    \caption{Requirements for network deployment and networking scenarios.}
    \begin{tabular}{lp{0.8\columnwidth}}
    \toprule
    \textbf{Req.} & \textbf{Requirement Description} \\ \midrule
    \textbf{NDS1} & Must demonstrate robust performance in a variety of environments \\
    \textbf{NDS2} & Must be able to integrate several wired and wireless technologies \\
    \textbf{NDS3} & Must support macro, micro and blended deployment strategies depending on the needs \\
    \textbf{NDS4} & Must undergo a rigorous performance assessment \\
    \textbf{NDS5} & Must support both indoor and outdoor environments \\
    \bottomrule
    \end{tabular}
    \label{tab:network_deployment}
\end{table}

\begin{enumerate}[wide, labelwidth=!, labelindent=9pt, start=1,label={\bfseries NDS\arabic*}]
    \item \textbf{- Environmental Versatility:} The network architecture must demonstrate robust performance in various environmental types, including dense urban, urban, suburban, rural, and potentially unpopulated areas. This versatility ensures the network's adaptability to diverse geographical and infrastructural challenges.
    \item \textbf{- Technological Integration:} To remain at the forefront of technological advancements, the network must support and integrate various wireless and wired ATs. This includes but is not limited to, 5G, WiFi, LiFi, Terahertz (THz) communications, Millimetre Waves, Non-Terrestrial, and Free Space Optical (FSO) technologies.
    \item \textbf{- Deployment Strategies:} Different deployment strategies must be incorporated into the network design. These strategies include macrocell-focused deployments, intense densification for high-demand areas, and a blended approach that combines various strategies to optimise coverage and capacity.
    \item \textbf{- Performance Assessment:} The proposed architecture should undergo rigorous evaluations in different deployment scenarios. This assessment will cover various environmental types and the availability of different ATs and frequency bands, such as WiFi, 5G, LiFi, etc., to ascertain the network's performance in a multi-technology landscape. In Sec.~\ref{subsec:network_management}, we will get into more technical requirements related to the network performance.
    \item \textbf{- Indoor and Outdoor Network Focus:} Attention should be given to both indoor and outdoor network deployments and the integration with private networks, which are vital for specific use cases like industrial metaverse applications. The architecture must cater to these specific needs while ensuring seamless connectivity.
\end{enumerate}

\subsection{Network Performance and Management Requirements}\label{subsec:network_management}
This section addresses the critical requirements for network performance and service management in the REASON’s architecture. It emphasises the need for an open platform system that tackles various challenges and is scalable, with the ability to integrate new technologies effectively. These requirements are vital to ensure the network is robust, adaptable, and future-proof. Table~\ref{tab:network_management} briefly summarises these requirements.

\begin{table}[t]
\renewcommand{\arraystretch}{1.1}
\centering
    \caption{Requirements for network performance and management.}
    \begin{tabular}{lp{0.8\columnwidth}}
    \toprule
    \textbf{Req.} & \textbf{Requirement Description} \\ \midrule
    \textbf{NPM1} & Must address performance, functional, and non-functional aspects in the access network \\
    \textbf{NPM2} & Should support a scalable architecture with standardised open interfaces for multi-domain networking \\
    \textbf{NPM3} & Must enable enhancement of network capabilities through new technologies \\
    \textbf{NPM4} & Should allow seamless integration and intelligent control of various radio ATs \\
    \textbf{NPM5} & Should address the implications on Web 3.0 network routing in fixed and mobile networks \\
    \bottomrule
    \end{tabular}
    \label{tab:network_management}
\end{table}

\begin{enumerate}[wide, labelwidth=!, labelindent=9pt, start=1,label={\bfseries NPM\arabic*}]
    \item \textbf{- Access Network Challenges:}  The system must comprehensively address the access network's performance and functional and non-functional challenges. This includes ensuring efficient data throughput, low latency, robust security, and functional requirements like consistent coverage and reliability.
    \item \textbf{- Scalability and Open Interfaces:}  The architecture should be scalable, functioning on standardised open interfaces that enable multi-domain networking. Both the architecture and all the network components must be built with scalability in mind, adapting to fluctuating network demands and facilitating the integration of diverse network elements. This should be attached to the notion of multi-domain networking when choosing the interconnection model (hub, cascade, or hybrid). Similarly, all open interfaces must consider the increase in requests and adapt accordingly to the traffic needs.
    \item \textbf{- Network Capability Advancement:}  The network must be able to advance its current capabilities by incorporating new, wired and wireless technologies. This forward-thinking approach ensures continual improvement and adaptation to emerging technological trends.
    \item \textbf{- Radio Access Technologies Integration:} The system should allow for the seamless integration and intelligent control of various RATs. This integration is crucial for optimising network performance and efficiently managing the diverse spectrum of technologies.
    \item \textbf{- Web 3.0 Network Routing:}  The network must address the implications of Web 3.0 on network routing, especially in the context of fixed versus mobile networks. This involves adapting to the decentralized and dynamic nature of Web 3.0, ensuring efficient and effective routing strategies.
\end{enumerate}

\subsection{System Performance and Service Delivery Requirements}\label{subsec:system_performance}
This section addresses the critical requirements for system performance and service delivery. It emphasises the need to use and adhere to SLAs, provide uninterrupted connectivity across all services, even when they require adaptable QoS, promote edge processing, and finally enhance the network’s automation. Table~\ref{tab:system_performance} briefly summarises these requirements.

\begin{table}[t]
\renewcommand{\arraystretch}{1.1}
\centering
    \caption{Requirements for system performance and service delivery.}
    \begin{tabular}{lp{0.8\columnwidth}}
    \toprule
    \textbf{Req.} & \textbf{Requirement Description} \\ \midrule
    \textbf{SPSD1} & Must fulfil contracted SLA requirements in performance and service delivery \\
    \textbf{SPSD2} & Must provide capacity management functions to handle SLA requests \\
    \textbf{SPSD3} & Must provide traffic-engineering capabilities and options like network slicing \\
    \textbf{SPSD4} & Must support uninterrupted connectivity and mobility across ATs and networks \\
    \textbf{SPSD5} & Should enhance the application awareness with adaptive QoS and QoE \\
    \textbf{SPSD6} & Must enable mobility for devices and networks across various ATs \\
    \textbf{SPSD7} & Must integrate with intelligent controllers for efficient edge resource utilisation \\
    \textbf{SPSD8} & Should allow densification mechanisms to boost network capacity and resource utilisation \\
    \textbf{SPSD9} & Should enable decentralisation \\
    \bottomrule
    \end{tabular}
    \label{tab:system_performance}
\end{table}

\begin{enumerate}[wide, labelwidth=!, labelindent=9pt, start=1,label={\bfseries SPSD\arabic*}]
    \item \textbf{- SLA Compliance:} The system must consistently meet or exceed the performance and service delivery standards outlined (relevant KPIs) in Service Level Agreements (SLAs). This involves maintaining high reliability, availability, and responsiveness to meet the agreed-upon criteria, ensuring customer satisfaction and trust.
    \item \textbf{- SLA management:} The system must provide capacity management functions at intra-domain levels to handle SLA requests in an aggregated form at the inter-domain level and optimise overall network performance accordingly. The REASON orchestration layer must be able to decompose complex tasks (an SLA of multiple nature) and compose requests of the same nature before presenting them to the intra-domain controller to fulfil the requests.
    \item \textbf{- Traffic-Engineering and Connectivity:} The system should provide advanced traffic-engineering capabilities, including options for network slicing to manage and optimise the flow of data across the E2E chain of a multi-domain network.
    \item \textbf{- Continuous Connectivity:} The architecture must support seamless and continuous connectivity when moving both between and within different ATs and across various networks. This is critical for ensuring a consistent and uninterrupted user experience, particularly in mobile environments.
    \item \textbf{- Application Awareness and Adaptive QoS/QoE:} There should be an enhanced focus on application awareness within the network, coupled with adaptive Quality of Service (QoS) and Quality of Experience (QoE). This approach allows the network to dynamically adjust its parameters to suit specific application needs, thereby optimising the user experience.
    \item \textbf{- Device and Network Mobility:} The system should provide robust support for both device and network mobility, facilitating smooth transitions between and within different ATs. This requirement is crucial for maintaining uninterrupted service in a highly mobile and dynamic network environment.
    \item \textbf{- Intelligent Controller Integration:} The system should provide integration with intelligent controllers for making informed decisions to ensure the efficient utilisation of edge resources. AL/ML should be used for NI to automatically manage the composite mosaic of network functions and associated resources. This integration allows for dynamic and real-time optimisation of network resources, enhancing overall performance and efficiency.
    \item \textbf{- Densification Mechanisms:} The architecture should allow the densification of the network when required. Investigating and implementing densification mechanisms is key to enhancing network capacity and resource utilisation. This involves deploying more network nodes in dense areas to improve coverage and capacity, ensuring high-quality service even in high-demand scenarios.
    \item \textbf{- Decentralisation:} The system should promote decentralisation. REASON should study and define the concept of decentralisation and its impact on network architecture, management, and performance and incorporate these insights into the network design.

\end{enumerate}

\subsection{Orchestration and Control Requirements}\label{subsec:orchestration}
This section outlines the requirements for orchestration and control in the proposed network architecture. It emphasises efficient resource management, secure interconnectivity, E2E orchestration, edge autonomy, and the integration of AI for effective network operation. The orchestrator is responsible for allocating resources and providing services based on the users/application demands of the required service types. These requirements are fundamental to building a network that is agile, responsive, and capable of meeting diverse and evolving service demands. Table~\ref{tab:orchestration} briefly summarises these requirements.

\begin{table}[t]
\renewcommand{\arraystretch}{1.1}
\centering
    \caption{Requirements for orchestration and control.}
    \begin{tabular}{lp{0.8\columnwidth}}
    \toprule
    \textbf{Req.} & \textbf{Requirement Description} \\ \midrule
    \textbf{OC1} & Must provide resource management and Computing as a Service capabilities \\
    \textbf{OC2} & Must support secure, open, and technology-neutral network and application interconnections \\
    \textbf{OC3} & Must incorporate a central unified E2E orchestrator for network-wide control and decision-making \\
    \textbf{OC4} & Should enable decentralised, autonomous edge computing within the network \\
    \textbf{OC5} & Must allow dynamic network and application function optimisations for varying use cases \\
    \textbf{OC6} & Must support collaborative and hierarchical intra- and inter-domain service control and orchestration \\
    \textbf{OC7} & Should define interfaces between E2E orchestration and MEC servers \\
    \textbf{OC8} & Must optimise resource allocation in multi-ATs \\
    \textbf{OC9} & Must provide E2E monitoring and orchestration across multiple domains and tenants \\
    \textbf{OC10} & Should establish interfaces between AI management and network domains for orchestration \\
    \textbf{OC11} & Must provide intelligent profiling to predict optimum resources assigned \\
    \textbf{OC12} & Must be automatically configurable and allow human-machine interactions  \\
    \bottomrule
    \end{tabular}
    \label{tab:orchestration}
\end{table}

\begin{enumerate}[wide, labelwidth=!, labelindent=9pt, start=1,label={\bfseries OC\arabic*}]
    \item \textbf{- Resource Management and Computing as a Service:} The orchestration/controllers should efficiently manage network resources, providing computing capabilities as a service available anywhere in the network. The intra-domain controller/orchestrator should have a way to receive the available resources at the edges, available ATs, and current resource utilisation. This enables a more dynamic and efficient allocation of resources, catering to the varying demands of network users and applications.
    \item \textbf{- Secure and Open Interconnectivity:} The architecture must support secure, open, and technology-neutral interconnectivity of networks and applications. This ensures a flexible and secure environment for diverse technological integrations and interactions.
    \item \textbf{- Unified E2E Orchestrator:} This central orchestrator streamlines network management and ensures consistent policy implementation. The orchestrator is required to make intelligent decisions to ensure efficient utilisation of edge resources regarding the application placement, optimum resource allocation and dynamic adaptations
    \item \textbf{- Edge Autonomy:} The architecture should support decentralised, autonomous edge computing, fully integrated with the overall network system. This facilitates localised decision-making and reduces latency for edge-based applications.
    \item \textbf{- Dynamic Use Case Optimisations:} The system must allow for optimisations tailored to meet the changing requirements of different use cases, especially for disaggregated network and application functions.
    \item \textbf{- Hierarchical Service Control and Orchestration:} The architecture should facilitate the collaboration of hierarchical intra- and cross-domain service control and orchestration, driven by high-level objectives like QoE targets and business/customer-driven KPIs.
    \item \textbf{- E2E Orchestration and MEC Server Interfaces:} Well-defined interfaces between the E2E orchestration plane and MEC (Multi-access Edge Computing) servers are crucial. The orchestrator shall provide appropriate APIs to share the applications' optimum resource requirements (leveraging analytics/monitoring microservices and profiling), QoS constraints, required KPIs to allocate multi-access technology resources dynamically, manage interferences and ensure the applications receive the necessary radio coverage and capacity for their operation.
    \item \textbf{- Resource Optimisation in Multi-ATs:} Orchestration services must interact with all available access network technologies to optimise resource allocation in a multi-access environment. This enhances the efficiency and performance of the network across different ATs.
    \item \textbf{- Multi-Domain, Multi-Tenancy Orchestration and Monitoring:} The system must support multi-domain, multi-tenancy orchestration, providing capabilities for E2E monitoring and real-time updates of services running across different domains. This ensures comprehensive oversight and management of network services. Co-operation based on an ``agreed framework'' of domains is essential and must be provided as a functionality by the domain management system within a service delivery chain. Moreover, the orchestrator should monitor the performance of individual domain controllers in the service delivery chain and make the scaling and resource provisioning decisions.
    \item \textbf{- AI Management Layer Integration:} Establishing interfaces between the AI management layer and network domains is essential for effective orchestration. This integration allows for intelligent, data-driven decision-making, enhancing network performance and efficiency.
    \item \textbf{- Intelligent Profiling:} An intelligent profiling must select the edge/core domains and predict the optimum amount of resources to be assigned to network services at each edge/core domain as well as finding the optimum path between the network functions to address the requested Service Function Chains (SFCs). This will help the intra-domain orchestrator and the MEC Platform Controller assign optimum resource configuration to the requested network services, hosting them on appropriate edges and interconnecting them to meet the required predefined KPIs.
    \item \textbf{- Autonomous Configurability:} The system must be configurable autonomously and on the fly through specific parameter settings or policies. The directives received through higher-level AI/ML model orchestrators should also undertake configuration actions. Human directives should only be undertaken in special circumstances when there is an absolute need to bring the network to a steady-state situation.

\end{enumerate}

\subsection{AI/ML Requirements}\label{subsec:ai_ml}
This section outlines the essential requirements for integrating AI/ML into REASON’s architecture. It emphasises the importance of efficient data collection, cognitive and AI planes for optimisation, systematic AI model lifecycle management, cognitive functions for E2E AI challenges, and ensuring trust and explainability in AI implementations. These elements are crucial to leveraging AI/ML for enhancing network performance and resource utilisation. Table~\ref{tab:ai_ml} briefly summarises these requirements.

\begin{table}[t]
\renewcommand{\arraystretch}{1.1}
\centering
    \caption{Requirements for AI/ML.}
    \begin{tabular}{lp{0.8\columnwidth}}
    \toprule
    \textbf{Req.} & \textbf{Requirement Description} \\ \midrule
    \textbf{ML1} & Must facilitate distributed raw and processed data collection and management for network optimisation \\
    \textbf{ML2} & Should include E2E Cognitive and AI Planes for optimisation and resource management \\
    \textbf{ML3} & Should implement a systematic approach for the lifecycle management of AI models \\
    \textbf{ML4} & Must provide cognitive functions to address challenges in E2E AI implementations \\
    \textbf{ML5} & Should ensure trust, verification, and explainability in E2E AI implementations \\
    \textbf{ML6} & Should define Network Intelligence (NI) as a pipeline of effective AL/ML models  \\
    \bottomrule
    \end{tabular}
    \label{tab:ai_ml}
\end{table}

\begin{enumerate}[wide, labelwidth=!, labelindent=9pt, start=1,label={\bfseries ML\arabic*}]
    \item \textbf{- Efficient Data Collection:} The architecture must enable an efficient way of data collection in a distributed manner. It should be capable of storing and managing both raw and processed data to facilitate knowledge acquisition and enhance resource utilisation within the network. This requirement is vital for the continuous improvement of the network operations.
    \item \textbf{- Cognitive and AI Planes:} The architecture should incorporate an E2E Cognitive Plane and an AI Plane. These planes are responsible for task-oriented optimisation and system-level resource management. Their integration is key to making intelligent, data-driven decisions that improve network efficiency and service quality. Multiple instances of NI can run across the different domains in order to adhere to set KPIs including QoS for service differentiation or QoE guarantees.
    \item \textbf{- AI Model Lifecycle Management:} A systematic approach must be followed for managing the lifecycle of AI models, from their development to retirement. A single NI instance may comprise several AI/ML components operating at different network locations to accomplish an overarching goal. This process should be supported by an AI orchestrator, which oversees AI models' deployment, monitoring, updating, and decommissioning. Effective lifecycle management ensures that AI models remain relevant and efficient throughout their usage.
    \item \textbf{- Cognitive Functions for E2E AI:} The architecture must include cognitive functions to tackle the challenges associated with E2E AI implementation effectively. These functions enable the network to adapt, learn, and make decisions in complex and dynamic environments, enhancing the overall intelligence and responsiveness of the system.
    \item \textbf{- Trust and Explainability:} Trust, verification, and explainability must be considered for the whole E2E architecture. Stakeholders need to understand and trust the decisions made by AI systems. Providing explainable AI capabilities ensures transparency and accountability in automated decision-making processes.
    \item \textbf{- AI/ML Model pipeline:} An NI instance should be defined as a pipeline of effective AI/ML models that swiftly detect or anticipate new requests or fluctuations in network activities and then react to those by instantiating, relocating, or re-configuring network functions in an automated manner.

\end{enumerate}

\subsection{Security Requirements}\label{subsec:security}
This section addresses the key security requirements for the REASON’s architecture. It emphasises the integration of security at every development stage, ensuring the safety of digital identities, API security, data privacy, cross-domain cooperation, and the use of distributed transaction layers for integrity. These elements are critical to building a network that is not only efficient and functional but also secure and trustworthy. Table~\ref{tab:security} briefly summarises these requirements.

\begin{table}[t]
\renewcommand{\arraystretch}{1.1}
\centering
    \caption{Requirements for Security.}
    \begin{tabular}{lp{0.8\columnwidth}}
    \toprule
    \textbf{Req.} & \textbf{Requirement Description} \\ \midrule
    \textbf{SEC1} & Must integrate security into every stage of development \\
    \textbf{SEC2} & Must provide digital identities for all assets with a coherent policy framework \\
    \textbf{SEC3} & Must ensure the security of public and private APIs \\
    \textbf{SEC4} & Must guarantee data privacy both in transit and at rest \\
    \textbf{SEC5} & Must allow secure cooperation across different domains \\
    \textbf{SEC6} & Must establish a distributed transaction layer  \\
    \bottomrule
    \end{tabular}
    \label{tab:security}
\end{table}

\begin{enumerate}[wide, labelwidth=!, labelindent=9pt, start=1,label={\bfseries SEC\arabic*}]
    \item \textbf{- Security by Design:} Security must be an integral part of the development process, following the principles of ``security by design''. The security service capabilities must be provided to the tenants anywhere in the E2E chain of multi-domain networks. This approach ensures that security considerations are embedded from the outset and throughout the lifecycle of the network, reducing vulnerabilities and enhancing overall system integrity.
    \item \textbf{- Universal Digital Identity and Policy Framework:} The architecture must provide a universal digital identity for all assets and establish a coherent policy framework. This is essential to support secure mobile access across deployed instances and ensure consistent management and authentication of network resources.
    \item \textbf{- API Security:} The system must ensure the security of both public and private APIs. The architecture must safeguard these interfaces against unauthorised access and threats, thereby maintaining the integrity and confidentiality of the network services and data exchanged through these APIs.
    \item \textbf{- Data Privacy Assurance:} The architecture must uphold data privacy during transit and at rest. This involves implementing robust encryption and other security measures to protect sensitive information from unauthorised access or breaches, thereby maintaining user trust and compliance with data protection regulations. Some more test and measurement data requirements can be found in Sec.~\ref{subsec:test}.
    \item \textbf{- Cross-Domain Cooperation with Security:} The system should facilitate secure cooperation across different domains while preserving the authority, integrity, and confidentiality of each domain's resources. This requirement ensures that collaborative efforts do not compromise the security posture of individual network segments.
    \item \textbf{- Distributed Transaction Layer:} A distributed transaction layer, possibly utilising blockchain technology, must be established. This layer would maintain the integrity of transactions across the network, offering a secure and transparent mechanism for recording and validating transactions.
\end{enumerate}

\subsection{Sustainability/Energy Efficiency  Requirements}\label{subsec:energy}
This section focuses on the proposed network architecture's sustainability and energy efficiency requirements. The emphasis is on reducing energy consumption, developing energy-aware optimisation algorithms, and providing comprehensive energy measurement across the network. These requirements are key to ensuring that the network is not only technologically advanced but also environmentally responsible and sustainable. Table~\ref{tab:energy} briefly summarises these requirements.

\begin{table}[t]
\renewcommand{\arraystretch}{1.1}
\centering
    \caption{Requirements for energy efficiency and sustainability.}
    \begin{tabular}{lp{0.8\columnwidth}}
    \toprule
    \textbf{Req.} & \textbf{Requirement Description} \\ \midrule
    \textbf{SUS1} & Should decrease overall energy use in the network architecture \\
    \textbf{SUS2} & Should develop novel mechanisms and AI-powered algorithms for energy efficiency \\
    \textbf{SUS3} & Must measure the energy consumption of services and data exchange across domains \\
    \bottomrule
    \end{tabular}
    \label{tab:energy}
\end{table}

\begin{enumerate}[wide, labelwidth=!, labelindent=9pt, start=1,label={\bfseries SUS\arabic*}]
    \item \textbf{- Energy Consumption Reduction:} The architecture must prioritise reducing energy consumption and improving efficiency in its utilisation. This involves introducing energy-saving technologies and strategies to minimise the overall energy footprint of the network, ensuring it operates in an environmentally conscious manner.
    \item \textbf{- Innovative Energy Efficiency Mechanisms:} The architecture should allow integration with novel energy-aware optimisation algorithms. AI-powered algorithms can be developed to enhance the network's energy efficiency and sustainability. These innovations could include intelligent resource allocation, predictive maintenance, and automated adjustments to network operations based on real-time energy usage data, all contributing to a more sustainable network ecosystem.
    \item \textbf{- Comprehensive Energy Measurement:} Across all system components, the architecture must provide a method for accurately measuring the energy consumed by services running across different domains, as well as the data exchanged over various interfaces. This capability is essential for identifying high-energy-demand areas, enabling targeted improvements, and ensuring accountability in energy usage across the network.
\end{enumerate}

\subsection{Policy/Cooperation Requirements}\label{subsec:policy}
This section highlights the essential requirements related to policy and cooperation in the context of future network architecture. It emphasises the need for a comprehensive policy framework to support mobile access across deployed instances and strict adherence to data protection laws. These elements are critical for ensuring that the network operates within legal and regulatory boundaries while providing consistent and secure access. Table~\ref{tab:policy} briefly summarises these requirements.

\begin{table}[t]
\renewcommand{\arraystretch}{1.1}
\centering
    \caption{Requirements for policies and cooperation.}
    \begin{tabular}{lp{0.8\columnwidth}}
    \toprule
    \textbf{Req.} & \textbf{Requirement Description} \\ \midrule
    \textbf{PC1} & Must establish a policy framework to support mobile access across all instances \\
    \textbf{PC2} & Must adhere to data protection laws, such as the UK Data Protection Act \\
    \bottomrule
    \end{tabular}
    \label{tab:policy}
\end{table}

\begin{enumerate}[wide, labelwidth=!, labelindent=9pt, start=1,label={\bfseries PC\arabic*}]
    \item \textbf{- Policy Framework for Mobile Access:} The architecture must incorporate a robust policy framework designed to support mobile access across all deployed instances. This framework should facilitate seamless and secure connectivity for users, regardless of their location or the type of network they are accessing. It should also ensure consistent policy enforcement across various network domains, providing a uniform user experience and maintaining network integrity.
    \item \textbf{- Compliance with Data Protection Laws:} The system must rigorously comply with relevant data protection laws, including but not limited to the UK Data Protection Act. This compliance involves implementing measures to safeguard personal data against unauthorized access, processing, alteration, or disclosure. The architecture should embed privacy by design principles, ensuring that user data is protected at all stages of the data lifecycle and that the network operations align with legal and regulatory standards.
\end{enumerate}

\subsection{Test and Measurements Requirements}\label{subsec:test}
Many of the preceding sections identify a need for monitoring and measurement to support validation (does it work) and verification (does it do what we want it to) of proposals or results. Overall, as a support for benefits demonstration and realisation, the REASON architecture must demonstrate a coherent, consistent, overall approach to test and measurement (T\&M) that meets those needs. Briefly, these requirements are summarised in Table~\ref{tab:test}.

\begin{table}[t]
\renewcommand{\arraystretch}{1.1}
\centering
    \caption{Requirements for test and measurements.}
    \begin{tabular}{lp{0.8\columnwidth}}
    \toprule
    \textbf{Req.} & \textbf{Requirement Description} \\ \midrule
    \textbf{TMP1} & Should be able to collect, store, process, originate or terminate data for all architectural elements \\
    \textbf{TMP2} & Should be able to collect, store, process, originate or terminate data for all functions implemented \\
    \textbf{TMP3} & Executed T\&M functions must not impact the system’s operation \\
    \textbf{TMP4} & T\&M functions must be authenticated and authorised \\
    \bottomrule
    \end{tabular}
    \label{tab:test}
\end{table}

\begin{enumerate}[wide, labelwidth=!, labelindent=9pt, start=1,label={\bfseries TMP\arabic*}]
    \item \textbf{- Collect, store and process for architecture elements:} The architecture should allow for functions to collect, store and process all data coming from various architectural elements. Depending on the element, different functionalities may be considered (e.g., periodic or on-demand retrieval of data, being retrieved by stateful or stateless APIs, etc.). Moreover, the data may originate or be terminated, again depending on the architecture element and the functionality required.
    \item \textbf{- Collect, store and process for functions:} Similarly to the above, all functions provided within an architecture component should allow collection, storage and processing of the data generated. The same considerations should be taken with respect to the direction of data, the periodicity, and the retrieval mechanisms for this requirement.
    \item \textbf{- Execution of T\&M functions:} All monitoring, test and measurement functions implemented should not impact the operation of the deployed system. The functionality implemented should be validated for the overhead introduced and be optimised or replaced with more lightweight approaches when bottlenecks are identified.
    \item \textbf{- Authentication and authorisation of T\&M functions:} Finally, security and privacy considerations must be taken into account. Originating data must be authenticated and authorised, and terminating data must be authenticated and authorised subject to access control policies that define what data can be received.
\end{enumerate}

\section{Reference Architecture}\label{sec:reference_architecture}
The preceding sections present a diverse range of requirements for devices (large and small, simple and complex, fixed and mobile, personal and enterprise), connectivity (fixed, mobile, dynamic, private, public—and various combinations of these), information, and service enablers that provide the advanced digital infrastructure of the REASON platform to support the use cases and platform management services.

The remainder of the paper details the REASON architecture and the envisaged digital infrastructure. It provides a detailed layered and functional block view for the proposed architecture, considering all aspects from the hardware to the security and explaining the reasoning behind each decision. The proposed framework addresses the ever-increasing demands of modern connectivity, accommodating the complex interplay between technology, data, and user experiences.

 \begin{figure*}[t]
    \centering
    \includegraphics[width=0.7\textwidth]{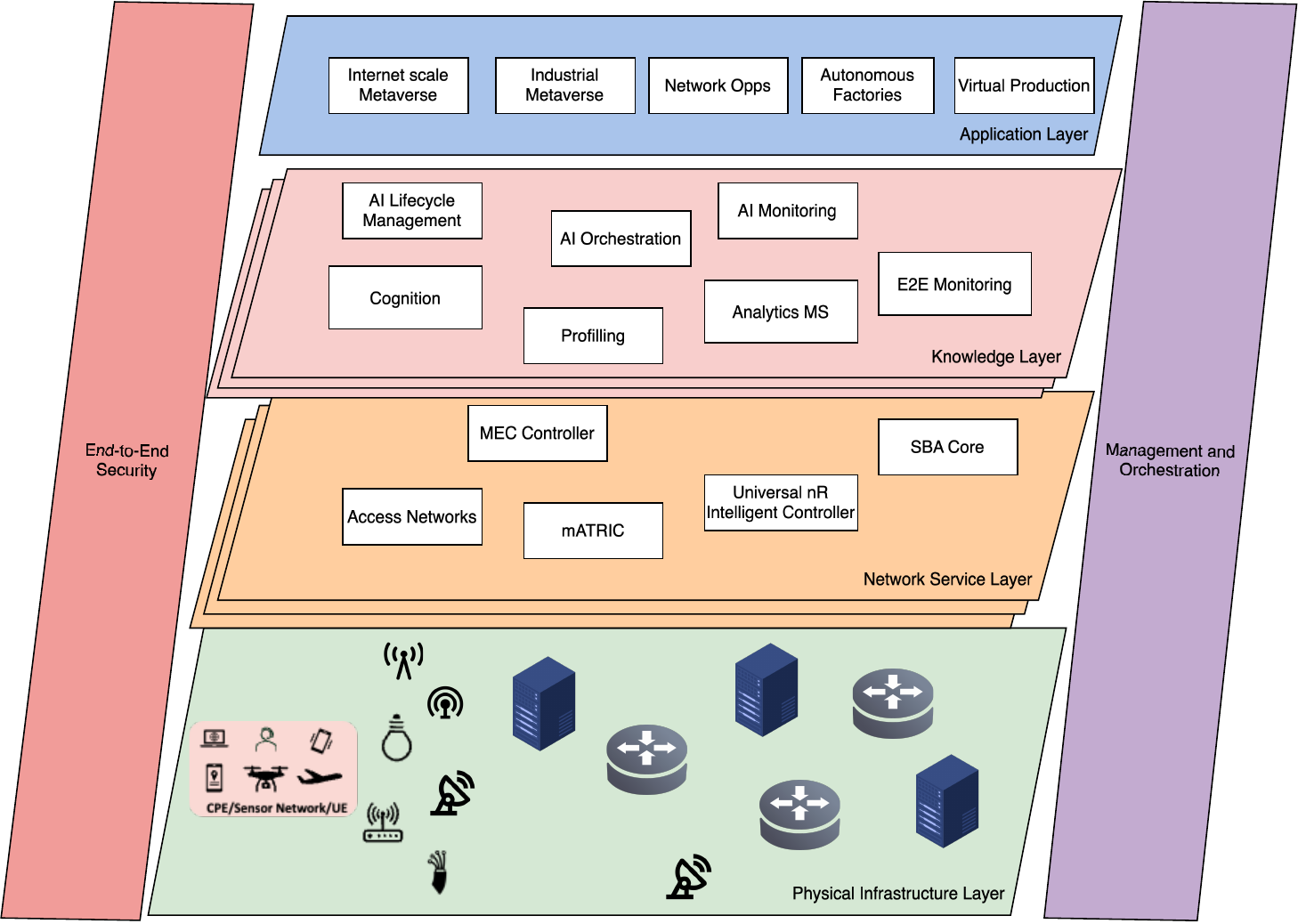}
    \caption{REASON Layer Architecture. Each layer serves a unique purpose, providing a structured approach that simplifies the function block design and management.}
    \label{fig:REASONLayer}
\end{figure*}

\subsection{Layered View}
We start with a high-level layered view of the system. Such an architecture design paradigm offers a holistic perspective on the structure and functionality of future open networks and allows for:
\begin{enumerate}
    \item \textbf{Modularity:} Layered architectures break the network into manageable, discrete layers, each responsible for specific functions. This modularity allows for easier development, maintenance, and updating of each layer independently without affecting the others.
    \item \textbf{Interoperability:} By standardising the functions of each layer, different systems and devices can communicate effectively. This is crucial in a diverse ecosystem like 6G, where multiple devices and technologies must interact seamlessly.
    \item \textbf{Scalability:} As network demands grow, a layered architecture can more easily accommodate scaling. You can upgrade or modify specific layers to handle more traffic or add new functionalities without overhauling the entire network.
    \item \textbf{Simplified Troubleshooting:} In a layered architecture, identifying and rectifying network issues becomes more straightforward. Problems can be isolated to specific layers, allowing for targeted diagnostics and solutions.
    \item \textbf{Flexibility and Innovation:} Different layers can evolve at their own pace, allowing for innovation in specific areas without requiring complete system overhauls. This encourages experimentation and adoption of new technologies.
    \item \textbf{Enhanced Security:} Security measures can be implemented at various layers, providing a more robust and multi-faceted defence against threats. This layered security approach is particularly important in 6G networks, where security concerns are paramount.
    \item \textbf{Standardisation and Compatibility}: Layered models often lead to industry-wide standards, ensuring compatibility across different vendors and technologies. This standardisation is crucial for widely adopting and integrating 6G technologies.
    \item \textbf{Easier Implementation and Training}: With clearly defined layers, it's easier for engineers and technicians to understand, implement, and manage network architectures. Training personnel becomes more straightforward, as the focus can be on individual layers rather than the entire network.
\end{enumerate}

The proposed architecture comprises four distinct horizontal layers—namely, the Physical Infrastructure Layer, Network Service Layer, Knowledge (Cognitive and AI Plane) Layer, and End-User Application Layer— and two vertical ``layers'' – Management and Orchestration and E2E Security, as illustrated in Fig.~\ref{fig:REASONLayer}.

Each layer serves a unique purpose, from the foundational hardware and infrastructure to the intelligence and applications that empower end users. By providing a structured, layered approach, this framework not only simplifies network design and management but also paves the way for innovation, scalability, and adaptability in an era where connectivity is more vital than ever before. For traceability, we provide a cross-reference to each layer's requirements.

\subsubsection{Physical Infrastructure Layer}
In our framework, the Physical Infrastructure Layer represents the foundational layer of the architecture. This layer comprises all the tangible hardware and facilities that form the network's backbone addressing \textbf{Environmental Versatility (Req. NDS1)} comprising robust hardware and facilities. It integrates \textbf{Technological Integration (Req. NDS2)} through servers, switches, routers, data centres, cables, and other physical assets. The role of this layer is crucial for tackling \textbf{Access Network Challenges (Req. NPM1)}, ensuring \textbf{Continuous Connectivity (Req. SPSD4)}, and providing the necessary resources and connectivity to support network operations. It lays the groundwork for \textbf{Resource Management and Computing as a Service (Req. OC1)} by facilitating reliable data transmission, storage, and processing, and finally contributes to \textbf{Data Privacy Assurance (Req. SEC4)}, ensuring that data is managed and processed reliably and efficiently.

\subsubsection{Network Service Layer}
The Network Service Layer focuses on defining the logical design and configuration of network services, addressing \textbf{Interfaces and Network Function Disaggregation (Req. ON4)} and \textbf{Scalability and Open Interfaces (Req. NPM2)}. This layer establishes the protocols, standards, and specifications governing data governance within the network. It includes elements such as network protocols, data formats, and service interfaces, addressing \textbf{Application Awareness and Adaptive QoS/QoE (Req. SPSD5)} by understanding the needs of different applications (awareness) and adapting the network’s response, ensuring \textbf{Continuous Connectivity (Req. SPSD4)} ensuring seamless handover and communication without interactions and maintaining \textbf{API Security (Req. SEC3)} through authentication, authorisation, encryption for the service interfaces. The Network Service Layer ensures that the network infrastructure can deliver the required services effectively, including connectivity, security, and QoS, addressing \textbf{Secure and Open Interconnectivity (Req. OC2)}.

\subsubsection{Knowledge Layer:}
The Knowledge Layer encompasses technologies like AI and ML that leverage data analytics to extract insights and optimise network operations. This is based on \textbf{Efficient Data Collection (Req. ML1)}, where raw and processed data are used for intelligent decisions. It involves developing and deploying intelligent algorithms and models to enhance network performance, security, and decision-making, addressing the \textbf{Cognitive and AI planes (Req. ML2)}. This layer contributes to the network's ability to adapt, self-optimise, and predict issues.

Targeting an AI-native approach that aligns with what is envisaged in 6G intelligent systems, the REASON platform and deployed networks will need to systematically control the lifecycle of AI models. The AI lifecycle, addressing \textbf{AI Model Lifecycle Management (Req. ML3)}, encompasses the stages involved in developing, deploying, maintaining, and withdrawing AI-enhanced systems. Processes in these stages include problem definition, data acquisition and pre-processing, model development and training, model evaluation and validation, deployment, monitoring and maintenance, retraining and improvement, and ends with retirement or replacement.

Within \textbf{AI Model Lifecycle Management (Req. ML3)}, the accuracy and effectiveness of AI models should also be considered. Starting with the data fed to a model, mechanisms that detect drifts and irregularities must be implemented. Various methods have been proposed in the literature for that. For example,~\cite{le3dDataDrift} proposes an ensemble data drift detector for time-series data,~\cite{driftDetectionWeakData} provides a framework for when data labelling is infeasible, or~\cite{dataGeneration,dataGeneration2} present various strategies for synthetic data generation, that could be used for training detection algorithm. Enabling proactive drift detection can prevent disruptions and data bias in the state of an application.

Moreover, the concept drift is of paramount importance in \textbf{AI Model Lifecycle Management (Req. ML3)}. From the ML model's perspective, the underlying relationship between the input data and expected output (target) changes over time, leading to an underfitted model. This behaviour is called concept drift~\cite{smartcities4010021}. It can occur for several reasons, e.g., long-term changes in the data (e.g., environmental changes), faulty hardware (sensor drift), or even adversarial actions, such as data poisoning attacks. These scenarios can be detrimental to ML predictions' quality when considering large-scale systems with misbehaving ``production'' models. Solutions in the literature have used either supervised~\cite{BAYRAM2022108632} or unsupervised solutions~\cite{FLAME}, usually providing ways to monitor the performance of a model and metrics for evaluating the drift. As in the data drift, proactive detection can enhance the model's performance and overall AI lifecycle.

To support native AI, a view will be required at the E2E system level and the service level to:
\begin{itemize}
    \item Analyse when a new model is required (i.e., what triggers the AI Orchestration?)
    \item How to optimise (distributed) data collection, compression and caching that feed AI models.
    \item What the domain and purpose of the deployed ML models are.
    \item How to analyse the performance of a model in service delivery and its impact on the system.
    \item How to reuse already trained and deployed models to target different purposes in the system.
    \item Coordinate and synchronise the process stages in the AI lifecycle, including conflict resolution and validation.
\end{itemize}

As part of the Knowledge Layer, two logical cross-cutting planes are being defined: the AI and Cognitive planes. Both planes ensure data-driven decision-making while addressing \textbf{Data collection for architectural elements (Req. TMP1)}. The manages tasks such as model placement, chaining, and computational resource allocation, ensuring that AI models are in line with privacy, security, and regulatory requirements, and allowing human intelligence to augment automated decisions when necessary, as discussed at \textbf{Autonomous Configurability (Req. OC12)}.

\paragraph{AI Plane}
The AI Orchestrator will manage and control the AI lifecycle. This orchestrator will maintain a catalogue of registered AI models, capturing key details such as versioning, descriptions, and domain-specific requirements. It will enable version control, automate training pipelines, manage data, and support the deployment and scaling of models across various environments. Additionally, it monitors model performance, ensures alerts for anomalies, and enforces robust security and access controls. The AI Orchestrator facilitates model deprecation and retirement while providing scalability and high availability mechanisms.

MLOps is a set of practices and principles aimed at streamlining and automating the E2E ML lifecycle, from model development to deployment and maintenance.  MLOps can provide a formal framework and practices to manage AI models and workflows required for AI orchestration, addressing \textbf{AI Management Layer Integration (Req. OC10)}, where seamless coordination, scalability, and trustworthy deployment of AI solutions are key as per \textbf{Trust and Explainability (Req. ML5)}.

XAI offers justifications for AI model outputs in human-understandable terms, improving trust and verification, particularly in complex systems like 6G networks. To ensure trustworthiness, AI explanations must be clear and accurate and include uncertainty quantification to inform users about output reliability. XAI methods fall into three categories: inherently interpretable models (e.g., linear regression), model-agnostic techniques (e.g., LIME, SHAP), and model-specific methods for particular models like deep learning.

We will use Shapley value-based explanations to assess the contribution of each feature to a model’s predictions. Originating from cooperative game theory, Shapley values provide a fair and consistent method for attributing feature importance~\cite{pmlr-v119-sundararajan20b}. Calculating these values breaks down complex models into interpretable components, helping clarify why specific predictions are made. This approach offers insights into model performance, potential biases, and areas for improvement, ensuring AI systems remain interpretable and trustworthy throughout their lifecycle. In REASON, we will demonstrate how model-agnostic explanation modules can be integrated into an AI model's lifecycle as part of AI Orchestration (with the help of the Cognitive Plane) across various use cases, enhancing transparency and decision-making.

Additional tasks of the AI Orchestrator include:

\begin{itemize}
    \item Selection of AI models (and the right AI models with the help of the cognitive plane).
    \item Placement of AI models.
    \item Chaining AI models, as per \textbf{AI/ML Model pipeline (Req. ML6)}.
    \item Computational resource allocation (CPUs, GPUs, TPUs).
    \item Monitoring AI models (through AI monitoring).
    \item Split learning and distributed computing.
\end{itemize}

Finally, this plane is responsible for the security and compliance of the provided models, thereby supporting \textbf{Universal Digital Identity and Policy Framework (Req. SEC2)} concerning the deployment and management of AI models.

\paragraph{Cognitive Plane}
We introduce cognitive functionality introduced in \textbf{Cognitive functions for E2E AI (Req. ML4)} as part of the cognition component to tackle the challenges of E2E AI in 6G networks mentioned above. The cognitive plane should have direct communication with the service delivery entities (e.g., intent engines, task translators, business mapping) to evaluate the performance of AI models as part of a system trying to deliver a service with different stakeholders involved. These stakeholders may require explanations about the system's behaviour to verify that it is behaving as expected. The cognitive plane will help AI orchestration ensure compliance and governance with ethical and regulatory standards and provide organisations with a holistic approach to maximise AI model value while mitigating operational challenges and risks.

The main tasks of the cognitive plane include:
\begin{itemize}
    \item Reason about ML models and their context in the system.
    \item Ensure the model is satisfying privacy and security requirements.
    \item Ensure that the model satisfies the required explainability, interpretability, trustworthiness and verifiability requirements.
    \item Ensure that the model is satisfying regulatory or compliance requirements.
    \item Check (in conjunction with the AI Orchestrator) for any data drift that might require unscheduled retraining of the model.
    \item Reactive and proactive conflict mitigation and resolution.
    \item Human intelligence, when automated decisions are impossible, the cognitive plane should allow user input to allow a human-in-the-loop approach.
\end{itemize}

\paragraph{Implications of E2E AI}\label{implications}
AI models are only as good as the data they are trained on. In REASON, we are focused on implementing a highly detailed and real-time AI monitoring system as a foundational element of the AI-native network architecture. This monitoring system enables the efficient use of network performance and effectiveness across various layers of the network stack. We propose a disaggregated, distributed micro-service-based approach that allows data producers to advertise their data pipelines and enables consumers to subscribe to the data they are interested in. This approach provides E2E data observability, which is crucial for a comprehensive approach to native intelligence. Successfully implementing E2E AI requires careful consideration of observability, distributed data, and appropriate controls and access for different nodes within the network. Furthermore, deploying models across the existing diverse fabric of the network requires clear context and standardised metadata regarding the model's purpose, inputs and outputs, execution environment, and runtime data monitoring. We aim to consider all these aspects in our end-to-end AI approach, further described in Sec.~\ref{subsection:key_functional_blocks}.

Lifecycle management of trustworthy AI models is another critical focus for REASON, emphasising the training, deployment, and continuous updating of AI models used within the network. As network conditions and requirements evolve, developing methodologies for continuous improvement ensures that these models remain effective and relevant. Integrating AI into 6G networks brings new challenges related to trust, reliability, and ethical considerations. It is crucial to address issues such as fairness, transparency, privacy, and verifiability, especially when supporting essential infrastructures like future networks. Through prioritising these aspects, we aim to build a robust framework that fosters confidence in AI technologies.

\subsubsection{End-User Application Layer}
At the topmost layer of the architecture is the End-User Application Layer. This layer is the interface where users engage, demanding \textbf{Application Awareness and Adaptive QoS/QoE (Req. SPSD5)} to ensure an intuitive and frictionless experience tailored to the diverse needs of consumers, tenants, enterprise customers, and content providers. It includes web applications, mobile apps, software solutions, and content delivery platforms. This layer also encapsulates the essence of \textbf{Platforms and Products (Req. ON1)} by allowing vendors and service providers to provide services aligned with Open RAN principles, standard compliance, and demonstrating interoperability and implementation neutrality. The End-User Application Layer aims to provide a seamless and intuitive user experience, meeting the diverse needs of consumers, tenants, enterprise customers, and content providers. It aligns with business objectives and ensures that the network serves as an enabler for various user-facing functionalities. Finally, the efficient delivery of services at this layer also contributes indirectly to \textbf{Energy Consumption Reduction (SUS1)}, as it optimises the delivery of network resources to end-user applications, thereby playing a vital role in reducing the overall energy footprint. This layer ensures the network not only meets technical specifications but also serves as a cornerstone for value-adding, user-centric functionalities. This layer should adhere to a \textbf{Policy Framework for Mobile Access (PC1)}, enabling consistent and reliable user interaction.

\subsubsection{Network Management and Orchestration}
Network Management and Orchestration is a critical component in our framework, serving as the backbone for \textbf{Resource Management and Computing as a Service (Req. OC1)}. It is the component responsible for handling Service Level Agreements (SLAs) requests as per \textbf{SLA Management (Req. SPSD2)}, that will adapt the performance and delivery of the system, necessary for \textbf{SLA Compliance (Req. SPSD1)}. It performs operations on assets in the four layers, embodying strategies for deployment as per \textbf{Deployment Strategies (Req. NDS3)}, and is responsible for managing, controlling, and orchestrating network resources and services. Through the sophisticated network management, orchestration platforms, and automation tools provided, this layer is instrumental in Hierarchical \textbf{Service Control and Orchestration (Req. OC6)} and \textbf{Multi-Domain, Multi-Tenancy Orchestration and Monitoring (Req. OC9)}, facilitating the seamless scaling and provisioning of services. Part of the monitoring should be a comprehensive suite of energy measurement tools for both data and services, adhering to \textbf{Comprehensive Energy Measurement (Req. SUS3)}. It is also critical for \textbf{Traffic-Engineering and Connectivity (Req. SPSD3)}, optimising network traffic to maintain constant connectivity. The layer ensures that network resources are allocated efficiently, services are provisioned and scaled as needed, and performance is monitored and optimised. Importantly, the NI provided should consider energy optimisation mechanisms across all domains in the system, as per Innovative \textbf{Energy Efficiency Mechanisms (Req. SUS2)}. All these are achieved by integrating intelligent controllers, which denotes an advanced \textbf{Intelligent Controller Integration (Req. SPSD7)}. It plays a pivotal role in maintaining the health and agility of the network.

\subsubsection{End-to-End Security }
The E2E Security Layer in REASON is the cornerstone for maintaining \textbf{Security by Design (Req. SEC1)} throughout the network, ensuring the integrity, confidentiality, and resilience of data and communications across the entire network. In the context of 6G, where data volumes and critical applications are prevalent, robust security is paramount. This layer is responsible for implementing cutting-edge encryption, authentication, and access control mechanisms to safeguard data as it traverses from user devices through the network's multiple domains and services, ensuring \textbf{Data Privacy Assurance (Req. SEC4)}, while also contributing to a robust \textbf{API Security (Req. SEC3)} framework. It also leverages advanced AI and ML to detect and mitigate emerging threats and vulnerabilities in real-time, supporting the cognition within the network as discussed in \textbf{Cognitive Functions for E2E AI (Req. ML4)}. The monitoring functionality implemented should ensure that all functions are authenticated and authorised as per \textbf{Authentication and Authorisation (Req. TMP4)}. Moreover, the E2E Security Layer not only guards against external attacks but also ensures the trustworthiness of the entire network infrastructure and across all the different domains as per \textbf{Cross-Domain Cooperation with Security (Req. SEC5)}, including hardware and software components, or employing strategies as \textbf{Distributed Transaction Layer (Req. SEC6)}. In 6G, E2E security is not just a feature; it's an essential foundation for enabling the broad range of innovative and sensitive applications that 6G networks are envisioned to support.

\bigbreak

In summary, this architecture framework provides a structured approach to understanding and designing a future open network architecture. The Physical Infrastructure Layer forms the foundation, the Network Service Layer defines logical services, the Knowledge Layer introduces intelligence and analytics, the Network Management and Orchestration Layer ensures efficient resource management, and the End-User Application Layer delivers the final user experience. This holistic approach helps solution and network architects create robust and adaptable network architectures that align with business goals and technological advancements.

\subsection{Key Function Blocks}\label{subsection:key_functional_blocks}
\begin{figure*}[t]
    \centering
    \includegraphics[width=0.9\textwidth]{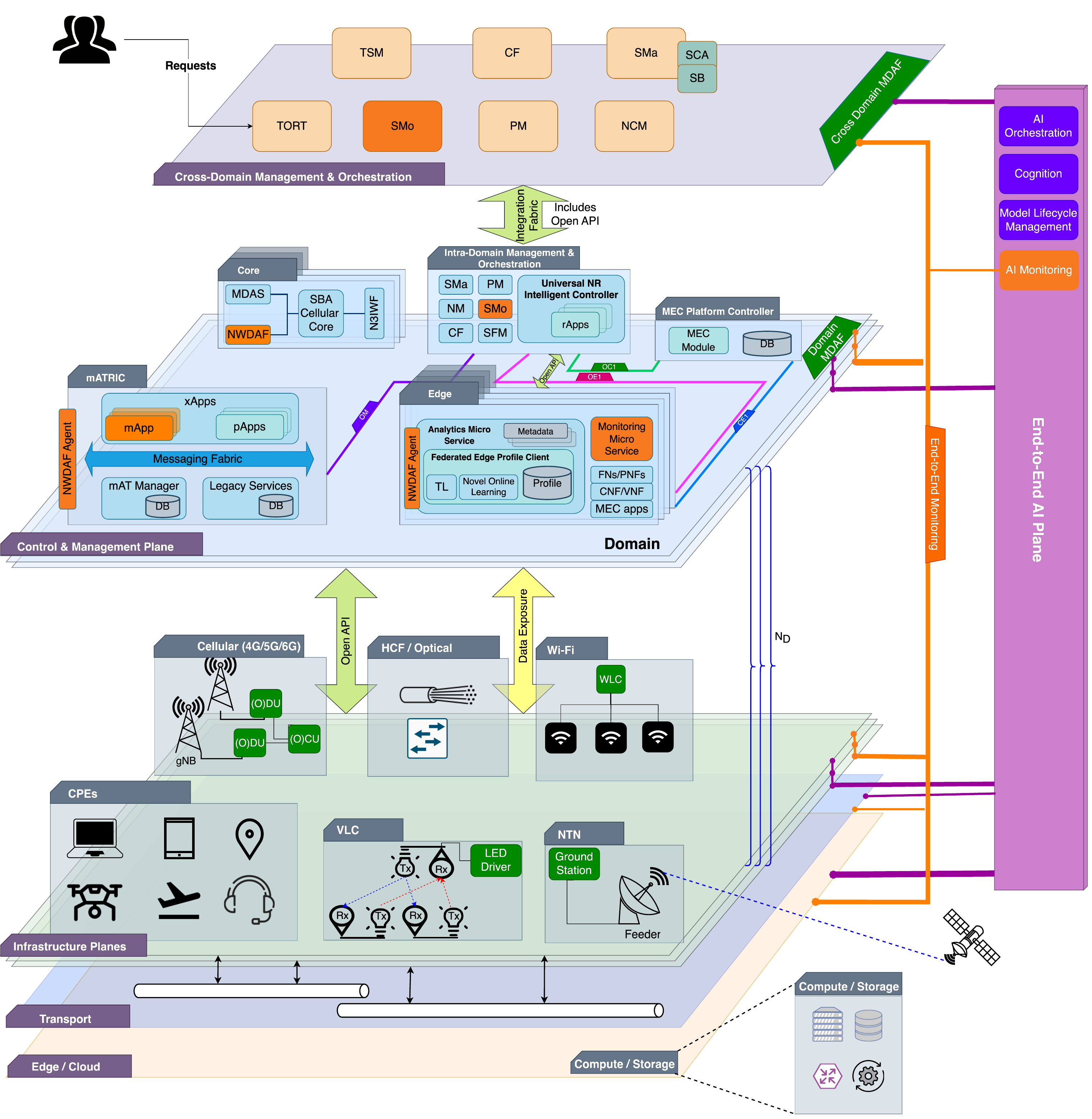}
    \caption{REASON Architecture: All function blocks resemble features and functionality envisaged to be found within the various planes of the system.}
    \label{fig:REASONArch}
\end{figure*}

The REASON architecture is informed by various standard reference architectures, including ORAN, ETSI NVF/MANO, ETSI MEC, ETSI ZSM and 3GPP. We present a detailed function block diagram of the envisaged architecture in Fig.~\ref{fig:REASONArch}. This diagram shows the innovative and forward-looking approaches introduced by REASON, designed to meet the evolving needs of future open networks. In an era characterised by rapid technological advancement, increasing connectivity demands, and the emergence of disruptive technologies such as AI/ML and Open RAN, architectures must adapt to ensure flexibility, scalability, and adaptability. Each block is critical in shaping the network's functionality and capabilities. As before, we will align these key blocks with the requirements outlined earlier in the paper.

All functional blocks presented in Fig.~\ref{fig:REASONArch} are key for REASON’s architecture. As shown in the figure, we have:
\paragraph{Access Technologies}
ATs in REASON serve as the entry points for user devices, providing high-speed and diverse connectivity options. ATs functional block is very important for the \textbf{Technological Integration (Req. NDS2)} as it enables the integration of several wired or wireless interfaces. These could range from terahertz communication to advanced satellite links, ensuring ubiquitous and ultra-reliable connectivity for various use cases and devices, as per \textbf{Environmental Versatility (Req. NDS1)}. Finally, the technologies considered should support indoor and outdoor deployments, as per \textbf{Indoor and Outdoor Network Focus (Req. NDS5)}. A collection of ATs forms an Infrastructure Plane (Fig.~\ref{fig:REASONArch}).

Different ATs depend on different situations and KPIs and can be used in parallel or adaptively. By continuously monitoring network KPIs, the network orchestrator can dynamically switch between or combine these technologies to maintain predefined QoS levels and meet specific network requirements. This enables the efficient and flexible use of network resources to support a wide array of use cases, from consumer applications to critical industrial processes. Some example ATs considered as part of REASON are the following:

\textbf{Cellular Technologies}: Cellular technologies are crucial for the REASON project, providing the backbone for advanced connectivity solutions. These technologies support diverse services such as eMBB, URLLC and mMTC. With the advent of 6G, cellular networks will support applications like holographic communications, digital twins, and pervasive AI. Key features like network slicing and MEC are essential for real-time processing and quick response times needed in autonomous driving and augmented reality. The REASON project leverages cellular technologies to ensure robust, high-performance, and adaptable connectivity for next-generation applications and services.

\textbf{LiFI}: LiFi is a revolutionary wireless technology using light (visible, ultraviolet, and infrared) to transmit data. It can be very important for REASON as it achieves data rates up to 100 Gbps~\cite{liFi} and the potential for higher speeds in the future. LiFi is ideal for data-intensive applications like high-definition video streaming, AR, and VR. Its confined nature enhances data security, which is crucial for environments such as hospitals, military installations, and financial institutions. Its immunity to electromagnetic interference makes it suitable for sites where Radio Frequency (RF) communications are impractical or unreliable, such as industrial sites, hospitals with sensitive equipment, and aircraft cabins. Integrating LiFi into the REASON project allows for seamless incorporation with other ATs, enhancing connectivity in RF-challenged environments or those needing extra bandwidth.

\textbf{WiFi}: WiFi remains a cornerstone for wireless connectivity across various residential, commercial and public environments. The latest standards, WiFi 6 (802.11ax) and the upcoming WiFi 7 (802.11be), enhance data rates, capacity, and efficiency. Such features can support the REASON project's goal of reliable high-speed connectivity in dense user areas. WiFi’s continuous advancements, such as beamforming and target wake time,  and widespread use are integral to the REASON architecture, providing high-speed internet in diverse settings and supporting latency-sensitive applications like video conferencing and IoT. These features align with REASON's objectives of seamless networking and advancing towards 6G technology.

\textbf{Satellite}: Satellite communication is vital for coverage in remote, rural, and underserved areas with limited terrestrial networks. Modern systems, including Geostationary Orbit (GEO), Medium Earth orbit (MEO), and Low Earth Orbit (LEO) satellites, provide essential capabilities for comprehensive network coverage. LEO constellations like Starlink and OneWeb offer low-latency (20-40 ms) and high-bandwidth connectivity comparable to terrestrial broadband services~\cite{satellite}. In the REASON project, satellite technology ensures ubiquitous connectivity and adds resilience and redundancy. In cases of terrestrial network failures due to natural disasters or other disruptions, satellites maintain connectivity, supporting emergency response and critical communications, aligning with REASON’s goal of a robust network.

\textbf{Fibre Optics}: Fibre optic communication is crucial for modern telecommunications, providing high-speed, reliable, and low-latency connectivity. It's ideal for bandwidth-intensive applications like HD video streaming, cloud computing, and data centre connectivity. The low-latency nature of fibre optics is essential for real-time applications such as online gaming, financial transactions, and telemedicine. Fibre optic networks are highly scalable, accommodating increasing bandwidth demands by upgrading optical equipment. This scalability supports the REASON project's network infrastructure and future-proofs it against evolving technology requirements.

\bigbreak

While the above technologies are envisaged to be part of REASON's architecture and will be integral for 6G networks, further extensions may not be limited to just them. The way we plan to interface with all ATs (briefly described in Sec.~\ref{par:matric}) will provide a way for new technologies to be incorporated within the architecture when they mature or when future use cases will require them.

\paragraph{Transport} The Transport Network in REASON networks is the backbone for ultra-fast and reliable transmission of massive data volumes across various geographical locations. It's designed to support the unprecedented throughput and low-latency requirements of 6G applications. This block ensures seamless, high-capacity data transfer across the network infrastructure by leveraging advanced technologies such as terahertz communication, optical networking, holocore-fibre, and advanced protocols. Different infrastructure planes can communicate with one or multiple other infrastructure planes through data pipes formed in the transport layer. The transport network plays a crucial role in facilitating the interconnection between different components of the network, as per the \textbf{Interconnect and Network Hierarchy (Req. ON2)}, providing the necessary infrastructure for the transmission of data between edge devices, edge compute resources, and the core cloud, forming the underlying fabric that supports the diverse requirements of 6G services and applications.

\paragraph{Edge/Core Cloud}
The Edge/Core Cloud is the decentralised, high-performance computing and storage hub within 6G networks, as required from \textbf{Decentralisation (Req. SPSD9)}. The core infrastructure hosts and orchestrates a wide array of services and applications. This block supports complex and resource-intensive applications that require significant computational power and scalability, as per \textbf{Edge Autonomy (Req. OC4)}. It manages network resources, enables seamless connectivity between various edge components, and supports the orchestration of services across the network. It acts as the foundation for the network's intelligent and adaptive capabilities, providing the necessary computing infrastructure for the diverse range of 6G services and applications. The Edge/Core Cloud continuum moving away from the monolithic network architectures and moving towards modular, flexible configurations means that functions that were traditionally bundled together, like routing, switching, and application services, can now be disaggregated. This partially addresses the requirement for \textbf{Network dis-integration (Req. ON6)}.

\paragraph{Domain}
The domain represents a logical and administrative boundary within the network, allowing for the efficient management of resources and services. It enables the isolation and allocation of resources to specific service domains, ensuring security, QoS, and resource optimisation for applications like augmented reality and autonomous vehicles, as per \textbf{Cross-Domain Cooperation with Security (Req. SEC5)} \textbf{and Resource Management and Computing as a Service (Req. OC1)}. A domain, as shown in Fig.~\ref{fig:REASONArch}, spans an entire Infrastructure plane and an entire Control and Management Plane, combining both to form a single domain.

\paragraph{Multi-access Technology Real-Time Intelligent Controller (mATRIC)}\label{par:matric}
REASON introduces a ground-breaking way to encompass current and future technologies, providing a unified approach to intelligent control and orchestration. This approach, namely Multi-access Technology Real-Time Intelligent Controller (mATRIC), aims to operate and intelligently control multiple ATs seamlessly. Such a framework is critical in applications like the one described in~\cite{cheeky_citation_2}, where distributed foundational models are used for multi-modal learning in a 6G network. As these data streams may come from different devices and technologies, providing a way of controlling the various ATs seamlessly is paramount. mATRIC plays a pivotal role in optimising and coordinating radio resources in REASON, as it is necessitated by \textbf{Resource Optimisation in Multi-ATs (Req. OC8)}. It utilises AI and ML to dynamically allocate spectrum and manage network resources across various ATs, enhancing the network's capacity and responsiveness, as per \textbf{Network Capability Advancement (Req. NPM3)}. mATRIC is responsible for the integration of various ATs necessary for \textbf{Radio ATs Integration (Req. NPM4)}, taking into account their authorisation and registration, as per \textbf{Universal Digital Identity and Policy Framework (Req. SEC2)}, and also provides monitoring capabilities as per \textbf{Multi-Domain, Multi-Tenancy Orchestration and Monitoring (Req. OC9)}.

 \begin{figure}[t]
    \centering
    \includegraphics[width=\columnwidth]{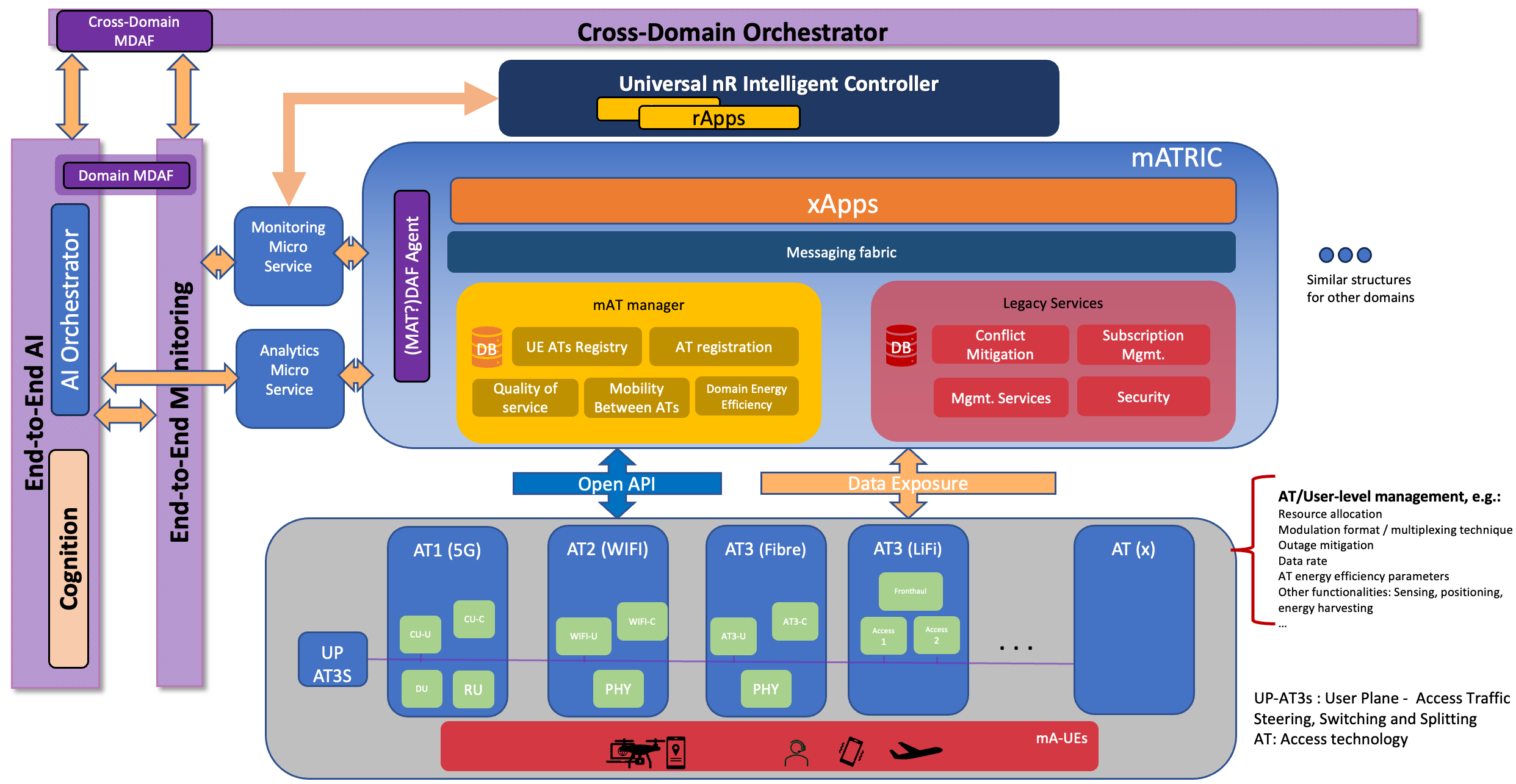}
    \caption{The Multi-Access Technology Intelligent Controller Architecture.}
    \label{fig:matric}
\end{figure}

mATRIC extends the functionality of ORAN's near-real-time RIC. As depicted in Fig.~\ref{fig:matric}, connections are established between mATRIC, the universal non-real-time intelligent controller within the edge, and analytical functions within the core network. These connections facilitate data exchange, enabling advanced analytics and decision-making capabilities. Moreover, mATRIC can expose monitoring data collected from different ATs to E2E monitoring functions for further analysis and profiling, thereby enhancing overall network visibility and performance monitoring.

Each administrative domain encompasses one mATRIC, serving one or more ATs and corresponding edge nodes. The multi-access Technology manager, a central software component, enables xApps to enforce policies across ATs and defines internal APIs to facilitate communication between various elements. Internal components of mATRIC include:
\begin{itemize}
    \item \textbf{Monitoring}: Facilitates closed-loop control by fetching performance metrics from ATs, enabling real-time tracking of critical events.
    \item \textbf{UE and AT Registration}: Supports registration, discovery, and deletion of endpoints, enabling authentication and authorization of requests.
    \item \textbf{QoS}: Manages resource allocation services to ensure QoS across ATs, automatically modifying parameters to meet SLAs.
    \item \textbf{Mobility between ATs}: Optimises handover, load balancing, and QoS between access networks based on predefined mobility configurations.
    \item \textbf{Database}: Collects and stores real-time telemetry data from multi-ATs, facilitating data extraction for MATRIC components.
\end{itemize}

The above achieve \textbf{Densification Mechanisms (Req. SPSD8)} by utilising more resources from existing available ATs or integrating new nodes and ATs. Finally, \textbf{Device and Network Mobility (Req. SPSD6)} is enabled with mATRIC, with the intelligent control of the resources across the different available technologies. The REASON architecture's innovative approach to managing and orchestrating different ATs through mATRIC enhances network intelligence, scalability, and adaptability, laying the foundation for advanced connectivity solutions in the 6G era.

\paragraph{Edge}
The edge block is a critical component in REASON, facilitating low-latency data processing and services at the network edge and integrating ATs as per \textbf{Access Technologies Integration (Req. NPM4)}. It empowers applications such as real-time IoT data analysis, virtual reality, and edge AI by reducing round-trip data transmission times, resulting in ultra-responsive and immersive user experiences. It encompasses and enhances the functionalities of ETSI Mobile Edge Computing (MEC), and an edge analytics platform expanding 3GPP NWDAF taking into account the requirement for \textbf{Intelligent Profiling (Req. OC11)} and \textbf{Data Collection for Functions Implemented (Req. TMP2)}, and, finally, implements the interfaces required by \textbf{E2E Orchestration and MEC Server Interfaces (Req. OC7)}.

\paragraph{Core}
The core block is the network's backbone, offering high-capacity, ultra-reliable connectivity and network services. It facilitates the routing of data and services across the network/domain as per \textbf{Multi-Domain, Multi-Tenancy Orchestration and Monitoring (Req. OC9)}, ensuring efficient communication between various edge and core resources. This refers to, but is not limited to, 3GPP 5G SBA Core.

\paragraph{Intelligent Intra-domain Orchestration} Intra-domain orchestration focuses on efficiently managing and coordinating resources and services within a specific network domain. This block partially addresses the requirement for \textbf{Unified E2E Orchestrator (Req. OC3)}, providing orchestration within a specific domain. An inter-domain orchestrator will collaboratively complete the E2E requirements. The intra-domain orchestration also ensures the dynamic allocation of resources and enforces policies to optimise the performance and resource utilisation within that domain. It interacts with mATRIC and Edge in the southbound and the intra-domain orchestration in the northbound. It extends the functionalities of ETSI MANO and ETSI ZSM and provides service management and Service Monitoring (SMo) capabilities and functions.

\paragraph{End-to-End Monitoring}
E2E monitoring, required as part of \textbf{Multi-Domain, Multi-Tenancy Orchestration and Monitoring (Req. OC9)} is a crucial component for ensuring the quality and reliability of services in REASON. E2E monitoring is a logical entity that is implemented with multiple functions spread across the different functions’ blocks within the different planes and domains. Each function monitors specific services, data, or functions associated with a given function block. Broadly, E2E monitoring is responsible for continuously observing and assessing network performance, from ATs to the core, providing real-time feedback for network optimisation, fault detection, and security management. This is aligned with the concepts of Observability and utilises functionality from NWDAF. The data collected should align with the \textbf{Data collection for architectural elements (Req. TMP1)} and \textbf{Data collection for functions implemented (Req. TMP2)} and make sure that does not affect the system’s operation, as per Function Execution (Req. TMP3).

\paragraph{E2E AI (End-to-End Artificial Intelligence)} E2E AI is at the heart of REASON, driving intelligent decision-making, resource allocation, and service optimisation. It harnesses the power of AI and ML to adapt to changing network conditions, automate network management tasks, and deliver personalized, context-aware services to users. The E2E AI, addressing requirements \textbf{Cognitive and AI planes (Req. ML2)}, \textbf{AI Model Lifecycle Management (Req. ML3)}, and \textbf{Cognitive functions for E2E AI (Req. ML4)} will serve to optimise system-level resources and meet the QoS requirements for embedded intelligence within network systems. The E2E AI will consist of three components: AI Orchestration, Cognition, AI Monitoring, and Model Lifecycle Management. AI monitoring refers to probes distributed across the network system to collect information about running ML models.

\paragraph{Inter-domain Orchestration}
Inter-domain orchestration is responsible for seamless coordination and communication between different network domains, ensuring the smooth flow of data and services across the entire 6G network. This is the second component required for the \textbf{Unified E2E Orchestrator (Req. OC3)}, following the intra-domain orchestration. It enables E2E Service Management (SMa) and delivery by orchestrating resources (through a Network Control \& Management (NCM) function block) and services across multiple administrative boundaries, enhancing the overall efficiency and connectivity of the network. Inter-domain orchestration is also the ingress point for user requests that are later used for Task-Oriented Networking. Similar to intra-domain orchestration, it extends the principles of ETSI MANO and ETSI ZSM with concepts of Task-Oriented Networking. Briefly, the components found in the inter-domain orchestrator are as follows:

\textbf{Policy Management (PM)}: It deals with policies to be issued by different organisations or entities, including government, regulators, businesses, and customers. The policies can influence the functions and algorithms' behaviour at both design time and run time. Policies are expected to impact the system's service management and resource and route management, integral for \textbf{Policy Framework for Mobile Access (Req. PC1)} and are also responsible for the \textbf{Compliance with Data Protection Laws (Req. PC2)}, therefore compliance functionality that should also be provisioned.

\textbf{Service Management (SMa)}: It encompasses all the higher-level business- and customer-related activities responsible for defining the services that the providers should offer to the inter-domain orchestrator for offering to its customers. These are specified according to the business intents and customer demands. Partially addresses \textbf{Hierarchical Service Control and Orchestration (Req. OC6)} and is an important functional block for all requirements in the System Performance and Service Delivery category, e.g.,  through a Service Broker (SB) it addresses \textbf{SLA Compliance (Req. SPSD1)} and \textbf{SLA Management (Req. SPSD2)}. Finally, the Task-Oriented Request Translator (TORT) is responsible for receiving high-level tasks and translating them into well-defined network KPIs that can be used intent-based.

\textbf{Service Capabilities Exchange}: Before establishing any SLA with a provider, the inter-domain orchestrator discovers the service capabilities, capacities, destination prefixes, and costs of potential services providers offer. Once the services have been defined and engineered (by domain providers), the inter-domain Service Capabilities Advertisement (SCA) block is responsible for promoting the services offered to its customers adhering to \textbf{Autonomous Configurability (Req. OC12)}. The interfaces defined for this functionality should adhere to the openness and security requirements presented, e.g., \textbf{Secure and Open Interconnectivity (Req. OC2)}.

\textbf{Cognitive Functions (CF)}: These functions allow the creation of intelligent orchestration leveraging AI/ML model-based intelligence for dynamic orchestration of services and resources and have processes across some of the inter-domain orchestrator functional blocks that can benefit from acceleration and automation in the service planning and delivery. The \textbf{AI Management Layer Integration (Req. OC10)} and \textbf{Intelligent Profiling (Req. OC11)} are fulfilled by this function block, as well as multiple AI/ML (e.g., \textbf{Cognitive functions for E2E AI (Req. ML4)}).

\textbf{Trust and Security Management (TSM)}: While a single network segment can be highly controlled, opening the network for inter-domain operations exposes it to various attacks, such as communication disruption, packet interception, and data modification. These attacks can impact critical services like remote device control. To mitigate risks, TSM should incorporate security strategies to choose the best set of Security Controls. Some examples could be \textit{Defense in Depth (DiD) mechanisms}, where multiple, unrelated layers of control are used to provide protection -- if one control is compromised, another acts as a defence, and \textit{Strength of Control (SoC)} mechanisms where a specific type of control (e.g., rigorous access control) is enhanced, requiring attackers to have more expertise, tools, and time.

\paragraph{Open APIs}
As mentioned in the previous section, one of the strong requirements for REASON is ``openness'' with a requirement for \textbf{Interfaces and network function disaggregation (Req. ON4)}, that are scalable and open as per \textbf{Scalability and Open Interfaces (Req. NPM2)}. This is manifested with the use of open APIs that interconnect the different function blocks and expose services/capabilities/data in a controlled and managed way to third parties. The primary example is OpenRAN, which opened the cellular RAN and is adopted in REASON, as per open \textbf{Platforms and products (Req. ON1)}. Similarly, there are other open standards, either for the ATs or for the upper layers, such as the management with TM Forum Open API for service enablement. The open interfaces also partially facilitate the modularity and interoperability required for \textbf{Network dis-integration (Req. ON6)}.

\bigbreak
In summary, these architectural blocks in REASON work together to provide a highly flexible, responsive, and intelligent network infrastructure capable of meeting the diverse and demanding requirements of future applications and services. They play a crucial role in shaping the future open network ecosystem's performance, scalability, and quality, enabling the delivery of groundbreaking technologies and services.

\section{Realising energy sustainability}\label{sec:energy_sustainability}

REASON, through this architecture proposal, unlocks several energy-saving capabilities that transverse all aspects of the network and contribute to the sustainability goal. Below, we present a list of how REASON will realise these capabilities, grouped into four major categories - \textbf{equipment efficiencies}, \textbf{RAN-based techniques}, \textbf{core and edge cloud}, and \textbf{artificial intelligence}. To support our claim, we will create a model that will calculate the E2E energy consumption of such a network, that is, considering RAN, edge, transmission and core, and compare it with a network based on the current technology.

\subsection{Equipment efficiencies}
\textbf{Power amplifiers} could consume up to 60\% of the energy in the radio unit. Gallium Nitride fabrication has shown significant improvements~\cite{pa_gallium} recently. REASON aims to extend on that developing novel architectures for power-efficient RF power amplifiers. \textbf{Softwarisation of the network} through network virtualisation and the abstraction of the physical hardware brings two benefits in terms of energy efficiency: 1) reducing the amount of hardware equipment (repurposing existing equipment), and 2) more efficient use of the underlying resources (utilising cloud computing)~\cite{energy_softwarisation}. REASON aims to incorporate softwarisation across the entire architecture and functional blocks for better network and application management, to allow fine-grained monitoring of the energy consumption of various components and to unlock avenues for optimisations.

\subsection{RAN-based techniques}
Service providers usually overprovision resources on network deployment, leading to resources being underutilised or idle most of the time. This could be tackled by various approaches across the entire RAN. \textbf{Cell switch-off/beam tracking} reduces energy waste by switching off carriers and beams and their associated PAs when not needed~\cite{cell_on_off}. Combined with \textbf{Advanced spectrum sharing} between the wireless services providers, utilising an architecture that promotes and facilitates sharing in an easy and tailored manner~\cite{spectrum_sharing,cheeky_citation_1} allows the resources to be managed more efficiently, achieving higher energy benefits overall.

Moving on to the different RAN technologies utilised, \textbf{High data-rate low-power (i.e., high bit/Joule) optical wireless links for x-haul and access} utilise optical transceiver devices that are generally more efficient and require less power to meet the communication KPIs compared to radio components~\cite{x_haul_optical}. \textbf{XL-MIMO/Cell-free MIMO} is an extremely large-scale MIMO (XL-MIMO) technique~\cite{xl_mimo}, which offers significantly more degrees of freedom than previous generations of MIMO-based systems. This will have many benefits, including the ability to reduce transmit power and allow base stations to tolerate even greater hardware imperfections (lower cost, more energy efficient antennas and RF chains). \textbf{Hollow core fibre} offers lower latency (similar to propagation in air), and it thus enables the physical topology of latency-sensitive networks, such as 5/6G access systems, to be less densely populated~\cite{hollowCore}. Simplifying the processing requirements of transmitted optical signals through hollow core fibres also contributes to energy savings since sophisticated and energy-hungry digital signal processing is avoided. All the above techniques are considered energy-saving measures within REASON and will be investigated as part of the project.

\subsection{Core and edge cloud (Data centre)}
The core and edge cloud utilisation and corresponding energy consumption are expected to increase due to the higher virtualisation and containerisation of the network functions as microservices (Virtual Network Functions (VNF)/Container Network Functions (CNF)). \textbf{Monitoring, analysing and reporting} relevant metrics, computing the energy consumption and reporting it in the profile of each VNF can enable optimisation mechanisms that will optimise the consumption~\cite{energySFC}. For example, REASON will investigate SFC placement with Federated Deep Reinforcement Learning (FDRL), minimising the E2E energy consumption and latency.

\textbf{Task-orientation for data filtering and reduced data requirement} is a novel approach which aims to optimise services and resources by defining well-structured policies and performing actions at the network and application level while monitoring task-specific metrics. Examples of tasks include in-network learning, streaming and image processing in the network and computation offloading, among others. One of the core principles of the task-oriented approach is to avoid using network and computing resources to perform activities that do not contribute significantly to the execution of an overall task, application, or service. \textbf{Edge-cloud optimisation} techniques apply data-thinning and task-oriented computation offloading by designing solutions on scheduling and task distribution, computation offloading and migration considering service-related, real-time, network resource metrics, and MEC-obtained metrics (e.g., UE location), as well as hosting applications and services closer to the user, which reduces unnecessary or avoidable data transmission to the core and conserve energy from the transmission and core part of the network.

\subsection{Artificial Intelligence}
REASON proposes an AI-native approach, having AI/ML as integral components of its architecture. However, AL/ML pipelines tend to be energy-intensive. REASON will investigate \textbf{monitoring approaches for the energy consumption of the E2E AI plane} and the AI-enabled components within its architecture~\cite{energy_ml}. \textbf{Hardware and software optimisations} specifically targeting MLOps (e.g., power capping strategies for reduced energy consumption~\cite{frost}) will be devised and integrated within the solutions investigated as part of REASON. Moreover, as REASON will develop many solutions in an FL-fashion, considerations around the data parallelism (e.g., as the one described in~\cite{cheeky_citation_2}) and how power efficient strategies can be optimised in a distributed fashion will also be considered.

Moreover, with mATRIC being a focal point of REASON (Sec.~\ref{par:matric}), \textbf{energy efficiency optimisation of multi-technology access by intelligent resource allocation} will be investigated. Within mATRIC, an AI-powered \textbf{joint optimisation computation and communication optimisation} emerges as a way to prove the efficacy of solutions. This approach significantly strides towards meeting the critical KPIs of latency, scalability, flexibility, and privacy preservation required for state-of-the-art edge frameworks~\cite{cheeky_citation_1} \textbf{Network service resource optimisation}, with the introduction of a multi-objective FDRL framework that finds the optimum resources needed to assign to network services while minimising the end-to-end energy consumption as well as the end-to-end latency. \textbf{RAN operation optimisation} is achieved by developing a strategy that jointly switches off carriers and performs MIMO muting – zero-touch network optimisation.

\section{Conclusion}\label{sec:conclusion}

In conclusion, this paper has provided an overview of the evolving landscape of 6G networks, highlighting its transformative potential and key technological advancements. As the world eagerly anticipates the development of 6G networks, it is evident that this next generation of communication will revolutionise various sectors through unprecedented speed, reliability, and connectivity. However, while the promise of 6G is tantalising, it also presents significant challenges, such as ensuring security, managing spectrum allocation, and addressing environmental concerns.
To this end, REASON proposes a novel open architecture based on concepts such as multi-ATs, Native-AI networking and intelligent edge/cloud service orchestration. From the physical layer, multiple ATs (wired and wireless) are considered controlled by mATRIC, a novel intelligent controller and orchestrator. A new cognitive and AI plane is introduced as an essential part of the architecture, considering the important role that data and AI/ML optimisation will play in designing and operating future networks. Openness and agility are key to REASON, enabling the network to support different workloads and services efficiently. Task-oriented networking will be the basis of REASON's network and service orchestration framework.
Moving forward, collaborative efforts among stakeholders will be essential to navigate these complexities and harness the full potential of 6G, thereby shaping a more connected and innovative future.

\section*{Acknowledgments}
This work is a contribution by Project REASON, a UK Government funded project under the Future Open Networks Research Challenge (FONRC) sponsored by the Department of Science Innovation and Technology (DSIT).

\bibliographystyle{IEEEtran}
\bibliography{bib.bib}

\begin{IEEEbiography}[{\includegraphics[width=1in,height=1.25in,clip,keepaspectratio]{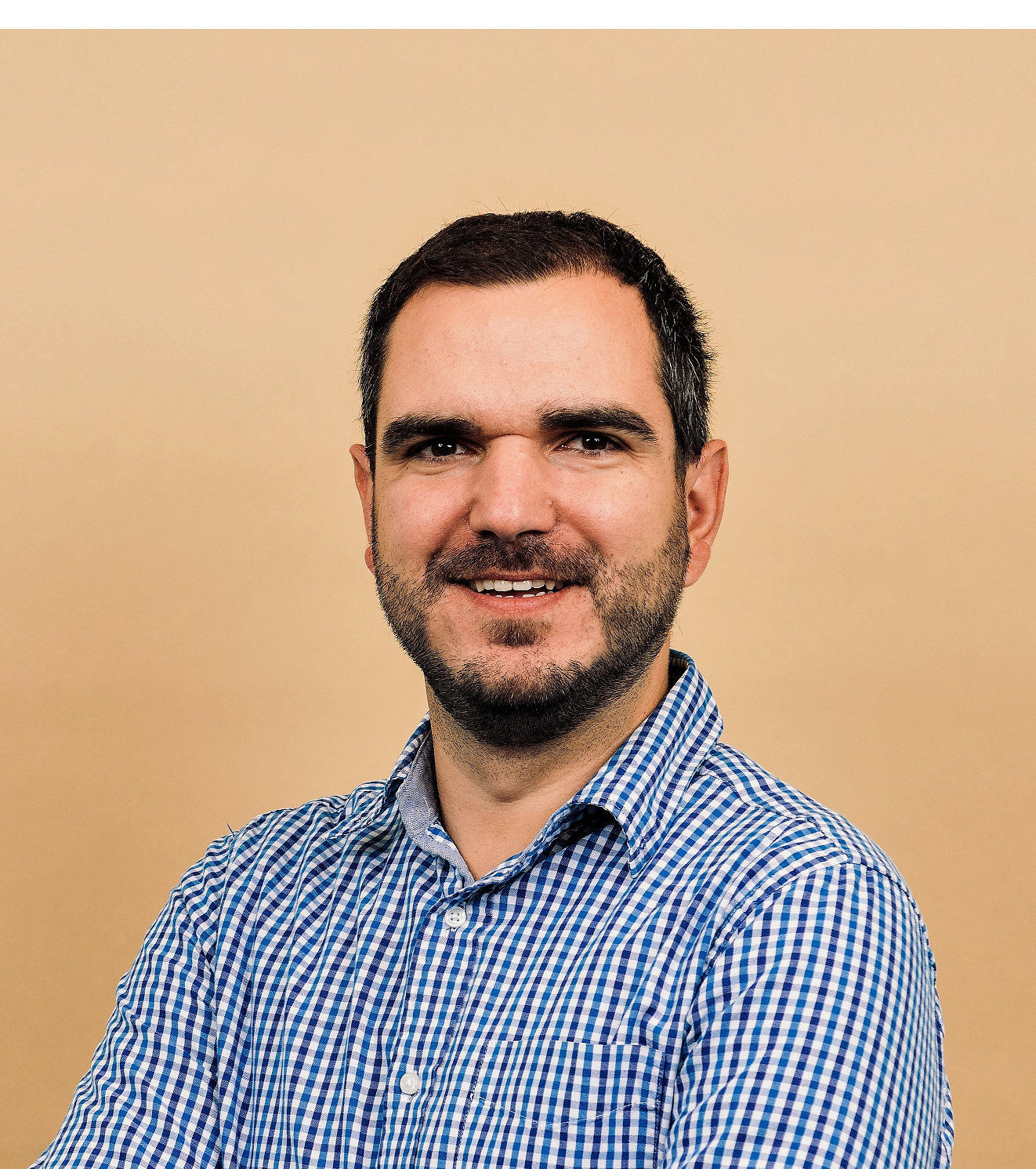}}]{Konstantinos Katsaros} has 14+ years of experience in wireless communications across industry and academia. At Digital Catapult, he provides oversight for 5G and beyond 5G system architecture and subsystem implementations. Specializing in new architectures using virtualisation and edge computing across industrial sectors. Has led research into the use of mobile edge computing to assist immersive applications and connected vehicles. He has co-authored more than 20 peer-reviewed technical articles and conference papers and has two patents on vehicular communication systems.
\end{IEEEbiography}

\begin{IEEEbiography}[{\includegraphics[width=1in,height=1.25in,clip,keepaspectratio]{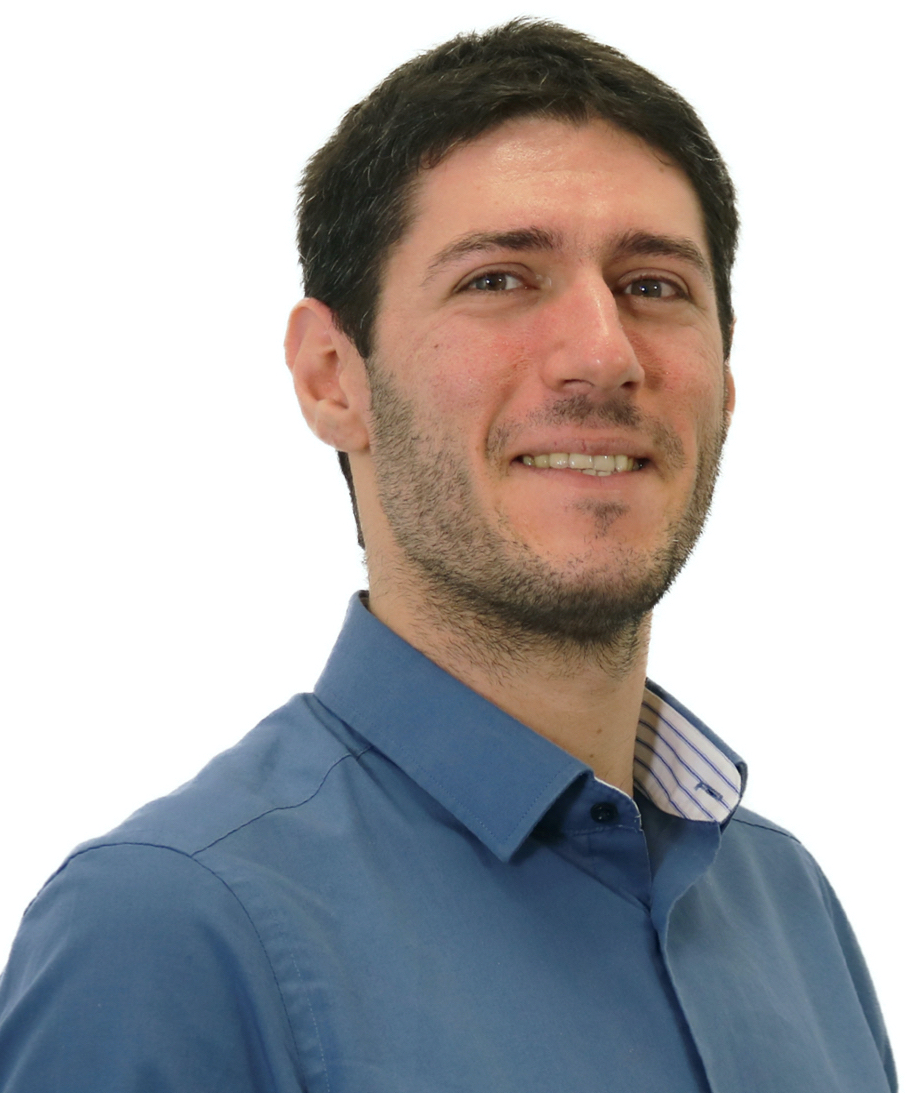}}]{Ioannis Mavromatis} is a Lead 5G/Future Networks Technologist at Digital Catapult, London, UK. He has extensive experience in 5G-and-beyond technologies, cloud-native computing, testbed deployments, wireless networking, software architecture and development. Dr Mavromatis received his PhD in ``5G Connected and Automated Vehicles'' in 2018 from the University of Bristol. He was the lead backend architect of the award-winning UMBRELLA framework, and, in the past, while working at Bristol Research and Innovation Laboratory of Toshiba Europe Ltd. and the University of Bristol, he was involved in several publicly and privately funded projects (SYNERGIA, CAVShield, BEACON-5G, FLOURISH, VENTURER, etc.). His research interests span the areas of 5G-and-beyond Communications, Cloud-native Computing, Cybersecurity, Machine Learning \& Federated Learning, and Sustainability. Dr Mavromatis received the IEEE Popularity Award from IEEE VNC 2018 and the IEEE Best Paper Award from VTC-Spring 2019.
\end{IEEEbiography}

\begin{IEEEbiography}[{\includegraphics[width=1in,height=1.25in,clip,keepaspectratio]{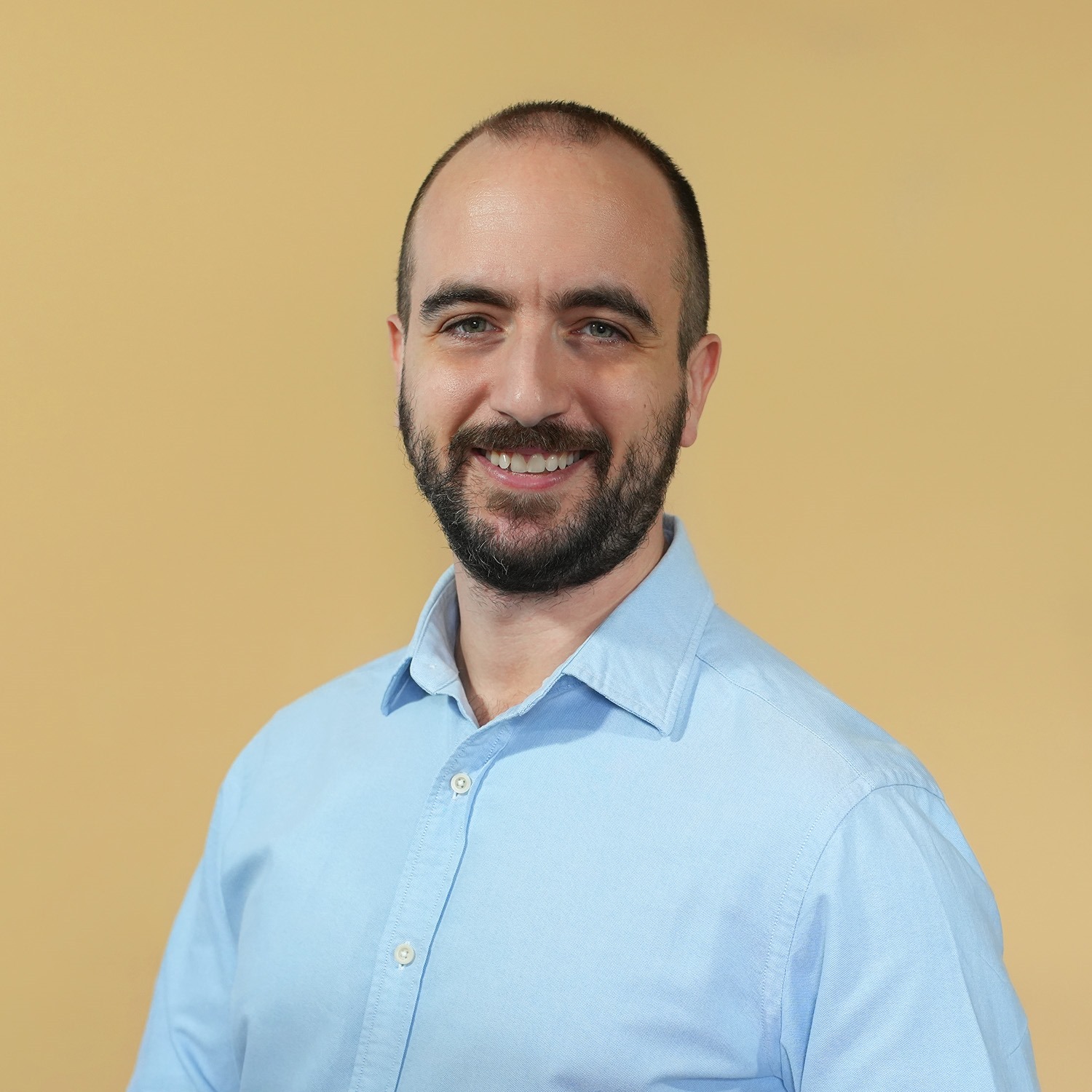}}]{Kostantinos Antonakoglou} is a Senior Future Networks Technologist at Digital Catapult where he contributes in research and development of 5G and beyond 5G systems focusing on network service orchestration. In the past he has worked as a Research Associate at King’s College London at the Centre for Telecommunications Research after completing a PhD there. Overall, he has worked on EU and EPSRC-funded projects such as 5GVICTORI, INITIATE, 5G-CAR and Primo-5G and is interested in a variety of topics including haptics and bilateral teleoperation over networks, inter-domain management and orchestration of network services as well as quantum communication and clock synchronisation.
\end{IEEEbiography}

\begin{IEEEbiography}[{\includegraphics[width=1in,height=1.25in,clip,keepaspectratio]{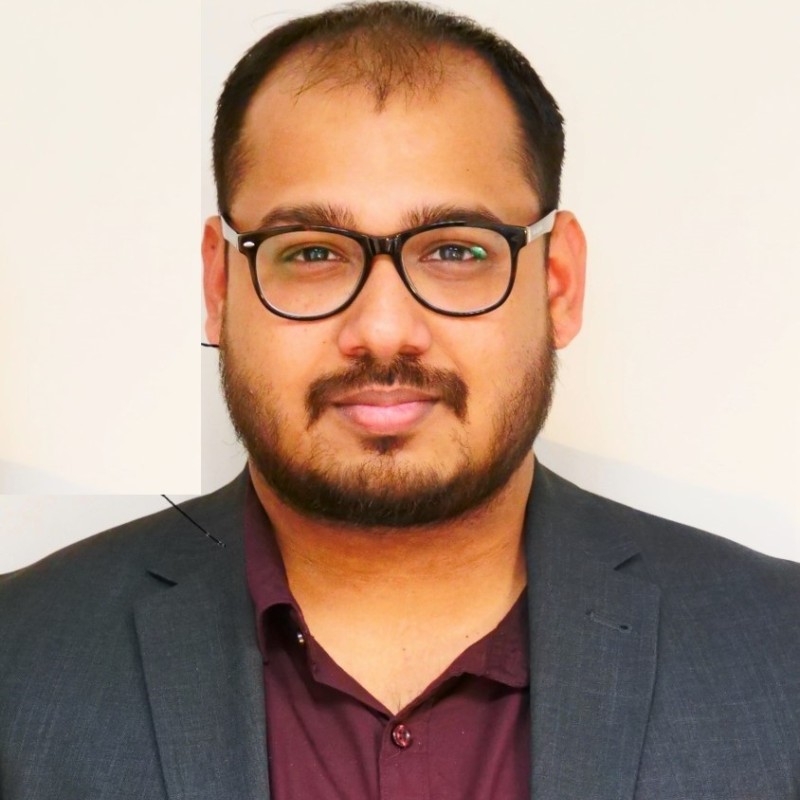}}]{Saptarshi Ghosh} is a Future Network Technologist in Orchestration at Digital Catapult with over five years of experience in B5G technologies. He received his M.E. in Software Engineering from Jadavpur University, India (2016), M.Sc. in Smart Networks from the University of the West of Scotland, UK (2017) and PhD in Computer Science and Informatics from London South Bank University, UK (2021) with GATE, Erasmus-Mundus and Marie Skłodowska-Curie Fellowships respectively. Saptarshi has contributed to the knowledge focusing on network softwarisation, automation, orchestration and intelligence through several EU/UK projects funded by Horizon 2020, Innovate UK, Erasmus+, DSTL-DASA, EPSRC, and DSIT and has obtained industrial certifications like JNCIA(DevOps) and CCNP (Enterprise Infrastructure). Formerly, he was associated with London South Bank University as a sessional lecturer and Senior Research Fellow. His domain of research includes network orchestration, knowledge-defined networking, IP routing, 6G Self-Organised-Networking and Graph Theory.

\end{IEEEbiography}

\begin{IEEEbiography}[{\includegraphics[width=1in,height=1.25in,clip,keepaspectratio]{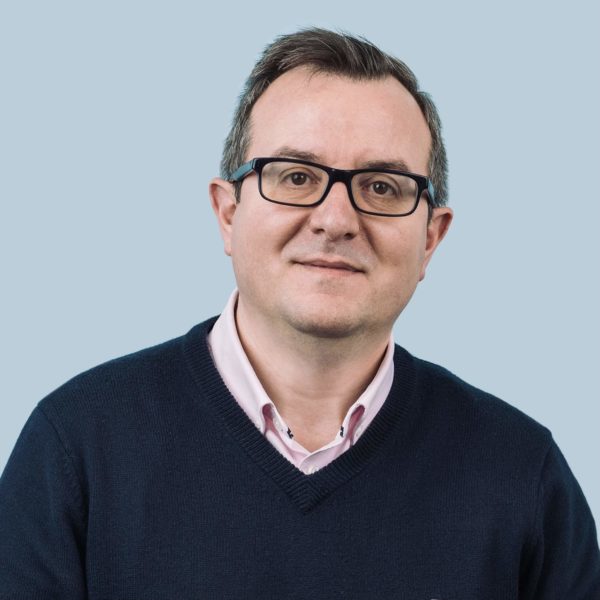}}]{Dritan Kaleshi} has 30+ years experience as a technologist and researcher in communication networks, distributed system design and interoperability. Dritan is Director of 5G Technology at Digital Catapult and Co-Director of SONIC Labs. Dritan has established and leads the Future Networks 5G and Digital Infrastructure programs in Digital Catapult, delivering technology development and innovation in advanced ICT systems and their adoption in enterprise networks in the UK. He was a founder of the SONIC Labs and of UK Telecommunication Network (UKTIN).
An acknowledged thought leader in his field, Dritan has published over 75 papers, holds three patents and has edited two international standards on interoperability, and is often invited to speak on advanced digital infrastructure and communication systems. He has represented the UK in international standardisation bodies and served on/chaired international conferences and technical committees, and provided technical advice to industry and the UK government.
His focus is on national coordination and how to best support advanced digital infrastructure and connectivity technology uptake, growing UK capabilities in telecommunications R\&D, as well as developing in Digital Catapult technical advances on new 5G/6G network architectures, network and service orchestration, edge computing and IoT support. Prior to Digital Catapult he was at University of Bristol, where he led a 15-strong research team, strategically combining network and distributed systems research with industrial collaboration, making a wide range of contributions and practical implementations in networking, interoperability, smart energy, cities and digital health.
\end{IEEEbiography}

\begin{IEEEbiography}[{\includegraphics[width=1in,height=1.25in,clip,keepaspectratio]{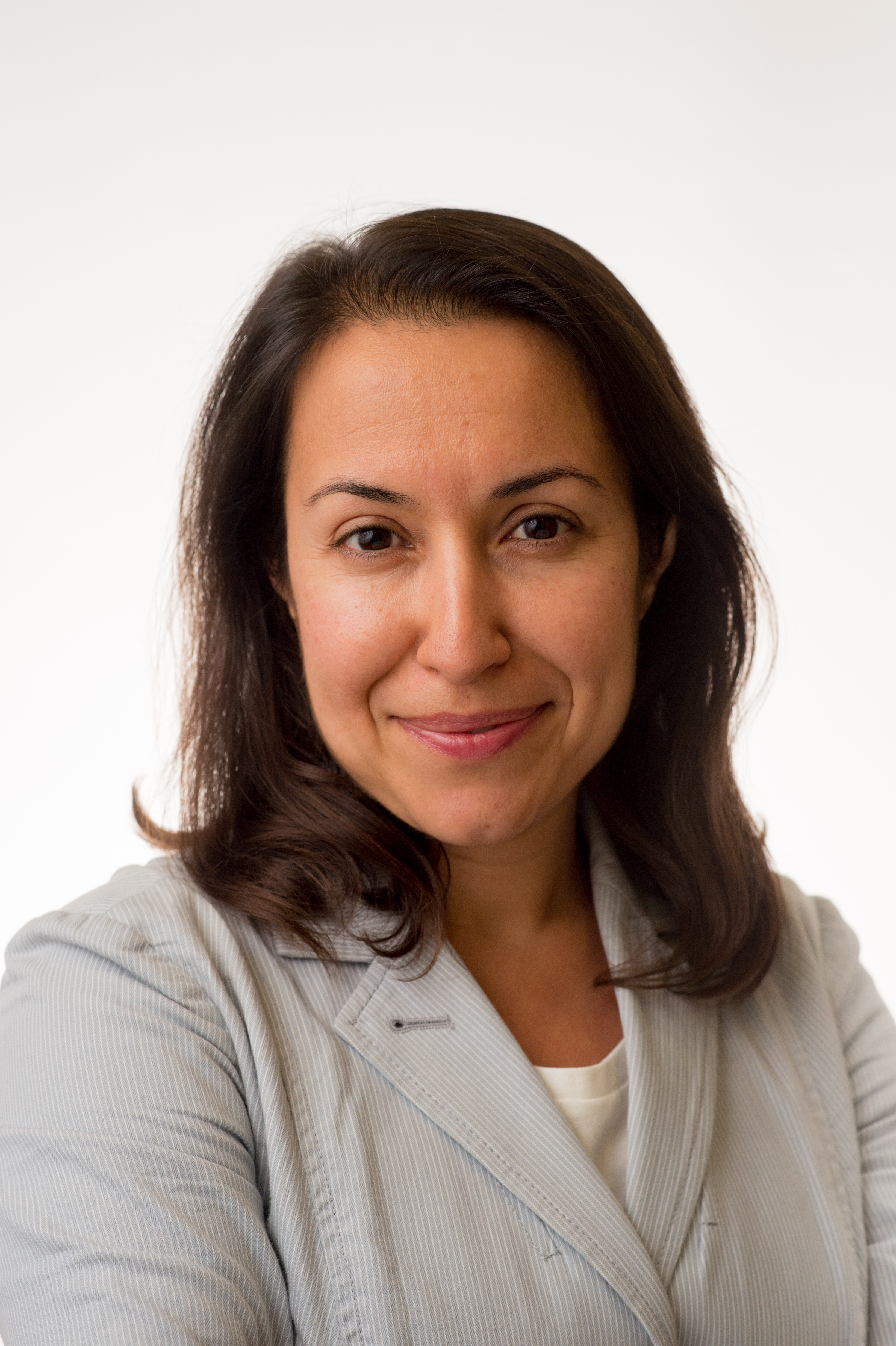}}]{Toktam Mahmoodi} is a Professor of Communication Engineering and head of the Centre for Telecommunications Research (CTR) in the Department of Engineering at King’s College London. She received B.Sc. degree in electrical engineering from Sharif University of Technology, Iran, in 2002, and the Ph.D. degree in telecommunications from King’s College London, U.K, in 2009. She was a Visiting Research Scientist with F5 Networks, San Jose, CA, in 2013, a Post-Doctoral Research Associate with the ISN Group, Electrical and Electronic Engineering Department, Imperial College, 2010 to 2011, and a Mobile VCE Researcher, 2006 to 2009. She has also worked in the mobile and personal communications industry from 2002 to 2006. Her research interests include network intelligence, and mission critical networking, with impact in healthcare, automotive, smart cities and emergency services.
\end{IEEEbiography}

\begin{IEEEbiography}[{\includegraphics[width=1in,height=1.25in,clip,keepaspectratio]{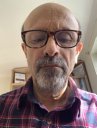}}]{Hamid Asgari} has been with Thales UK Research, Technology, Solution \& Innovation (RTSI) since 1996 and is a Thales Expert. He is also a Visiting Professor at King’s College London since 2013. He is currently leading both Future Network and Verification \&Validation research activities at RTSI. Hamid in highly experienced and been has directing research, leading R\&D teams internally at Thales and externally in national and European collaborative projects since year 2000. Hamid has also been liaising and coordinating external collaborations with industry and academia. He has been very active in transferring technology resulted from R\&D, linking the technologies with application areas, and contributing towards their use. He has a proven track record of publications in highly-valued journals and peer-reviewed conferences. He is a senior member of IEEE and has been involved in Technical Committees and standardisation Working Groups.
\end{IEEEbiography}

\begin{IEEEbiography}[{\includegraphics[width=1in,height=1.25in,clip,keepaspectratio]{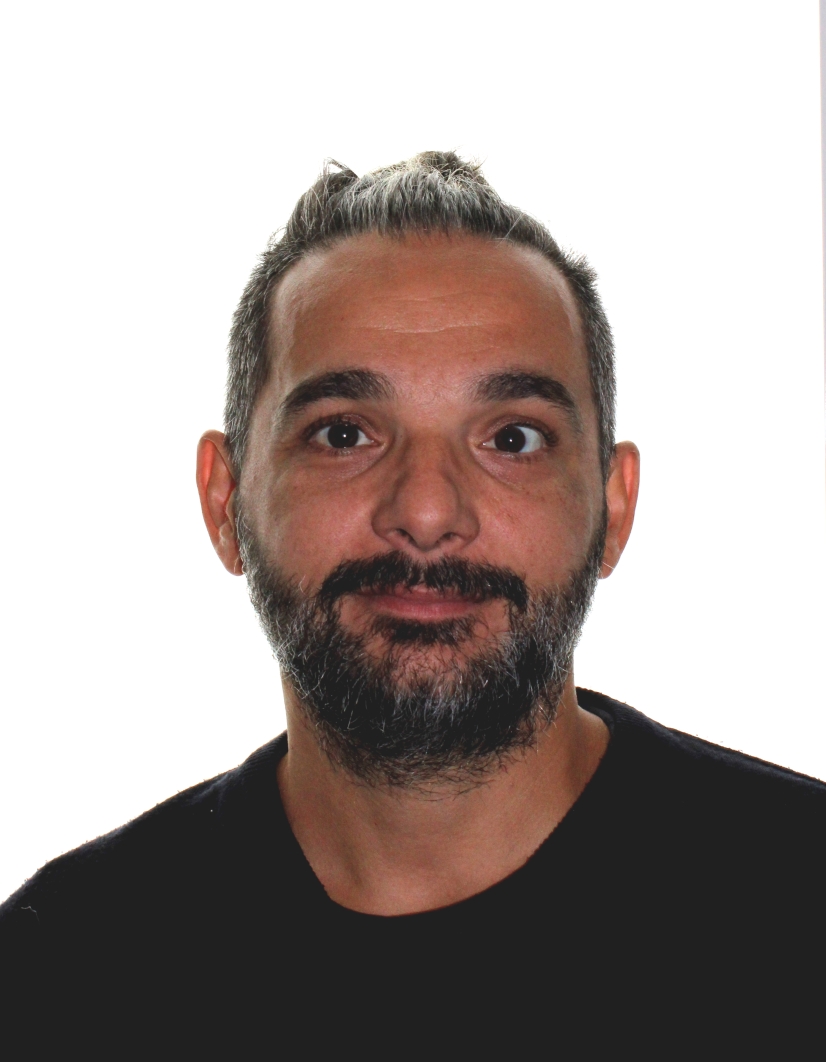}}]{Anastasios Karousos} brings over 15 years of experience in the wireless industry. He has managed and delivered a number of projects around wireless connectivity, its challenges and its impact, and he specialises in creating complex models for understanding wireless systems, from the network feasibility perspective, technology evolution and the techno-economic view. As Head of Network Modelling and Simulations, Anastasios works with Real Wireless experts across all client projects matching their needs to existing capabilities or creating new capabilities when required, through strategic market assessments. He defines platform strategy to enable viable business performance and implements a business/product line management approach to the tools developed in the company.
\end{IEEEbiography}

\begin{IEEEbiography}[{\includegraphics[width=1in,height=1.25in,clip,keepaspectratio]{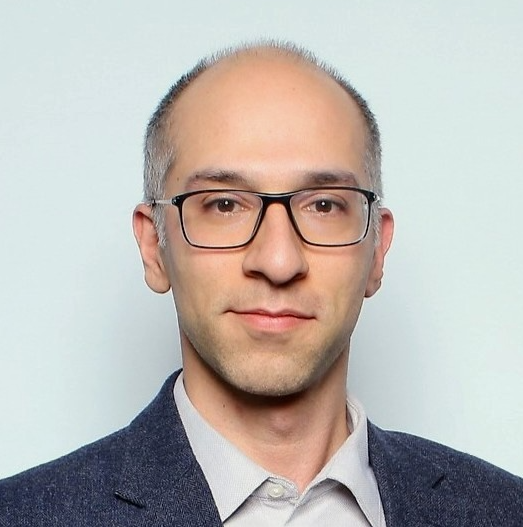}}]{Iman Tavakkolnia} is an Assistant Professor at the Electrical Engineering division at the University of Cambridge, UK. His research focuses on developing a fundamental understanding of the energy-efficiency of current and future telecommunication systems and lies on the frontier of communication theory, advanced materials, signal processing, and optical communications. He also works on low-complexity optical space communication systems for small satellites. He is a co-investigator on two EPSRC Future Communication Hubs in the UK (TITAN and HASC) as well as the project REASON under the Future Open Networks Research Challenge grant funded by the UK’s Department of Science, Innovation and Technology. He was a working group member of the European COST Action, CA19111 NEWFOCUS, and an associate editor of the IEEE Communications Letters. He has been a co-chair of the optical wireless communication workshops in WCNC 2023, WCNC 2024, and GLOBECOM 2024, a local organizing committee member of ECOC 2023, and TPC member of several workshops and conferences. Iman Tavakkolnia obtained his PhD degree from the University of Edinburgh in 2018. He was a research associate at the University of Edinburgh until 2020 and then at the University of Strathclyde until September 2021, before being appointed as the Strathclyde Chancellor's Fellow (Lecturer) until February 2024.
\end{IEEEbiography}

\begin{IEEEbiography}[{\includegraphics[width=1in,height=1.25in,clip,keepaspectratio]{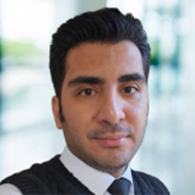}}]{Hossein Safi} is currently a Post-Doctoral Research Associate at the LiFi Research Centre, Department of Engineering, University of Cambridge, UK. Before his tenure at Cambridge University, he served as a Post-Doctoral Research Associate at the University of Strathclyde, Glasgow, UK, from October 2023 to April 2024. His research interests include optical wireless communication and statistical signal processing, with a focus on non-terrestrial platforms.
\end{IEEEbiography}

\begin{IEEEbiography}[{\includegraphics[width=1in,height=1.25in,clip,keepaspectratio]{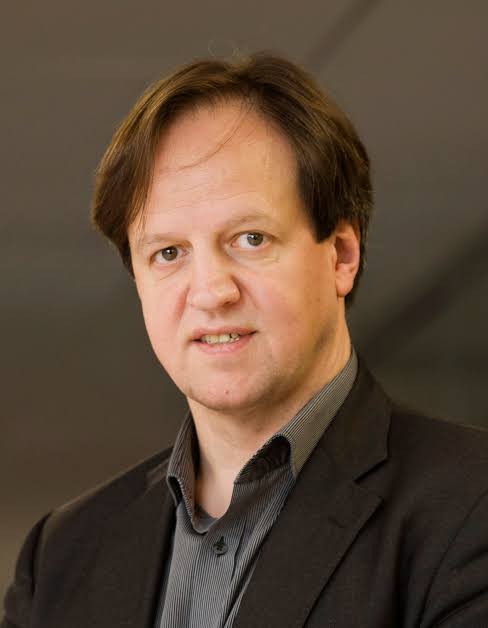}}]{Harald Hass} (Fellow, IEEE) received the Ph.D. degree from the University of Edinburgh, Edinburgh, U.K., in 2001. He currently is the Van Eck Professor of Engineering at the University of Cambridge and the Director of the LiFi Research and Development Centre. He founded pureLiFi Ltd. and holds the position of Chief Scientific Officer (CSO). His most recent research interests lie in integrating physics and communication theory to design secure, highspeed wireless multi-user access networks and distributed x-haul networks utilising the optical spectrum toward building net-zero and pervasive wireless networks. He has coauthored more than 650 conference and journal papers with more than 55,000 citations and holds more than 45 patents. He has been listed as highly cited researcher by Clarivate/Web of Science since 2017. Prof. Haas has delivered two TED talks and one TEDx talk which have been watched online more than 5.5 million times. In 2016, he was the recipient of the Outstanding Achievement Award from the International Solid State Lighting Alliance. Prof Haas was awarded a Royal Society Wolfson Research Merit Award in 2017. In 2019 he received the IEEE Vehicular Society James Evans Avant Garde Award and the Enginuity The Connect Places Innovation Award in 2021. In 2022 he was the recipient of a Humboldt Research Award for his research achievements. In 2023, Prof Haas among the three shortlisted candidates for a European Inventor Award. He is a Fellow of the IEEE, a Fellow of the Royal Academy of Engineering (RAEng), a Fellow of the Royal Society of Edinburgh (RSE) and a Fellow of the Institution of Engineering and Technology (IET).
\end{IEEEbiography}

\begin{IEEEbiography}[{\includegraphics[width=1in,height=1.25in,clip,keepaspectratio]{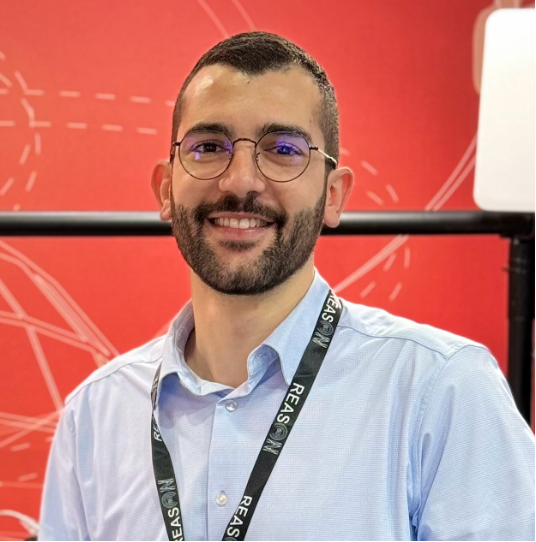}}]{Constantinos Vrontos}is an enthusiastic telecommunications engineer with a Ph.D. in "mmWave Spectrum Sharing Between Mobile Operators" from the University of Bristol, UK. During his tenure as a Senior Research Associate at the Smart Internet Lab, University of Bristol, he honed a comprehensive understanding of networking and telecommunications. In this role, he conceptualized and implemented E2E solutions encompassing 4G, 5G, and emerging technologies. His contributions extended to crafting Cloud architecture designs, employing various virtualization technologies such as Openstack and Kubernetes. Notably, he contributed to the development of multi-RAT (Radio Access Technology) Customer Premises Equipment (CPE) devices, while also managing a private 5G network with diverse multi-access technologies deployed across numerous sites within Bristol city center. In his current role as a Telco Specialist at the Smart Internet Lab, Constantinos Vrontos is steering efforts to pioneer the first research platform for 6G. This involves spearheading the delivery of a national testbed as part of the JOINER project, contributing to the evolution of telecommunications.
\end{IEEEbiography}

\begin{IEEEbiography}[{\includegraphics[width=1in,height=1.25in,clip,keepaspectratio]{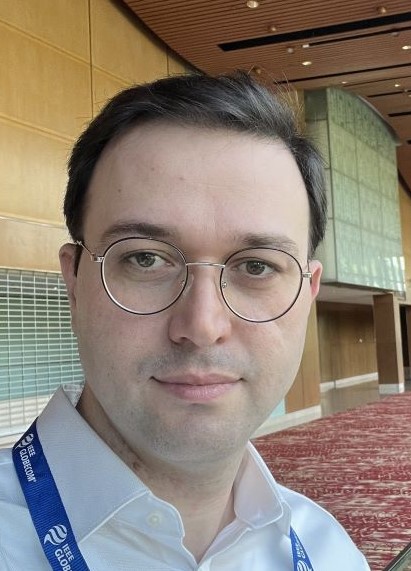}}]{Amin Emani} is a Telco Specialist - Systems and Solutions Architect at the Smart Internet Lab, the University of Bristol. Holding an M.Sc. degree in Electrical and Electronics Engineering, he possesses a wealth of technical expertise and practical skills, coupled with strong interpersonal abilities. This enabled him to make significant contributions to a diverse array of European and UK national research projects, including notable ones like 5GCLARITY, 5GVICTORI, 5GASP, REASON, and 5G Smart Tourism. Currently, he leads the technical team for the JOINER project, which will pave the way for the first 6G national scale testbed. With a robust background spanning 19 years in both academia and industry, he stands as a proficient IT professional. His areas of research focus on the realms of beyond 5G, and 6G networks, O-RAN, cloud computing and hybrid telco clouds, SD-WAN, SDN, NFV, MANO, and their applications in NextG of telecommunication technologies. \end{IEEEbiography}

\begin{IEEEbiography}[{\includegraphics[width=1in,height=1.25in,clip,keepaspectratio]{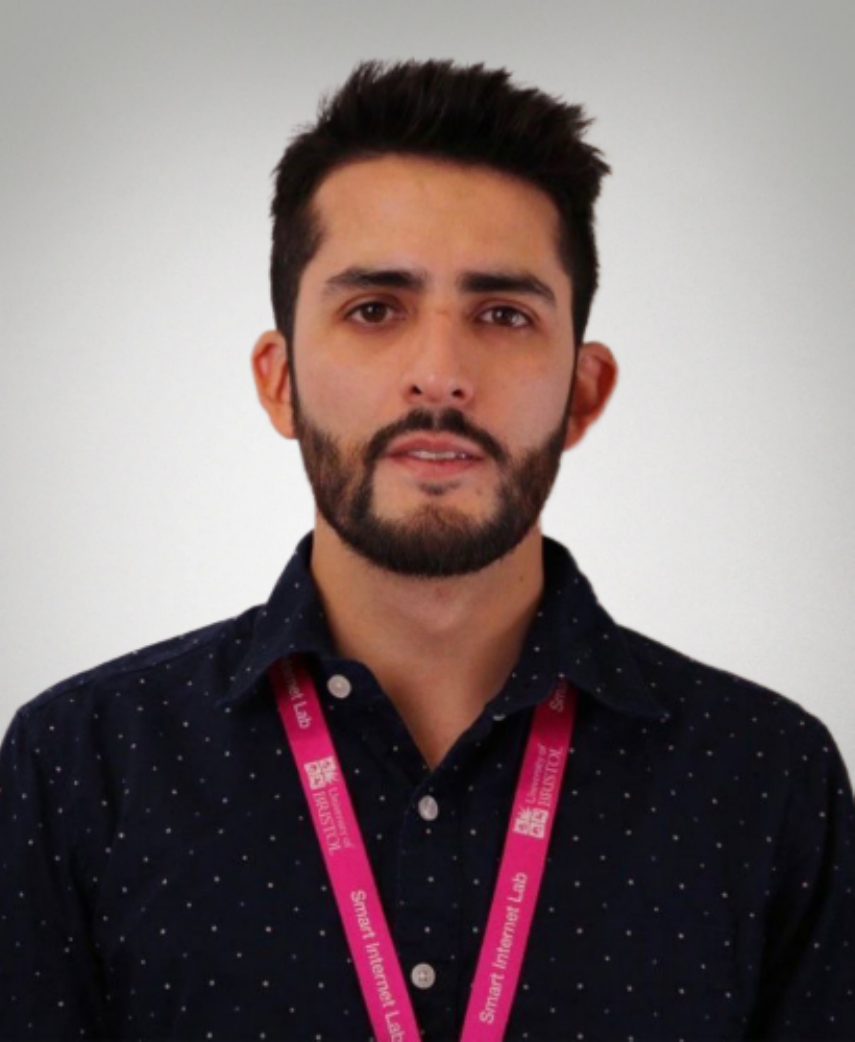}}]{Juan Marcelo Parra-Ullauri} earned his bachelor's degree in Electronics and Telecommunications Engineering from the Universidad de Cuenca, Ecuador in 2017. In 2022, he is PhD in Computer Science from Aston University in the UK. He currently works as a Senior Research Associate at the Smart Internet Lab, focusing on researching AI/ML for Networked Systems. His research interests include the Internet of Things, Distributed Machine Learning, Cloud Computing, Explainability in Autonomous Systems, and Data Engineering. He is the author of more than 20 articles. He actively collaborates on academy-industry projects, contributes to open-source FL projects, and leads research activities within EU and UK-funded projects.
\end{IEEEbiography}

\begin{IEEEbiography}[{\includegraphics[width=1in,height=1.25in,clip,keepaspectratio]{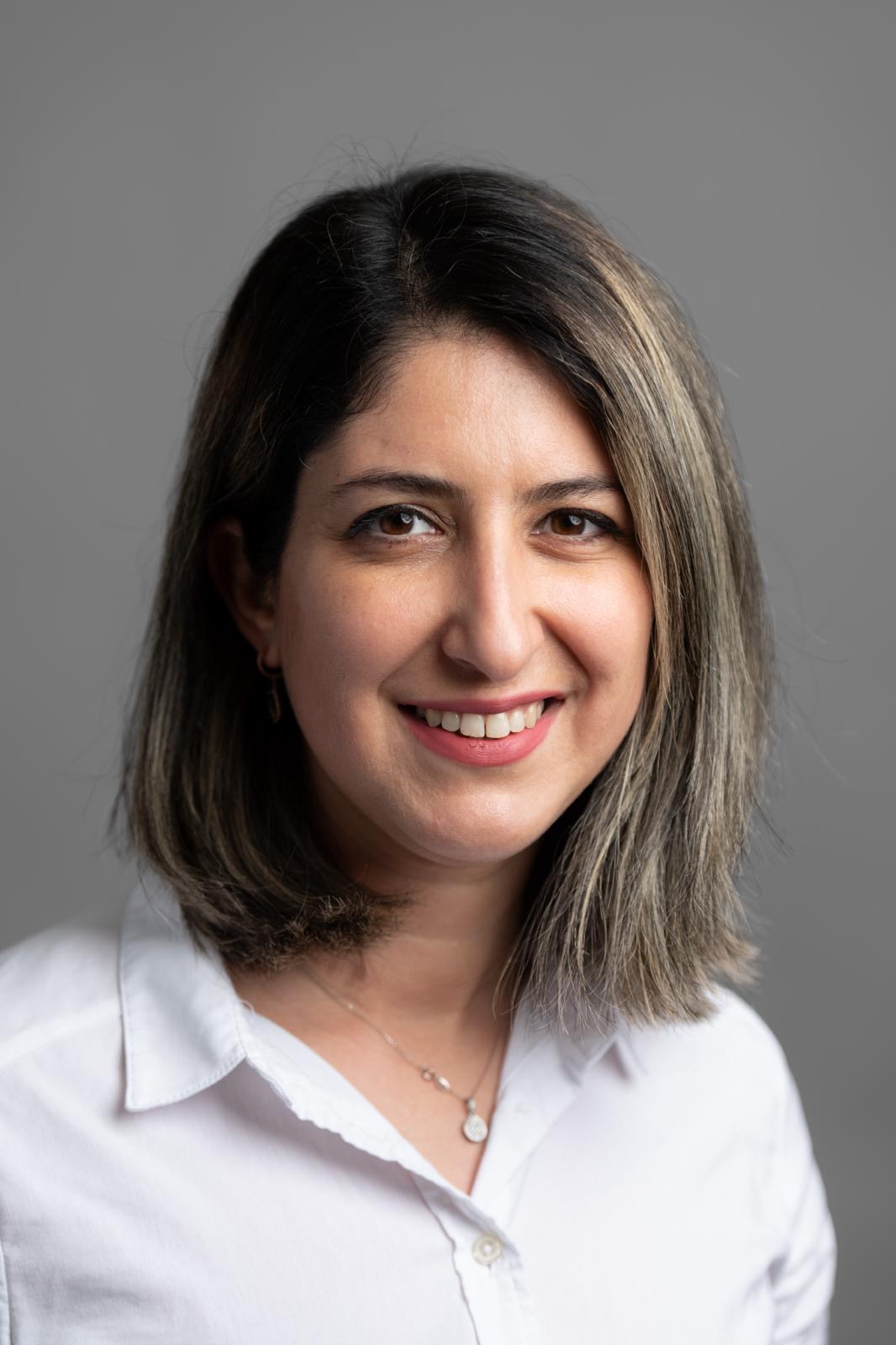}}]{Shadi Moazzeni} is a Lecturer in Networks at the University of Bristol, Bristol, UK, and a member of the Smart Internet Lab and Bristol Digital Futures Institute (BDFI). She previously served as a Senior Research Associate and later as a Research Fellow at the University of Bristol. She is the project investigator for the Innovate-UK funded nCOMM+ project and has collaborated as Co-Investigator on various UK projects such as TITAN, REASON, and UK-TIN. Additionally, she was the cluster lead researcher for the EU Horizon 2020 5G-VICTORI project. She received her M.Sc. degree in Computer Architecture Engineering from Amirkabir University of Technology (Tehran Polytechnic), Tehran, Iran, in 2010, and her PhD in Computer Architecture Engineering from the University of Isfahan, Iran, in 2018. From July 2016 to February 2017, she was a visiting PhD researcher at the University of Bologna, Italy.

She specialises in AI-Native 6G networks, focusing on the orchestration and optimisation of next-generation intelligent networks, multi-access edge computing, and intelligent multi-objective profiling towards zero-touch network and service management. Her work aims to enhance future network performance and develop innovative, human-centric networking solutions.
\end{IEEEbiography}

\begin{IEEEbiography}[{\includegraphics[width=1in,height=1.25in,clip,keepaspectratio]{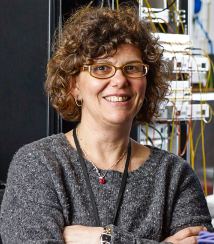}}]{Dimitra Simeonidou} is a Full Professor at the University of Bristol, the Co-Director of the Bristol Digital Futures Institute and the Director of Smart Internet Lab. Her research is focusing on the fields of high-performance networks, programmable networks, Future Internet, wireless-optical convergence, 5G/6G and smart city infrastructures. In the past few years, she is increasingly working with Social Sciences and Humanities on topics of climate change and digital transformation for society and businesses. Dimitra has been the Technical Architect and the CTO of the smart city project Bristol Is Open. She is currently leading the Bristol City/Region 5G and Open RAN pilots.
Dimitra is a member of the UK Government Supply Chain Diversification Advisory Council, a founding member of the UK Telecoms Innovation Network and member of the OFCOM Spectrum Advisory Board. She has led major research projects funded by UK Government and the EC. She is currently coordinating the DSIT REASON project developing  blueprint architectures and technologies for 6G and the EPSRC JOINER project, aiming to establish a UK-wide experimentation platform for 6G research and innovation.

She is the author and co-author of over 700 publications, numerous patents and several contributions to standards.
She has been co-founder of three spin-out companies developing solutions for connected smart infrastructures.
Dimitra is a Fellow of the Royal Academy of Engineering (FREng), a Fellow of the IEEE (FIEEE), Fellow of WWRF and member of UKCRC.

\end{IEEEbiography}

\EOD

\end{document}